\documentclass[amsmath,amssymb,twocolumn,prl,longbibliography,superscriptaddress]{revtex4-1}
\usepackage{graphicx,bm,times}
\usepackage[utf8]{inputenc}
\usepackage{braket}
\usepackage{amsmath}

\usepackage[breaklinks]{hyperref}
\hypersetup{
    bookmarks=false,
    pdfstartview={FitH},
    colorlinks=true,
    linkcolor=blue,
    citecolor=blue,
    urlcolor=blue
}

\newcommand{\nocontentsline}[3]{}
\newcommand{\tocless}[2]{\bgroup\let\addcontentsline=\nocontentsline#1{#2}\egroup}

\begin{document}

\title{Quantum Confinement Induced Metal-Insulator Transition in Strongly
Correlated Quantum Wells of SrVO$_3$ Superlattices}

\author{A.~D.~N.~James}
\affiliation{H.\ H.\ Wills Physics Laboratory,
University of Bristol, Tyndall Avenue, Bristol, BS8 1TL, United Kingdom}

\author{M.~Aichhorn}
\affiliation{Institute of Theoretical and Computational Physics, TU Graz, NAWI
Graz, Petersgasse 16, 8010 Graz, Austria}

\author{J.~Laverock}
\affiliation{H.\ H.\ Wills Physics Laboratory,
University of Bristol, Tyndall Avenue, Bristol, BS8 1TL, United Kingdom}

\date{\today}

\begin{abstract}
Dynamical mean-field theory (DMFT) has been employed in conjunction with density
functional theory (DFT+DMFT) to investigate the metal-insulator transition (MIT)
of strongly correlated $3d$ electrons due to quantum confinement. We shed new
light on the microscopic mechanism of the MIT and previously reported anomalous
subband mass enhancement, both of which arise as a direct consequence of the
quantization of V~$xz(yz)$ states in the SrVO$_3$ layers.
We therefore show that quantum confinement can sensitively tune the strength of
electron correlations, leading the way to applying such approaches in other
correlated materials.
\end{abstract}

\maketitle

In strongly correlated quantum materials, the interaction energy between
electrons is comparable to their kinetic energy, leading to many-body behavior
and the emergence of qualitatively new phenomena \cite{tokura2017}. For example,
transition metal oxides host properties as diverse as colossal
magnetoresistance, high-temperature superconductivity, and Mott insulating
phases, each of which have huge potential for future device and technology
applications \cite{brahlek2017,*mannhart2010,kotliar2006}.  More recently,
substantial advances in the quality and control of layer-by-layer growth methods
have facilitated designed transition metal oxide heterostructures and
superlattices (SL), often focusing on emergence at interfaces and/or surfaces
\cite{izumi2001,*ohtomo2004,*hwang2012,*chakhalian2014}.  Here, we show that
correlated electronic behavior may be delicately tuned by quantum confinement,
which narrows the effective bandwidth of the correlated quantum well (QW)
subbands, and establishes another means to tune physical properties to suit
applications. We illustrate this by driving the prototypical correlated metal,
SrVO$_3$, through a metal-insulator transition (MIT) to a Mott insulating phase,
in excellent quantitative agreement with the experimental spectral function
extracted from recent spectroscopic measurements \cite{laverock2017}.

In the bulk, SrVO$_3$ is a well-characterized correlated metal
\cite{nekrasov2005}. Sharp $3d$ quasiparticle (QP) bands at low excitation
energies lead to a well-defined Fermi surface \cite{yoshida2010,*aizaki2012},
while localized states form incoherent Hubbard sidebands at an energy scale
comparable with the Coulomb repulsion parameter, $U$
\cite{sekiyama2004,pen1999,*laverock2013b}. Together, these yield the familiar
three-peaked spectral function \cite{georges1996,kotliar2006}. While density
functional theory (DFT) often adequately describes QP states (once
renormalization is accounted for), it is
not capable of capturing the many-body behavior, e.g.\ Hubbard sidebands are
completely absent. Dynamical mean-field theory (DMFT), on the other hand, is
able to describe all of the on-site local correlations
\cite{georges1996,kotliar2006}, and has been well-tested on SrVO$_3$ with
very good results, including the energetics and spectral weight of Hubbard
sidebands and QP renormalization
\cite{sekiyama2004,nekrasov2006,kotliar2006,byczuk2007,*tomczak2012}.

The importance of electron correlations in a system may be gauged in terms of
the ratio, $U/W$, where $W$ is the bandwidth. This ratio is known to be
significantly enhanced at the SrVO$_3$ surface
\cite{maiti2001,*liebsch2003,*laverock2015b,ishida2006}, and in few layer
systems \cite{yoshimatsu2010,zhong2015}. However, such systems have reduced
coordination at the surface, which is often complicated by reconstruction and
relaxation \cite{ishida2006}. For example, an insulating phase has been observed
in SrVO$_3$ thin films for thicknesses below approximately 6 unit cells, which
was attributed to dimensional crossover due to reduced coordination
\cite{yoshimatsu2010,gu2014}. In bilayer SrVO$_3$, a DFT+DMFT study has
established that a Mott insulator emerges due to crystal field (CF) effects
\cite{zhong2015}; similar CF-induced insulating phases are predicted in other
vanadates \cite{bhandary2016,schuler2018,hampel2019,beck2018,*sclauzero2016}.
Although QW structures are well-known in semiconducting \cite{colakerol2006} and
free electron-like \cite{chiang2000} materials, their application to correlated
$3d$ metals has only recently been realized in a select few transition metal
oxides \cite{liu2012,*jeong2020,kawasaki2018}, including in few-layer SrVO$_3$
thin films \cite{yoshimatsu2011}. However, the precise nature of these states,
including their unusual subband renormalization
\cite{yoshimatsu2011,kobayashi2015}, is not yet well understood, although this
is essential in order to exploit their properties.

In this Letter, we show that quantum confinement can be employed to precisely
control strongly correlated electron behavior. We reveal the microscopic
mechanism involved using a combination of DFT, DFT+DMFT and tight-binding models
applied to the prototypical correlated oxide SrVO$_3$, which we embed within
layers of SrTiO$_3$ in a SL for direct comparison with previous experimental
results \cite{laverock2017}. Our results accurately reproduce the trends of the
experimental data, and reveal that the microscopic mechanism for the MIT in
SrVO$_3$/SrTiO$_3$ SLs is due to quantum confinement. Our
results also shed light on previous observations of anomalous mass enhancement
in SrVO$_3$ QWs \cite{yoshimatsu2011,kobayashi2015}, which can be
naturally explained as consequences of quantization. These results underline the
efficacy of SL engineering in tuning strongly correlated behavior, and leads the way to harnessing functional correlated electron properties in other
materials \cite{adler2019}.

\begin{figure}[t!]
\includegraphics[width=1.0\linewidth]{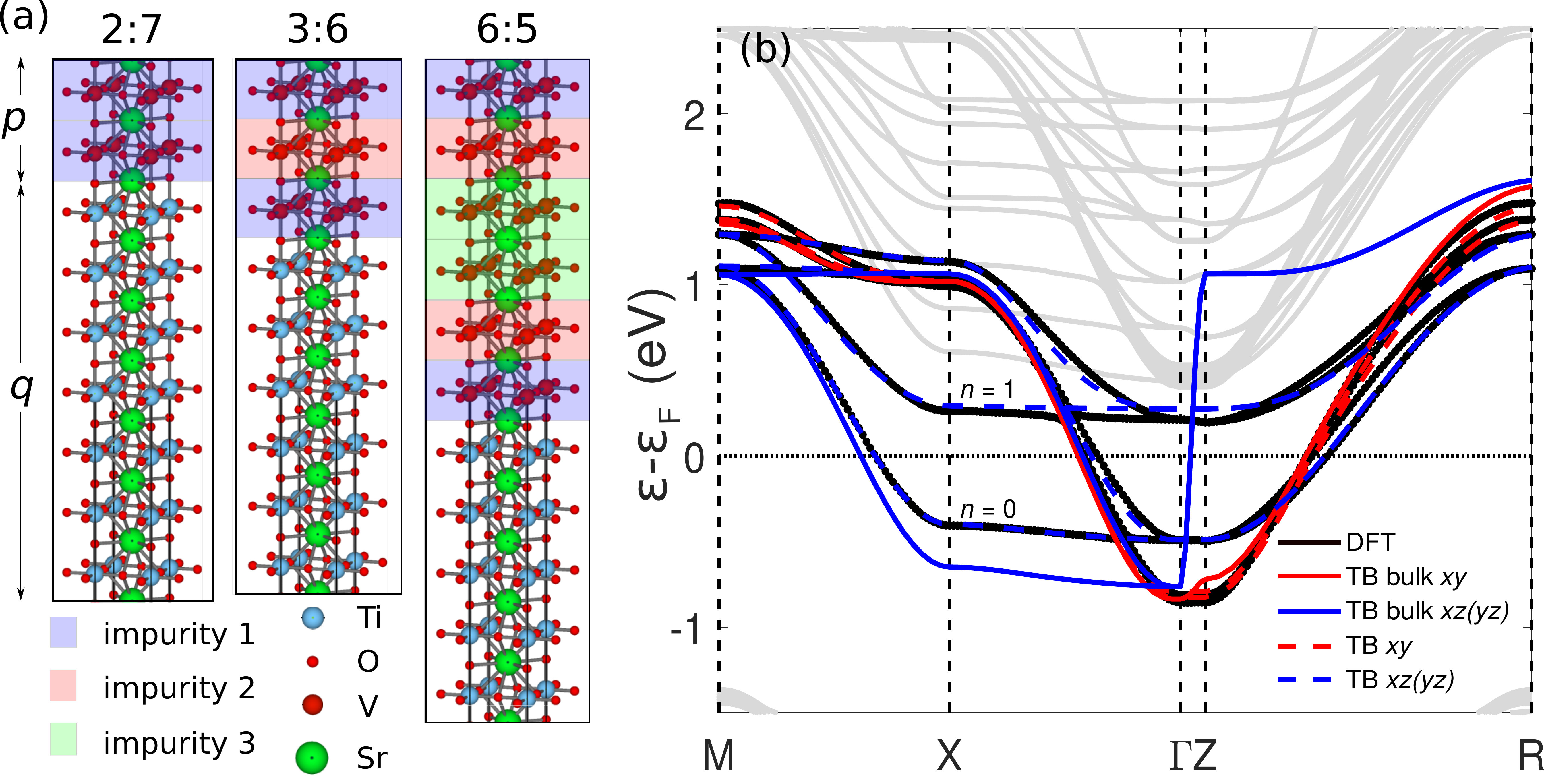}
\caption{(a) Schematic diagram of the SL structures, where $p$ and $q$
represent the number of SrVO$_3$ and SrTiO$_3$ layers, respectively. The SL
unit cells used in the DFT
are indicated by the black boxes. The color coded layers illustrate the
impurities used in the DMFT calculation. (b) Quantized tight-binding bands for
the 2:7 SL. The colored dashed lines indicate the intrinsic bulk bands from
which the quantized bands are derived (see text). The discrete energies from the
DFT calculation are shown for comparison (both V and Ti $t_{2g}$ bands are
given). The grey bands are non-V $t_{2g}$.}
\label{f:slstruc}
\end{figure}

DFT calculations were performed using the all-electron full potential augmented
plane wave {\sc Elk} package \cite{elk} within the local density approximation
(LDA). The results are in excellent agreement with previous pseudopotential
calculations \cite{laverock2017}. The structures
of the (SrVO$_3$)$_p$/(SrTiO$_3$)$_q$
SLs are shown in Fig.~\ref{f:slstruc}(a) for the three SLs investigated here,
with $p$:$q = 2$:7, 3:6 and 6:5. These were chosen for direct comparison with
previous experimental results \cite{gu2018,laverock2017}.  Experimental lattice
parameters were used: the in-plane parameters were those of the
(LaAlO$_3$)$_{0.3}$(Sr$_2$TaAlO$_6$)$_{0.7}$ (LSAT) substrate, $a = b =
3.868$~{\AA}, and out-of-plane parameters were $c_{2:7} = 4.00$~{\AA},
$c_{3:6} = 3.97$~{\AA} and
$c_{6:5} = 3.92$~{\AA}. Even in the absence of the SL heterostructure,
the different in- and out-of-plane lattice parameters weakly break the V
$t_{2g}$ degeneracy into $3d$ $xy$ and $xz(yz)$ orbitals.

In order to interpret our results, the LDA results for each SL were
parameterized using a quantized Bohr-Sommerfeld tight-binding (QTB) model
\cite{sup,chiang2000}. In this model, the {\em shape} of the TB bands was fixed
to that of bulk SrVO$_3$, and the free parameters represent the band centers,
bandwidths and quantization parameters. In this way, CF
splitting and band narrowing between $xy$ and $xz(yz)$ orbitals can be fully
captured. The results of fitting this model to the full {\sc Elk} calculation is
shown in Fig.\ \ref{f:slstruc}(b) for the 2:7 SL, demonstrating the excellent
agreement between the two. The quantized nature of the V $xz(yz)$ orbitals is
clearly evident, leading to two subbands ($n = 0,1$) originating from the two
SrVO$_3$ layers. In
Fig.\ \ref{f:slstruc}(b), we also show the ``intrinsic'' TB bands, which
correspond to bulk-like bands before the quantization conditions are applied and
represent the intrinsic 3D electronic structure from which the
QW states emerge.

\begin{figure}[t!]
\centerline{\includegraphics[width=1.0\linewidth]{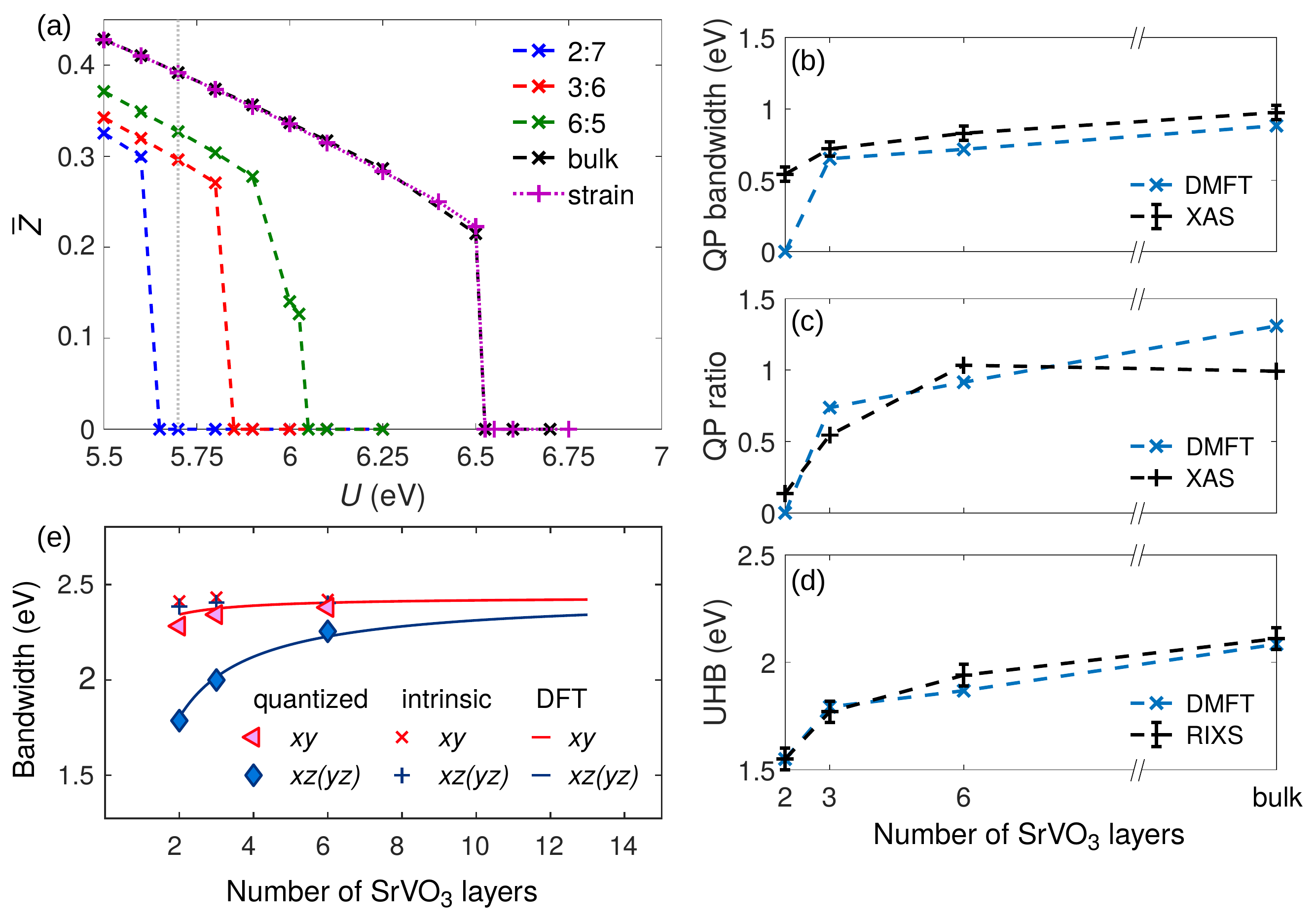}}
\caption{(a)~The orbitally-averaged QP residue, $\bar{Z}$, for each of
the $p$:$q$ SLs. Also shown are the
results of bulk and 1\% strained calculations (see
text). The grey dotted line indicates the value of $U = 5.7$~eV used for
subsequent calculations. (b-d)~Comparison with experimental quantities from
x-ray absorption spectroscopy (XAS) and resonant inelastic x-ray scattering
(RIXS) data \cite{laverock2017,sup}; (b)~QP bandwidth; (c)~Spectral weight
within QP states compared with upper Hubbard band states (UHB); (d)~Energy of
UHB (the DMFT results have been rigidly shifted to match at the 2:7 SL).
(e)~The bandwidths of the SLs from the DFT calculation and QTB model. Also
shown are the intrinsic (bulk) bandwidths of the QTB model.}
\label{f:exp}
\end{figure}

For the DMFT calculation,
the TRIQS/CTHYB continuous-time quantum Monte Carlo (CTQMC) solver \cite{seth2016}
and the TRIQS library \cite{triqs} were used with the Hubbard–Kanamori
interaction Hamiltonian, $\beta = 40$~eV$^{-1}$ (290~K), $J = 0.75$~eV and the
fully localized limit double counting term, as used previously
\cite{zhong2015,bhandary2016,schuler2018}. The
DMFT cycle requires multiple impurities depending on the SL structure, as
illustrated schematically in Fig.~\ref{f:slstruc}(a). The systems investigated
correspond to a single impurity for the 2:7 SL and bulk calculations, and two
and three impurities for the 3:6 and 6:5 SLs, respectively. In each case, the
impurities are considered to be independent of one another.
Only the V $t_{2g}$ bands were
projected (using Wannier projectors) to construct the LDA Hamiltonian in Wannier
space to be used in the DMFT calculation \cite{aichhorn2009}. The results
presented use the fully charge self-consistent DFT+DMFT technique as
implemented in the TRIQS/DFTTools library
\cite{aichhorn2011,*aichhorn2016}. We obtain similar results with the one-shot
approach with the exception of the orbital charges \cite{sup}, which is
consistent with other studies \cite{bhandary2016,schuler2018,hampel2019}. For
each SL structure, $U$ was varied in the range 5.5 to 6.25~eV, and $U_{\rm MIT}$
was located, corresponding to the $U$ at which the SL becomes insulating. In
Fig.\ \ref{f:exp}(a), we show the orbitally-averaged QP residue,
$\bar{Z} = (\Sigma_i Z_i) / N$,
for each SL, showing how $U_{\rm MIT}$ increases by $\sim 0.2$~eV for
each SL. Here, $Z_i$ was determined from the $i$th orbital self-energy
on the Matsubara frequency axis.

\begin{figure*}[t]
\centerline{\includegraphics[width=1.0\linewidth]{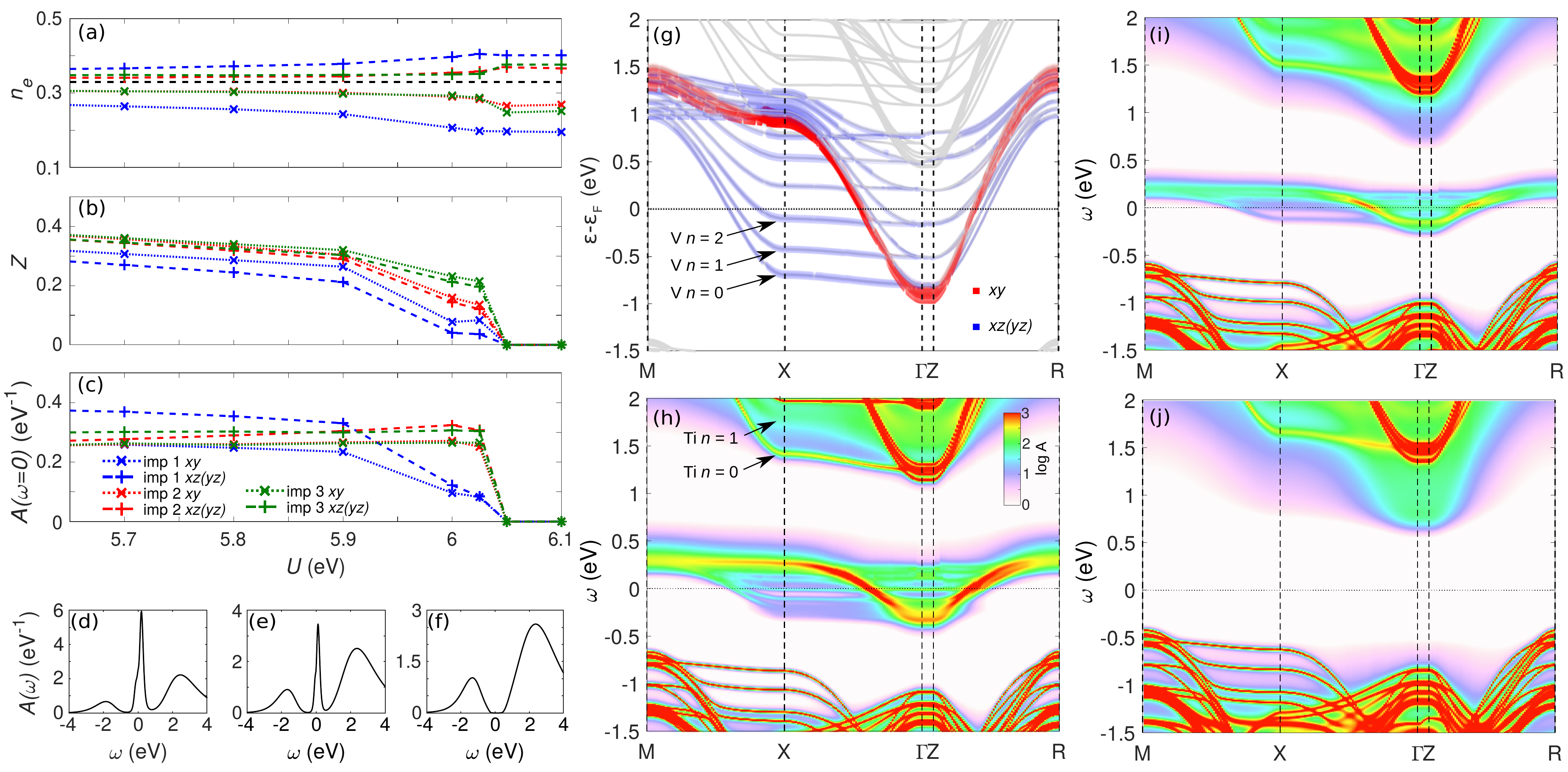}}
\caption{Correlated QW electronic structure. (a)~Occupation number, $n_e$,
(b)~QP residue, $Z$, and (c)~spectral function at the Fermi level ($A(\omega=0$)
of all impurity (imp)
correlated orbitals from DFT+DMFT across the MIT for the 6:5 SL. (d-f) The
momentum-integrated V $t_{2g}$ spectral function, $A(\omega)$, for $U=5.7$,
6.025 and 6.05~eV respectively. (g)~DFT bandstructure of the 6:5
SL, with V $t_{2g}$ band characters shown. (h)~DFT+DMFT momentum-resolved
spectral function, $A(\textbf{k},\omega)$, of the 6:5 SL for $U=5.7$~eV,
showing directly the renormalization of the correlated bands. (i-j)~DFT+DMFT
spectral function close to the MIT for $U=6.025$ and 6.05~eV, respectively.}
\label{f:spec}
\end{figure*}

We begin by ensuring that our DFT+DMFT calculations accurately describe the
experimental system. In Figs.\ \ref{f:exp}(b-d), we compare quantities extracted
from x-ray absorption spectroscopy (XAS) and resonant inelastic x-ray scattering
(RIXS) experiments \cite{laverock2017} with the corresponding theoretical
quantities from our DFT+DMFT calculations for $U = 5.7$~eV, chosen
to reproduce the phenomenological behavior of the experiments. As demonstrated
by Figs.\ \ref{f:exp}(b-d), our DFT+DMFT calculations not only capture the
qualitative behavior, but also yield excellent quantitative agreement with the
experiment trends.
We note that, although the 2:7 SL is macroscopically insulating, it has a
small QP spectral weight in the spectroscopic experiments due to properties of
the sample.

Insight into the microscopic mechanism for the MIT can be obtained by analyzing
the QTB model for each SL. In Fig.\ \ref{f:exp}(e), the bandwidths of the SLs
from the DFT calculations are shown alongside those from the QTB model. While
the in-plane $xy$ orbitals experience a slight narrowing for thinner SrVO$_3$
SLs, the overall band narrowing of the (quantized) out-of-plane $xz(yz)$ orbitals
is substantial, leading to a bandwidth reduction of $\sim 70$\% for the 2:7 SL
\cite{laverock2017}. As expected from the quality of the fits in Fig.\
\ref{f:slstruc}(b), good agreement is observed between the QTB and DFT results.
However, this behavior is not captured at all by the ``intrinsic'' bandwidths,
which correspond to the effective 3D bands of the QTB model before
quantization. These results demonstrate that the band narrowing in Fig.\
\ref{f:exp}(e) is due to quantization of the $xz(yz)$ orbitals, which has a more
pronounced impact for thinner SrVO$_3$ layers. The band narrowing of the thinner
SLs leads to a greater $U/W$ ratio, which results in stabilization of the
insulating phase, as illustrated by the DFT+DMFT calculations shown in Fig.\
\ref{f:exp}(a).

A previous DMFT study has attributed CF effects as being the principle factor
driving the MIT in bilayer SrVO$_3$
on SrTiO$_3$ \cite{zhong2015}, and it is
pertinent to ask what role, if any, the CF plays in our system. In Ref.\
\onlinecite{zhong2015}, strain
induced by the SrTiO$_3$ substrate led to a lowering of the $xy$ orbitals by
180~meV due to the CF. In contrast, we find a CF splitting of $\leq 51$~meV in
favor of the V $xz(yz)$ bands in our SLs, in part owing to the lower strain
imparted by the LSAT substrate.  To reproduce the effects of this CF, we have
calculated strained SrVO$_3$ with a volume-conserving strain of 1\%, which leads
to similar CF splitting of 53~meV \cite{sup}. This 3D system, which reproduces
the CF levels of our SLs but without the quantization effects, is shown in Fig.\
\ref{f:exp}(a), and shows very similar behavior to the bulk cubic system.
Therefore, we confindently rule out CF effects as a dominant factor in our SLs.

Finally, we discuss the correlated behavior of the quantized electron states,
taking the 6:5 SL as an example. Figure \ref{f:spec}(a-c) shows the orbital-
and layer-resolved occupation number, $n_e$, QP residue, $Z$, and
spectral weight at the Fermi
level, $A(\omega=0)$, as a function of $U$, illustrating the transition to the
insulating phase at $U=6.05$~eV for this SL. As presented in Fig.\
\ref{f:spec}(a), each layer (impurity) exhibits a sizeable orbital polarization
in favor of the $xy$ orbitals, which is exaggerated both at the outer
(interface) layer and in the insulating phase. This behavior is consistent
across all SLs, and originates from a small polarization in the DFT calculations
due to the local CF, which is subsequently amplified in the DMFT cycle
\cite{sup}. The outer layer
(impurity 1) is significantly ``more correlated'', experiencing a smaller QP
residue $Z$ than the other layers [Fig.\ \ref{f:spec}(b)], corresponding to a
greater renormalization factor, $1/Z$. Near
the MIT, this leads to a collapse in
the spectral weight at $\omega=0$ of the interface layer [Fig.\
\ref{f:spec}(c)]. In this sense, the more correlated interface layer simultaneously triggers
the MIT in the remaining layers, in much the same way as suggested for SrVO$_3$
bilayers in Ref.\ \onlinecite{zhong2015}.

\begin{figure}[t]
\centerline{\includegraphics[width=0.8\linewidth]{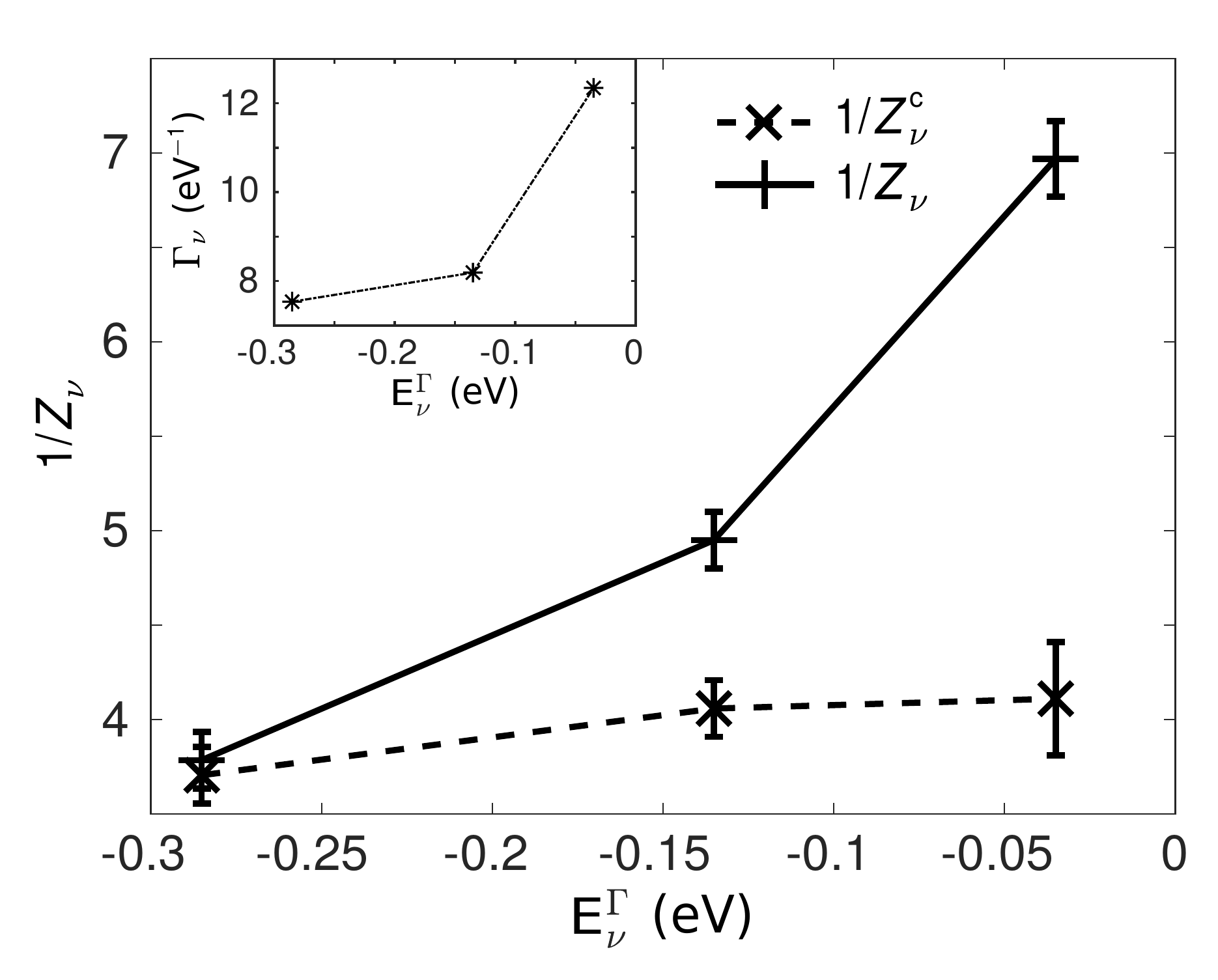}}
\caption{The total mass enhancement factor of each subband, $1/Z_{\nu}$, of the
6:5 SL at $U=5.7$~eV shown alongside the correlation-only mass enhancement
factor $1/Z^{\rm c}_{\nu}$, evaluated along $X$-$M$. The
abscissa represents the bottom energy of the subband.
Error bars indicate the estimated error in determining these quantities.
The inset shows the subband lifetime $\Gamma_{\nu}$.
All lines are guides to the eye.}
\label{f:zfig}
\end{figure}

The DMFT spectral functions, $A(\omega)$ and $A(\bf{k},\omega)$, are shown in
Fig.\ \ref{f:spec}(d-f) and
Fig.\ \ref{f:spec}(h-j) for $U=5.7$, 6.025 and 6.05~eV, respectively, and were
calculated using the maximum entropy method within TRIQS \cite{maxent}. The
renormalization of the V bands increases from a factor of
$1/\bar{Z} \approx 3.5$ at $U=5.7$~eV to $\approx 8$ just below
the MIT ($U=6.05$~eV) before the QP spectral weight vanishes in the insulating
phase [Fig.\ \ref{f:spec}(f)].  The renormalization of the V bands is
accompanied by a lowering in their energy with respect to O and Ti bands,
leading to pronounced energy separation of O/Ti bands and metallic V $t_{2g}$
QPs.  Nevertheless, the quantized subband structure remains clear in the
metallic solutions, leading to genuinely strongly correlated QWs. At $U=5.7$~eV,
the lowest three $yz$ subbands are occupied at $X$ and $\Gamma$, similar to the
DFT, however by $U=6.025$~eV only the lowest subband is occupied, indicating the
quantized system undergoes a correlation-induced Lifshitz transition prior to
the MIT. Although the correlated orbitals remain relatively sharp throughout
near $\omega=0$, implying long-lived QPs with a well-defined Fermi surface,
the lifetime rapidly broadens away from $\varepsilon_{\rm F}$, in line with
a strongly
correlated Fermi liquid (see inset to Fig.\ \ref{f:zfig}). We
also note that the incoherent upper Hubband band
(UHB), visible above 1~eV in [Fig.\ \ref{f:spec}(j)], exhibits rather strong
momentum dependence, visible in previous bulk DMFT calculations
\cite{nekrasov2006,bhandary2019}.  Unexpectedly, certain (uncorrelated)
Ti states also show
pronounced broadening, e.g.\ the $n=1$ Ti $yz$ orbital indicated in Fig.\
\ref{f:spec}(h), which is almost completely smeared out at $U=6.025$~eV [Fig.\
\ref{f:spec}(i)]. This arises due to the spatial penetration of the Ti
wavefunctions at the interface into the correlated SrVO$_3$ layers, leading to
broadening of those subbands that have signficant weight at the interface
\cite{sup}. This
surprising sensitivity of the Ti states may help to explain recent
RIXS results of the SrTiO$_3$ layers of SrVO$_3$/SrTiO$_3$ SLs, where a sudden
change in the delocalized Ti $3d$ carrier density was observed coincident with
the MIT of the SrVO$_3$ layers \cite{svo_sl_int}.

We finish by discussing some important implications of our results. Anomalous
mass enhancement has been reported in SrVO$_3$ QWs, whereby shallower occupied
subbands were found to have a larger mass enhancement
($m^{\star}=m/Z_{\nu}$) than subbands at deeper energies \cite{yoshimatsu2011}.
Subsequent theoretical
\cite{okamoto2011} and experimental \cite{kobayashi2015} studies argued the
anomalous enhancement was
a consequence of a combination of electron correlation effects and the reduced
dimensionality of the surface.
To address
this, we compare the total mass enhancement of a particular quantized
subband, $1/Z_{\nu}$, with that due solely to
electron correlation, $1/Z_{\nu}^{\rm c}$, both of which
are obtained through
analysis of the Fermi velocities \cite{sup}. The trend in the
total mass enhancement factor, $1/Z_{\nu}$, which includes both correlation
effects and band effects due to quantization, qualitatively reproduces the
experimental data \cite{yoshimatsu2011} rather well, but this behavior is only
very weakly present in $1/Z_{\nu}^{\rm c}$.
Therefore, in our SLs the origin of the mass enhancement is firmly due to $k_z$
sampling introduced by
quantization, as has been observed in other systems \cite{kawasaki2018}.
While reduced coordination at the surface
\cite{ishida2006,okamoto2011}, surface reconstruction
\cite{oka2018,*wang2019b}, or specific $k_z$ sampling due to an
asymmetric phase shift at the vacuum
\cite{matsuda2002,*yoshimatsu2013,kawasaki2018}
may play additional
roles in specific surface-terminated systems, our
results reveal quantization as the primary source of
anamalous mass enhancement in embedded (or capped) few-layer SrVO$_3$.

In summary, our results establish quantum confinement engineering as a sensitive
method to tune the correlated electron behavior of $3d$ electron systems. We
demonstrate our approach using SLs of few-layer SrVO$_3$ embedded in
SrTiO$_3$: this system exhibits an MIT due to a reduction in
bandwidth by varying the number of SrVO$_3$ layers. The microscopic mechanism
we reveal is a direct consequence of quantum confinement, and is distinct from
previous studies invoking CF effects \cite{zhong2015} or dimensionality
\cite{yoshimatsu2010}, demonstrating the excellent versatility of the MIT in
SrVO$_3$ with respect to different control parameters. Similar to Ref.\
\onlinecite{zhong2015}, we find that the outer layers at the interface of the
quantum well are more strongly correlated, and trigger the transition in the
rest of the layer. Our interpretation is also able to naturally explain the
anomalous mass enhancement previously reported in ARPES measurements
\cite{yoshimatsu2011,kobayashi2015}. Together, these results demonstrate the
potency of employing quantum confinement as a tuning parameter for correlated
electron behavior in engineered SLs. This approach also has the significant
benefits that
surface effects are avoided and that large (bulk-like) volumes of the system may
respond to external stimuli. Aside from the expected improved
performance of such a SrVO$_3$ device as a Mott transistor \cite{zhong2015},
quantized SLs made of other correlated materials are likely to show improved
properties, e.g.\ superconducting devices in cuprate SLs \cite{ahn1999},
conductivity in nickelate SLs \cite{chen2013} or spintronic devices
\cite{samanta2018}.

\begin{acknowledgments}
A.D.N.J.\ acknowledges funding and support from the
Engineering and Physical Sciences Research Council (EPSRC) Centre for Doctoral
Training in Condensed Matter Physics (CDT-CMP), Grant No.\ EP/L015544/1. M.A\
acknowledges financial support from the Austrian Science Fund (FWF), START
program Y746.
Calculations were performed using the computational facilities of the Advanced
Computing Research Centre, University of Bristol
(\href{http://bris.ac.uk/acrc/}{http://bris.ac.uk/acrc/}). The authors would
like to thank Dr.\ G.\ Kraberger and Prof.\ S.\ B.\ Dugdale for their useful
discussions and contributions related to this work. The VESTA package
(\href{https://jp-minerals.org/vesta/en/}{https://jp-minerals.org/vesta/en/})
has been used in the preparation of some figures.
\end{acknowledgments}


\begin{thebibliography}{14}%
\makeatletter
\providecommand \@ifxundefined [1]{%
 \@ifx{#1\undefined}
}%
\providecommand \@ifnum [1]{%
 \ifnum #1\expandafter \@firstoftwo
 \else \expandafter \@secondoftwo
 \fi
}%
\providecommand \@ifx [1]{%
 \ifx #1\expandafter \@firstoftwo
 \else \expandafter \@secondoftwo
 \fi
}%
\providecommand \natexlab [1]{#1}%
\providecommand \enquote  [1]{``#1''}%
\providecommand \bibnamefont  [1]{#1}%
\providecommand \bibfnamefont [1]{#1}%
\providecommand \citenamefont [1]{#1}%
\providecommand \href@noop [0]{\@secondoftwo}%
\providecommand \href [0]{\begingroup \@sanitize@url \@href}%
\providecommand \@href[1]{\@@startlink{#1}\@@href}%
\providecommand \@@href[1]{\endgroup#1\@@endlink}%
\providecommand \@sanitize@url [0]{\catcode `\\12\catcode `\$12\catcode
  `\&12\catcode `\#12\catcode `\^12\catcode `\_12\catcode `\%12\relax}%
\providecommand \@@startlink[1]{}%
\providecommand \@@endlink[0]{}%
\providecommand \url  [0]{\begingroup\@sanitize@url \@url }%
\providecommand \@url [1]{\endgroup\@href {#1}{\urlprefix }}%
\providecommand \urlprefix  [0]{URL }%
\providecommand \Eprint [0]{\href }%
\providecommand \doibase [0]{http://dx.doi.org/}%
\providecommand \selectlanguage [0]{\@gobble}%
\providecommand \bibinfo  [0]{\@secondoftwo}%
\providecommand \bibfield  [0]{\@secondoftwo}%
\providecommand \translation [1]{[#1]}%
\providecommand \BibitemOpen [0]{}%
\providecommand \bibitemStop [0]{}%
\providecommand \bibitemNoStop [0]{.\EOS\space}%
\providecommand \EOS [0]{\spacefactor3000\relax}%
\providecommand \BibitemShut  [1]{\csname bibitem#1\endcsname}%
\let\auto@bib@innerbib\@empty
\bibitem [{\citenamefont {Gu}\ \emph {et~al.}(2018)\citenamefont {Gu},
  \citenamefont {Wolf},\ and\ \citenamefont {Lu}}]{gu2018}%
  \BibitemOpen
  \bibfield  {author} {\bibinfo {author} {\bibfnamefont {M.}~\bibnamefont
  {Gu}}, \bibinfo {author} {\bibfnamefont {S.~A.}\ \bibnamefont {Wolf}}, \ and\
  \bibinfo {author} {\bibfnamefont {J.}~\bibnamefont {Lu}},\ }\bibfield
  {title} {\enquote {\bibinfo {title} {Transport phenomena in
  {SrVO$_3$}/{SrTiO$_3$} superlattices},}\ }\href {\doibase
  10.1088/1361-6463/aaabac} {\bibfield  {journal} {\bibinfo  {journal} {J.
  Phys. D: Appl. Phys.}\ }\textbf {\bibinfo {volume} {51}},\ \bibinfo {pages}
  {10LT01} (\bibinfo {year} {2018})}\BibitemShut {NoStop}%
\bibitem [{\citenamefont {Laverock}\ \emph {et~al.}(2017)\citenamefont
  {Laverock}, \citenamefont {Gu}, \citenamefont {Jovic}, \citenamefont {Lu},
  \citenamefont {Wolf}, \citenamefont {Qiao}, \citenamefont {Yang},\ and\
  \citenamefont {Smith}}]{laverock2017}%
  \BibitemOpen
  \bibfield  {author} {\bibinfo {author} {\bibfnamefont {J.}~\bibnamefont
  {Laverock}}, \bibinfo {author} {\bibfnamefont {M.}~\bibnamefont {Gu}},
  \bibinfo {author} {\bibfnamefont {V.}~\bibnamefont {Jovic}}, \bibinfo
  {author} {\bibfnamefont {J.~W.}\ \bibnamefont {Lu}}, \bibinfo {author}
  {\bibfnamefont {S.A.}\ \bibnamefont {Wolf}}, \bibinfo {author} {\bibfnamefont
  {R.~M.}\ \bibnamefont {Qiao}}, \bibinfo {author} {\bibfnamefont
  {W.}~\bibnamefont {Yang}}, \ and\ \bibinfo {author} {\bibfnamefont {K.E.}\
  \bibnamefont {Smith}},\ }\bibfield  {title} {\enquote {\bibinfo {title}
  {Nano-engineering of electron correlation in oxide superlattices},}\ }\href
  {\doibase 10.1088/2399-1984/aa8f39} {\bibfield  {journal} {\bibinfo
  {journal} {Nano Futures}\ }\textbf {\bibinfo {volume} {1}},\ \bibinfo {pages}
  {031001} (\bibinfo {year} {2017})}\BibitemShut {NoStop}%
\bibitem [{\citenamefont {Dewhurst}\ \emph {et~al.}(2019)\citenamefont
  {Dewhurst}, \citenamefont {Sharma}, \citenamefont {Nordstr\"{o}m},
  \citenamefont {Cricchio}, \citenamefont {Granas},\ and\ \citenamefont
  {Gross}}]{elk}%
  \BibitemOpen
  \bibfield  {author} {\bibinfo {author} {\bibfnamefont {J.~K.}\ \bibnamefont
  {Dewhurst}}, \bibinfo {author} {\bibfnamefont {S.}~\bibnamefont {Sharma}},
  \bibinfo {author} {\bibfnamefont {L.}~\bibnamefont {Nordstr\"{o}m}}, \bibinfo
  {author} {\bibfnamefont {F.}~\bibnamefont {Cricchio}}, \bibinfo {author}
  {\bibfnamefont {O.}~\bibnamefont {Granas}}, \ and\ \bibinfo {author}
  {\bibfnamefont {E.~K.~U.}\ \bibnamefont {Gross}},\ }\href@noop {} {}\bibinfo
  {howpublished} {\url{http://elk.sourceforge.net/}} (\bibinfo {year}
  {2019})\BibitemShut {NoStop}%
\bibitem [{\citenamefont {Yoshimatsu}\ \emph {et~al.}(2010)\citenamefont
  {Yoshimatsu}, \citenamefont {Okabe}, \citenamefont {Kumigashira},
  \citenamefont {Okamoto}, \citenamefont {Aizaki}, \citenamefont {Fujimori},\
  and\ \citenamefont {Oshima}}]{yoshimatsu2010}%
  \BibitemOpen
  \bibfield  {author} {\bibinfo {author} {\bibfnamefont {K.}~\bibnamefont
  {Yoshimatsu}}, \bibinfo {author} {\bibfnamefont {T.}~\bibnamefont {Okabe}},
  \bibinfo {author} {\bibfnamefont {H.}~\bibnamefont {Kumigashira}}, \bibinfo
  {author} {\bibfnamefont {S.}~\bibnamefont {Okamoto}}, \bibinfo {author}
  {\bibfnamefont {S.}~\bibnamefont {Aizaki}}, \bibinfo {author} {\bibfnamefont
  {A.}~\bibnamefont {Fujimori}}, \ and\ \bibinfo {author} {\bibfnamefont
  {M.}~\bibnamefont {Oshima}},\ }\bibfield  {title} {\enquote {\bibinfo {title}
  {Dimensional-crossover-driven metal-insulator transition in {SrVO}$_3$
  ultrathin films},}\ }\href {\doibase 10.1103/PhysRevLett.104.147601}
  {\bibfield  {journal} {\bibinfo  {journal} {Phys. Rev. Lett.}\ }\textbf
  {\bibinfo {volume} {104}},\ \bibinfo {pages} {147601} (\bibinfo {year}
  {2010})}\BibitemShut {NoStop}%
\bibitem [{\citenamefont {Liebsch}(2003)}]{liebsch2003}%
  \BibitemOpen
  \bibfield  {author} {\bibinfo {author} {\bibfnamefont {A.}~\bibnamefont
  {Liebsch}},\ }\bibfield  {title} {\enquote {\bibinfo {title} {{Surface versus
  Bulk {Coulomb} Correlations in Photoemission Spectra of {SrVO$_3$} and
  {CaVO$_3$}}},}\ }\href {\doibase 10.1103/PhysRevLett.90.096401} {\bibfield
  {journal} {\bibinfo  {journal} {Phys. Rev. Lett.}\ }\textbf {\bibinfo
  {volume} {90}},\ \bibinfo {pages} {096401} (\bibinfo {year}
  {2003})}\BibitemShut {NoStop}%
\bibitem [{\citenamefont {Chiang}(2000)}]{chiang2000}%
  \BibitemOpen
  \bibfield  {author} {\bibinfo {author} {\bibfnamefont {T.-C.}\ \bibnamefont
  {Chiang}},\ }\bibfield  {title} {\enquote {\bibinfo {title} {Photoemission
  studies of quantum well states in thin films},}\ }\href {\doibase
  10.1016/S0167-5729(00)00006-6} {\bibfield  {journal} {\bibinfo  {journal}
  {Surf. Sci. Rep.}\ }\textbf {\bibinfo {volume} {39}},\ \bibinfo {pages} {181}
  (\bibinfo {year} {2000})}\BibitemShut {NoStop}%
\bibitem [{\citenamefont {Parcollet}\ \emph {et~al.}(2015)\citenamefont
  {Parcollet}, \citenamefont {Ferrero}, \citenamefont {Ayral}, \citenamefont
  {Hafermann}, \citenamefont {Krivenko}, \citenamefont {Messio},\ and\
  \citenamefont {Seth}}]{triqs}%
  \BibitemOpen
  \bibfield  {author} {\bibinfo {author} {\bibfnamefont {O.}~\bibnamefont
  {Parcollet}}, \bibinfo {author} {\bibfnamefont {M.}~\bibnamefont {Ferrero}},
  \bibinfo {author} {\bibfnamefont {T.}~\bibnamefont {Ayral}}, \bibinfo
  {author} {\bibfnamefont {H.}~\bibnamefont {Hafermann}}, \bibinfo {author}
  {\bibfnamefont {I.}~\bibnamefont {Krivenko}}, \bibinfo {author}
  {\bibfnamefont {L.}~\bibnamefont {Messio}}, \ and\ \bibinfo {author}
  {\bibfnamefont {P.}~\bibnamefont {Seth}},\ }\bibfield  {title} {\enquote
  {\bibinfo {title} {{TRIQS}: A toolbox for research on interacting quantum
  systems},}\ }\href {\doibase 10.1016/j.cpc.2015.04.023} {\bibfield  {journal}
  {\bibinfo  {journal} {Comp. Phys. Commun.}\ }\textbf {\bibinfo {volume}
  {196}},\ \bibinfo {pages} {398} (\bibinfo {year} {2015})}\BibitemShut
  {NoStop}%
\bibitem [{\citenamefont {Aichhorn}\ \emph {et~al.}(2016)\citenamefont
  {Aichhorn}, \citenamefont {Pourovskii}, \citenamefont {Seth}, \citenamefont
  {Vildosola}, \citenamefont {Zingl}, \citenamefont {Peil}, \citenamefont
  {Deng}, \citenamefont {Mravlje}, \citenamefont {Kraberger}, \citenamefont
  {Martins}, \citenamefont {Ferrero},\ and\ \citenamefont
  {Parcollet}}]{aichhorn2016}%
  \BibitemOpen
  \bibfield  {author} {\bibinfo {author} {\bibfnamefont {M.}~\bibnamefont
  {Aichhorn}}, \bibinfo {author} {\bibfnamefont {L.}~\bibnamefont
  {Pourovskii}}, \bibinfo {author} {\bibfnamefont {P.}~\bibnamefont {Seth}},
  \bibinfo {author} {\bibfnamefont {V.}~\bibnamefont {Vildosola}}, \bibinfo
  {author} {\bibfnamefont {M.}~\bibnamefont {Zingl}}, \bibinfo {author}
  {\bibfnamefont {O.~E.}\ \bibnamefont {Peil}}, \bibinfo {author}
  {\bibfnamefont {X.}~\bibnamefont {Deng}}, \bibinfo {author} {\bibfnamefont
  {J.}~\bibnamefont {Mravlje}}, \bibinfo {author} {\bibfnamefont {G.~J.}\
  \bibnamefont {Kraberger}}, \bibinfo {author} {\bibfnamefont {C.}~\bibnamefont
  {Martins}}, \bibinfo {author} {\bibfnamefont {M.}~\bibnamefont {Ferrero}}, \
  and\ \bibinfo {author} {\bibfnamefont {O.}~\bibnamefont {Parcollet}},\
  }\bibfield  {title} {\enquote {\bibinfo {title} {{TRIQS/DFTTools}: A {TRIQS}
  application for ab initio calculations of correlated materials},}\ }\href
  {\doibase 10.1016/j.cpc.2016.03.014} {\bibfield  {journal} {\bibinfo
  {journal} {Comp. Phys. Commun.}\ }\textbf {\bibinfo {volume} {204}},\
  \bibinfo {pages} {200} (\bibinfo {year} {2016})}\BibitemShut {NoStop}%
\bibitem [{\citenamefont {Zhong}\ \emph {et~al.}(2015)\citenamefont {Zhong},
  \citenamefont {Wallerberger}, \citenamefont {Tomczak}, \citenamefont
  {Taranto}, \citenamefont {Parragh}, \citenamefont {Toschi}, \citenamefont
  {Sangiovanni},\ and\ \citenamefont {Held}}]{zhong2015}%
  \BibitemOpen
  \bibfield  {author} {\bibinfo {author} {\bibfnamefont {Z.}~\bibnamefont
  {Zhong}}, \bibinfo {author} {\bibfnamefont {M.}~\bibnamefont {Wallerberger}},
  \bibinfo {author} {\bibfnamefont {J.~M.}\ \bibnamefont {Tomczak}}, \bibinfo
  {author} {\bibfnamefont {C.}~\bibnamefont {Taranto}}, \bibinfo {author}
  {\bibfnamefont {N.}~\bibnamefont {Parragh}}, \bibinfo {author} {\bibfnamefont
  {A.}~\bibnamefont {Toschi}}, \bibinfo {author} {\bibfnamefont
  {G.}~\bibnamefont {Sangiovanni}}, \ and\ \bibinfo {author} {\bibfnamefont
  {K.}~\bibnamefont {Held}},\ }\bibfield  {title} {\enquote {\bibinfo {title}
  {{Electronics with Correlated Oxides: {SrVO$_3$}/{SrTiO$_3$} as a {Mott}
  Transistor}},}\ }\href {\doibase 10.1103/PhysRevLett.114.246401} {\bibfield
  {journal} {\bibinfo  {journal} {Phys. Rev. Lett.}\ }\textbf {\bibinfo
  {volume} {114}},\ \bibinfo {pages} {246401} (\bibinfo {year}
  {2015})}\BibitemShut {NoStop}%
\bibitem [{\citenamefont {Bhandary}\ \emph {et~al.}(2016)\citenamefont
  {Bhandary}, \citenamefont {Assmann}, \citenamefont {Aichhorn},\ and\
  \citenamefont {Held}}]{bhandary2016}%
  \BibitemOpen
  \bibfield  {author} {\bibinfo {author} {\bibfnamefont {S.}~\bibnamefont
  {Bhandary}}, \bibinfo {author} {\bibfnamefont {E.}~\bibnamefont {Assmann}},
  \bibinfo {author} {\bibfnamefont {M.}~\bibnamefont {Aichhorn}}, \ and\
  \bibinfo {author} {\bibfnamefont {K.}~\bibnamefont {Held}},\ }\bibfield
  {title} {\enquote {\bibinfo {title} {Charge self-consistency in density
  functional theory combined with dynamical mean field theory: $k$-space
  reoccupation and orbital order},}\ }\href {\doibase
  10.1103/PhysRevB.94.155131} {\bibfield  {journal} {\bibinfo  {journal} {Phys.
  Rev. B}\ }\textbf {\bibinfo {volume} {94}},\ \bibinfo {pages} {155131}
  (\bibinfo {year} {2016})}\BibitemShut {NoStop}%
\bibitem [{\citenamefont {Sch{\"u}ler}\ \emph {et~al.}(2018)\citenamefont
  {Sch{\"u}ler}, \citenamefont {Peil}, \citenamefont {Kraberger}, \citenamefont
  {Pordzik}, \citenamefont {Marsman}, \citenamefont {Kresse}, \citenamefont
  {Wehling},\ and\ \citenamefont {Aichhorn}}]{schuler2018}%
  \BibitemOpen
  \bibfield  {author} {\bibinfo {author} {\bibfnamefont {M.}~\bibnamefont
  {Sch{\"u}ler}}, \bibinfo {author} {\bibfnamefont {O.~E.}\ \bibnamefont
  {Peil}}, \bibinfo {author} {\bibfnamefont {G.~J.}\ \bibnamefont {Kraberger}},
  \bibinfo {author} {\bibfnamefont {R.}~\bibnamefont {Pordzik}}, \bibinfo
  {author} {\bibfnamefont {M.}~\bibnamefont {Marsman}}, \bibinfo {author}
  {\bibfnamefont {G.}~\bibnamefont {Kresse}}, \bibinfo {author} {\bibfnamefont
  {T.~O.}\ \bibnamefont {Wehling}}, \ and\ \bibinfo {author} {\bibfnamefont
  {M.}~\bibnamefont {Aichhorn}},\ }\bibfield  {title} {\enquote {\bibinfo
  {title} {Charge self-consistent many-body corrections using optimized
  projected localized orbitals},}\ }\href {\doibase 10.1088/1361-648X/aae80a}
  {\bibfield  {journal} {\bibinfo  {journal} {J. Phys.: Conden. Matter}\
  }\textbf {\bibinfo {volume} {30}},\ \bibinfo {pages} {475901} (\bibinfo
  {year} {2018})}\BibitemShut {NoStop}%
\bibitem [{\citenamefont {Aichhorn}\ \emph {et~al.}(2009)\citenamefont
  {Aichhorn}, \citenamefont {Pourovskii}, \citenamefont {Vildosola},
  \citenamefont {Ferrero}, \citenamefont {Parcollet}, \citenamefont {Miyake},
  \citenamefont {Georges},\ and\ \citenamefont {Biermann}}]{aichhorn2009}%
  \BibitemOpen
  \bibfield  {author} {\bibinfo {author} {\bibfnamefont {M.}~\bibnamefont
  {Aichhorn}}, \bibinfo {author} {\bibfnamefont {L.}~\bibnamefont
  {Pourovskii}}, \bibinfo {author} {\bibfnamefont {V.}~\bibnamefont
  {Vildosola}}, \bibinfo {author} {\bibfnamefont {M.}~\bibnamefont {Ferrero}},
  \bibinfo {author} {\bibfnamefont {O.}~\bibnamefont {Parcollet}}, \bibinfo
  {author} {\bibfnamefont {T.}~\bibnamefont {Miyake}}, \bibinfo {author}
  {\bibfnamefont {A.}~\bibnamefont {Georges}}, \ and\ \bibinfo {author}
  {\bibfnamefont {S.}~\bibnamefont {Biermann}},\ }\bibfield  {title} {\enquote
  {\bibinfo {title} {Dynamical mean-field theory within an augmented plane-wave
  framework: Assessing electronic correlations in the iron pnictide
  {LaFeAsO}},}\ }\href {\doibase 10.1103/PhysRevB.80.085101} {\bibfield
  {journal} {\bibinfo  {journal} {Phys. Rev. B}\ }\textbf {\bibinfo {volume}
  {80}},\ \bibinfo {pages} {085101} (\bibinfo {year} {2009})}\BibitemShut
  {NoStop}%
\bibitem [{\citenamefont {Kraberger}\ and\ \citenamefont
  {Zingl}(2019)}]{maxent}%
  \BibitemOpen
  \bibfield  {author} {\bibinfo {author} {\bibfnamefont {G.~J.}\ \bibnamefont
  {Kraberger}}\ and\ \bibinfo {author} {\bibfnamefont {M.}~\bibnamefont
  {Zingl}},\ }\href@noop {} {}\bibinfo {howpublished}
  {\url{https://github.com/TRIQS/maxent}} (\bibinfo {year} {2019})\BibitemShut
  {NoStop}%
\bibitem [{\citenamefont {Hampel}\ \emph {et~al.}(2019)\citenamefont {Hampel},
  \citenamefont {Beck},\ and\ \citenamefont {Ederer}}]{hampel2019}%
  \BibitemOpen
  \bibfield  {author} {\bibinfo {author} {\bibfnamefont {A.}~\bibnamefont
  {Hampel}}, \bibinfo {author} {\bibfnamefont {S.}~\bibnamefont {Beck}}, \ and\
  \bibinfo {author} {\bibfnamefont {C.}~\bibnamefont {Ederer}},\ }\bibfield
  {title} {\enquote {\bibinfo {title} {Charge self-consistency and
  double-counting in {DFT+DMFT} calculations for complex transition metal
  oxides},}\ }\href {https://arxiv.org/abs/1907.10339} {\bibfield  {journal}
  {\bibinfo  {journal} {arXiv:1907.10339}\ } (\bibinfo {year}
  {2019})}\BibitemShut {NoStop}%
\end{thebibliography}%


\begin{thebibliography}{58}%
\makeatletter
\providecommand \@ifxundefined [1]{%
 \@ifx{#1\undefined}
}%
\providecommand \@ifnum [1]{%
 \ifnum #1\expandafter \@firstoftwo
 \else \expandafter \@secondoftwo
 \fi
}%
\providecommand \@ifx [1]{%
 \ifx #1\expandafter \@firstoftwo
 \else \expandafter \@secondoftwo
 \fi
}%
\providecommand \natexlab [1]{#1}%
\providecommand \enquote  [1]{``#1''}%
\providecommand \bibnamefont  [1]{#1}%
\providecommand \bibfnamefont [1]{#1}%
\providecommand \citenamefont [1]{#1}%
\providecommand \href@noop [0]{\@secondoftwo}%
\providecommand \href [0]{\begingroup \@sanitize@url \@href}%
\providecommand \@href[1]{\@@startlink{#1}\@@href}%
\providecommand \@@href[1]{\endgroup#1\@@endlink}%
\providecommand \@sanitize@url [0]{\catcode `\\12\catcode `\$12\catcode
  `\&12\catcode `\#12\catcode `\^12\catcode `\_12\catcode `\%12\relax}%
\providecommand \@@startlink[1]{}%
\providecommand \@@endlink[0]{}%
\providecommand \url  [0]{\begingroup\@sanitize@url \@url }%
\providecommand \@url [1]{\endgroup\@href {#1}{\urlprefix }}%
\providecommand \urlprefix  [0]{URL }%
\providecommand \Eprint [0]{\href }%
\providecommand \doibase [0]{http://dx.doi.org/}%
\providecommand \selectlanguage [0]{\@gobble}%
\providecommand \bibinfo  [0]{\@secondoftwo}%
\providecommand \bibfield  [0]{\@secondoftwo}%
\providecommand \translation [1]{[#1]}%
\providecommand \BibitemOpen [0]{}%
\providecommand \bibitemStop [0]{}%
\providecommand \bibitemNoStop [0]{.\EOS\space}%
\providecommand \EOS [0]{\spacefactor3000\relax}%
\providecommand \BibitemShut  [1]{\csname bibitem#1\endcsname}%
\let\auto@bib@innerbib\@empty
\bibitem [{\citenamefont {Tokura}\ \emph {et~al.}(2017)\citenamefont {Tokura},
  \citenamefont {Kawasaki},\ and\ \citenamefont {Nagaosa}}]{tokura2017}%
  \BibitemOpen
  \bibfield  {author} {\bibinfo {author} {\bibfnamefont {Y.}~\bibnamefont
  {Tokura}}, \bibinfo {author} {\bibfnamefont {M.}~\bibnamefont {Kawasaki}}, \
  and\ \bibinfo {author} {\bibfnamefont {N.}~\bibnamefont {Nagaosa}},\
  }\bibfield  {title} {\enquote {\bibinfo {title} {Emergent functions of
  quantum materials},}\ }\href {\doibase 10.1038/nphys4274} {\bibfield
  {journal} {\bibinfo  {journal} {Nature Phys.}\ }\textbf {\bibinfo {volume}
  {13}},\ \bibinfo {pages} {1056} (\bibinfo {year} {2017})}\BibitemShut
  {NoStop}%
\bibitem [{\citenamefont {Brahlek}\ \emph {et~al.}(2017)\citenamefont
  {Brahlek}, \citenamefont {Zhang}, \citenamefont {Lapano}, \citenamefont
  {Zhang}, \citenamefont {Engel-Herbert}, \citenamefont {Shukla}, \citenamefont
  {Datta}, \citenamefont {Paik},\ and\ \citenamefont {Schlom}}]{brahlek2017}%
  \BibitemOpen
  \bibfield  {author} {\bibinfo {author} {\bibfnamefont {M.}~\bibnamefont
  {Brahlek}}, \bibinfo {author} {\bibfnamefont {L.}~\bibnamefont {Zhang}},
  \bibinfo {author} {\bibfnamefont {J.}~\bibnamefont {Lapano}}, \bibinfo
  {author} {\bibfnamefont {H.-T.}\ \bibnamefont {Zhang}}, \bibinfo {author}
  {\bibfnamefont {R.}~\bibnamefont {Engel-Herbert}}, \bibinfo {author}
  {\bibfnamefont {N.}~\bibnamefont {Shukla}}, \bibinfo {author} {\bibfnamefont
  {S.}~\bibnamefont {Datta}}, \bibinfo {author} {\bibfnamefont
  {H.}~\bibnamefont {Paik}}, \ and\ \bibinfo {author} {\bibfnamefont {D.G.}\
  \bibnamefont {Schlom}},\ }\bibfield  {title} {\enquote {\bibinfo {title}
  {Opportunities in vanadium-based strongly correlated electron systems},}\
  }\href {https://doi.org/10.1557/mrc.2017.2} {\bibfield  {journal} {\bibinfo
  {journal} {MRS Commun.}\ }\textbf {\bibinfo {volume} {7}},\ \bibinfo {pages}
  {27} (\bibinfo {year} {2017})}\BibitemShut {NoStop}%
\bibitem [{\citenamefont {Mannhart}\ and\ \citenamefont
  {Schlom}(2010)}]{mannhart2010}%
  \BibitemOpen
  \bibfield  {author} {\bibinfo {author} {\bibfnamefont {J.}~\bibnamefont
  {Mannhart}}\ and\ \bibinfo {author} {\bibfnamefont {D.~G.}\ \bibnamefont
  {Schlom}},\ }\bibfield  {title} {\enquote {\bibinfo {title} {Oxide interfaces
  — an opportunity for electronics},}\ }\href {\doibase
  10.1126/science.1181862} {\bibfield  {journal} {\bibinfo  {journal}
  {Science}\ }\textbf {\bibinfo {volume} {327}},\ \bibinfo {pages} {1607}
  (\bibinfo {year} {2010})}\BibitemShut {NoStop}%
\bibitem [{\citenamefont {Kotliar}\ \emph {et~al.}(2006)\citenamefont
  {Kotliar}, \citenamefont {Savrasov}, \citenamefont {Haule}, \citenamefont
  {Oudovenko}, \citenamefont {Parcollet},\ and\ \citenamefont
  {Marianetti}}]{kotliar2006}%
  \BibitemOpen
  \bibfield  {author} {\bibinfo {author} {\bibfnamefont {G.}~\bibnamefont
  {Kotliar}}, \bibinfo {author} {\bibfnamefont {S.~Y.}\ \bibnamefont
  {Savrasov}}, \bibinfo {author} {\bibfnamefont {K.}~\bibnamefont {Haule}},
  \bibinfo {author} {\bibfnamefont {V.~S.}\ \bibnamefont {Oudovenko}}, \bibinfo
  {author} {\bibfnamefont {O.}~\bibnamefont {Parcollet}}, \ and\ \bibinfo
  {author} {\bibfnamefont {C.~A.}\ \bibnamefont {Marianetti}},\ }\bibfield
  {title} {\enquote {\bibinfo {title} {Electronic structure calculations with
  dynamical mean-field theory},}\ }\href
  {https://doi.org/10.1103/RevModPhys.78.865} {\bibfield  {journal} {\bibinfo
  {journal} {Rev. Mod. Phys.}\ }\textbf {\bibinfo {volume} {78}},\ \bibinfo
  {pages} {865} (\bibinfo {year} {2006})}\BibitemShut {NoStop}%
\bibitem [{\citenamefont {Izumi}\ \emph {et~al.}(2001)\citenamefont {Izumi},
  \citenamefont {Ogimoto}, \citenamefont {Konishi}, \citenamefont {Manako},
  \citenamefont {Kawasaki},\ and\ \citenamefont {Tokura}}]{izumi2001}%
  \BibitemOpen
  \bibfield  {author} {\bibinfo {author} {\bibfnamefont {M.}~\bibnamefont
  {Izumi}}, \bibinfo {author} {\bibfnamefont {Y.}~\bibnamefont {Ogimoto}},
  \bibinfo {author} {\bibfnamefont {Y.}~\bibnamefont {Konishi}}, \bibinfo
  {author} {\bibfnamefont {T.}~\bibnamefont {Manako}}, \bibinfo {author}
  {\bibfnamefont {M.}~\bibnamefont {Kawasaki}}, \ and\ \bibinfo {author}
  {\bibfnamefont {Y.}~\bibnamefont {Tokura}},\ }\bibfield  {title} {\enquote
  {\bibinfo {title} {Perovskite superlattices as tailored materials of
  correlated electrons},}\ }\href {\doibase 10.1016/S0921-5107(01)00569-4}
  {\bibfield  {journal} {\bibinfo  {journal} {Mater. Sci. Eng. B}\ }\textbf
  {\bibinfo {volume} {84}},\ \bibinfo {pages} {53} (\bibinfo {year}
  {2001})}\BibitemShut {NoStop}%
\bibitem [{\citenamefont {Ohtomo}\ and\ \citenamefont
  {Hwang}(2004)}]{ohtomo2004}%
  \BibitemOpen
  \bibfield  {author} {\bibinfo {author} {\bibfnamefont {A.}~\bibnamefont
  {Ohtomo}}\ and\ \bibinfo {author} {\bibfnamefont {H.~Y.}\ \bibnamefont
  {Hwang}},\ }\bibfield  {title} {\enquote {\bibinfo {title} {A high-mobility
  electron gas at the {LaAlO$_3$}/{SrTiO$_3$} heterointerface},}\ }\href
  {\doibase 10.1038/nature02308} {\bibfield  {journal} {\bibinfo  {journal}
  {Nature}\ }\textbf {\bibinfo {volume} {427}},\ \bibinfo {pages} {423}
  (\bibinfo {year} {2004})}\BibitemShut {NoStop}%
\bibitem [{\citenamefont {Hwang}\ \emph {et~al.}(2012)\citenamefont {Hwang},
  \citenamefont {Iwasa}, \citenamefont {Kawasaki}, \citenamefont {Keimer},
  \citenamefont {Nagaosa},\ and\ \citenamefont {Tokura}}]{hwang2012}%
  \BibitemOpen
  \bibfield  {author} {\bibinfo {author} {\bibfnamefont {H.~Y.}\ \bibnamefont
  {Hwang}}, \bibinfo {author} {\bibfnamefont {Y.}~\bibnamefont {Iwasa}},
  \bibinfo {author} {\bibfnamefont {M.}~\bibnamefont {Kawasaki}}, \bibinfo
  {author} {\bibfnamefont {B.}~\bibnamefont {Keimer}}, \bibinfo {author}
  {\bibfnamefont {N.}~\bibnamefont {Nagaosa}}, \ and\ \bibinfo {author}
  {\bibfnamefont {Y.}~\bibnamefont {Tokura}},\ }\bibfield  {title} {\enquote
  {\bibinfo {title} {Emergent phenomena at oxide interfaces},}\ }\href
  {\doibase 10.1038/nmat3223} {\bibfield  {journal} {\bibinfo  {journal}
  {Nature Mater.}\ }\textbf {\bibinfo {volume} {11}},\ \bibinfo {pages} {103}
  (\bibinfo {year} {2012})}\BibitemShut {NoStop}%
\bibitem [{\citenamefont {Chakhalian}\ \emph {et~al.}(2014)\citenamefont
  {Chakhalian}, \citenamefont {Freeland}, \citenamefont {Millis}, \citenamefont
  {Panagopoulos},\ and\ \citenamefont {Rondinelli}}]{chakhalian2014}%
  \BibitemOpen
  \bibfield  {author} {\bibinfo {author} {\bibfnamefont {J.}~\bibnamefont
  {Chakhalian}}, \bibinfo {author} {\bibfnamefont {J.~W.}\ \bibnamefont
  {Freeland}}, \bibinfo {author} {\bibfnamefont {A.~J.}\ \bibnamefont
  {Millis}}, \bibinfo {author} {\bibfnamefont {C.}~\bibnamefont
  {Panagopoulos}}, \ and\ \bibinfo {author} {\bibfnamefont {J.~M.}\
  \bibnamefont {Rondinelli}},\ }\bibfield  {title} {\enquote {\bibinfo {title}
  {Colloquium: {E}mergent properties in plane view: {S}trong correlations at
  oxide interfaces},}\ }\href {\doibase 10.1103/RevModPhys.86.1189} {\bibfield
  {journal} {\bibinfo  {journal} {Rev. Mod. Phys.}\ }\textbf {\bibinfo {volume}
  {86}},\ \bibinfo {pages} {1189--1202} (\bibinfo {year} {2014})}\BibitemShut
  {NoStop}%
\bibitem [{\citenamefont {Laverock}\ \emph {et~al.}(2017)\citenamefont
  {Laverock}, \citenamefont {Gu}, \citenamefont {Jovic}, \citenamefont {Lu},
  \citenamefont {Wolf}, \citenamefont {Qiao}, \citenamefont {Yang},\ and\
  \citenamefont {Smith}}]{laverock2017}%
  \BibitemOpen
  \bibfield  {author} {\bibinfo {author} {\bibfnamefont {J.}~\bibnamefont
  {Laverock}}, \bibinfo {author} {\bibfnamefont {M.}~\bibnamefont {Gu}},
  \bibinfo {author} {\bibfnamefont {V.}~\bibnamefont {Jovic}}, \bibinfo
  {author} {\bibfnamefont {J.~W.}\ \bibnamefont {Lu}}, \bibinfo {author}
  {\bibfnamefont {S.A.}\ \bibnamefont {Wolf}}, \bibinfo {author} {\bibfnamefont
  {R.~M.}\ \bibnamefont {Qiao}}, \bibinfo {author} {\bibfnamefont
  {W.}~\bibnamefont {Yang}}, \ and\ \bibinfo {author} {\bibfnamefont {K.E.}\
  \bibnamefont {Smith}},\ }\bibfield  {title} {\enquote {\bibinfo {title}
  {Nano-engineering of electron correlation in oxide superlattices},}\ }\href
  {\doibase 10.1088/2399-1984/aa8f39} {\bibfield  {journal} {\bibinfo
  {journal} {Nano Futures}\ }\textbf {\bibinfo {volume} {1}},\ \bibinfo {pages}
  {031001} (\bibinfo {year} {2017})}\BibitemShut {NoStop}%
\bibitem [{\citenamefont {Nekrasov}\ \emph {et~al.}(2005)\citenamefont
  {Nekrasov}, \citenamefont {Keller}, \citenamefont {Kondakov}, \citenamefont
  {Kozhevnikov}, \citenamefont {Pruschke}, \citenamefont {Held}, \citenamefont
  {Vollhardt},\ and\ \citenamefont {Anisimov}}]{nekrasov2005}%
  \BibitemOpen
  \bibfield  {author} {\bibinfo {author} {\bibfnamefont {I.~A.}\ \bibnamefont
  {Nekrasov}}, \bibinfo {author} {\bibfnamefont {G.}~\bibnamefont {Keller}},
  \bibinfo {author} {\bibfnamefont {D.~E.}\ \bibnamefont {Kondakov}}, \bibinfo
  {author} {\bibfnamefont {A.~V.}\ \bibnamefont {Kozhevnikov}}, \bibinfo
  {author} {\bibfnamefont {Th.}\ \bibnamefont {Pruschke}}, \bibinfo {author}
  {\bibfnamefont {K.}~\bibnamefont {Held}}, \bibinfo {author} {\bibfnamefont
  {D.}~\bibnamefont {Vollhardt}}, \ and\ \bibinfo {author} {\bibfnamefont
  {V.~I.}\ \bibnamefont {Anisimov}},\ }\bibfield  {title} {\enquote {\bibinfo
  {title} {Comparative study of correlation effects in {CaVO}$_3$ and
  {SrVO}$_3$},}\ }\href {\doibase 10.1103/PhysRevB.72.155106} {\bibfield
  {journal} {\bibinfo  {journal} {Phys. Rev. B}\ }\textbf {\bibinfo {volume}
  {72}},\ \bibinfo {pages} {155106} (\bibinfo {year} {2005})}\BibitemShut
  {NoStop}%
\bibitem [{\citenamefont {Yoshida}\ \emph {et~al.}(2010)\citenamefont
  {Yoshida}, \citenamefont {Hashimoto}, \citenamefont {Takizawa}, \citenamefont
  {Fujimori}, \citenamefont {Kubota}, \citenamefont {Ono},\ and\ \citenamefont
  {Eisaki}}]{yoshida2010}%
  \BibitemOpen
  \bibfield  {author} {\bibinfo {author} {\bibfnamefont {T.}~\bibnamefont
  {Yoshida}}, \bibinfo {author} {\bibfnamefont {M.}~\bibnamefont {Hashimoto}},
  \bibinfo {author} {\bibfnamefont {T.}~\bibnamefont {Takizawa}}, \bibinfo
  {author} {\bibfnamefont {A.}~\bibnamefont {Fujimori}}, \bibinfo {author}
  {\bibfnamefont {M.}~\bibnamefont {Kubota}}, \bibinfo {author} {\bibfnamefont
  {K.}~\bibnamefont {Ono}}, \ and\ \bibinfo {author} {\bibfnamefont
  {H.}~\bibnamefont {Eisaki}},\ }\bibfield  {title} {\enquote {\bibinfo {title}
  {Mass renormalization in the bandwidth-controlled mott-hubbard systems
  {SrVO}$_3$ and {CaVO}$_3$ studied by angle-resolved photoemission
  spectroscopy},}\ }\href {\doibase 10.1103/PhysRevB.82.085119} {\bibfield
  {journal} {\bibinfo  {journal} {Phys. Rev. B}\ }\textbf {\bibinfo {volume}
  {82}},\ \bibinfo {pages} {085119} (\bibinfo {year} {2010})}\BibitemShut
  {NoStop}%
\bibitem [{\citenamefont {Aizaki}\ \emph {et~al.}(2012)\citenamefont {Aizaki},
  \citenamefont {Yoshida}, \citenamefont {Yoshimatsu}, \citenamefont
  {Takizawa}, \citenamefont {Minohara}, \citenamefont {Ideta}, \citenamefont
  {Fujimori}, \citenamefont {Gupta}, \citenamefont {Mahadevan}, \citenamefont
  {Horiba}, \citenamefont {Kumigashira},\ and\ \citenamefont
  {Oshima}}]{aizaki2012}%
  \BibitemOpen
  \bibfield  {author} {\bibinfo {author} {\bibfnamefont {S.}~\bibnamefont
  {Aizaki}}, \bibinfo {author} {\bibfnamefont {T.}~\bibnamefont {Yoshida}},
  \bibinfo {author} {\bibfnamefont {K.}~\bibnamefont {Yoshimatsu}}, \bibinfo
  {author} {\bibfnamefont {M.}~\bibnamefont {Takizawa}}, \bibinfo {author}
  {\bibfnamefont {M.}~\bibnamefont {Minohara}}, \bibinfo {author}
  {\bibfnamefont {S.}~\bibnamefont {Ideta}}, \bibinfo {author} {\bibfnamefont
  {A.}~\bibnamefont {Fujimori}}, \bibinfo {author} {\bibfnamefont
  {K.}~\bibnamefont {Gupta}}, \bibinfo {author} {\bibfnamefont
  {P.}~\bibnamefont {Mahadevan}}, \bibinfo {author} {\bibfnamefont
  {K.}~\bibnamefont {Horiba}}, \bibinfo {author} {\bibfnamefont
  {H.}~\bibnamefont {Kumigashira}}, \ and\ \bibinfo {author} {\bibfnamefont
  {M.}~\bibnamefont {Oshima}},\ }\bibfield  {title} {\enquote {\bibinfo {title}
  {Self-energy on the low- to high-energy electronic structure of correlated
  metal {SrVO}$_3$},}\ }\href {\doibase 10.1103/PhysRevLett.109.056401}
  {\bibfield  {journal} {\bibinfo  {journal} {Phys. Rev. Lett.}\ }\textbf
  {\bibinfo {volume} {109}},\ \bibinfo {pages} {056401} (\bibinfo {year}
  {2012})}\BibitemShut {NoStop}%
\bibitem [{\citenamefont {Sekiyama}\ \emph {et~al.}(2004)\citenamefont
  {Sekiyama}, \citenamefont {Fujiwara}, \citenamefont {Imada}, \citenamefont
  {Suga}, \citenamefont {Eisaki}, \citenamefont {Uchida}, \citenamefont
  {Takegahara}, \citenamefont {Harima}, \citenamefont {Saitoh}, \citenamefont
  {Nekrasov}, \citenamefont {Keller}, \citenamefont {Kondakov}, \citenamefont
  {Kozhevnikov}, \citenamefont {Pruschke}, \citenamefont {Held}, \citenamefont
  {Vollhardt},\ and\ \citenamefont {Anisimov}}]{sekiyama2004}%
  \BibitemOpen
  \bibfield  {author} {\bibinfo {author} {\bibfnamefont {A.}~\bibnamefont
  {Sekiyama}}, \bibinfo {author} {\bibfnamefont {H.}~\bibnamefont {Fujiwara}},
  \bibinfo {author} {\bibfnamefont {S.}~\bibnamefont {Imada}}, \bibinfo
  {author} {\bibfnamefont {S.}~\bibnamefont {Suga}}, \bibinfo {author}
  {\bibfnamefont {H.}~\bibnamefont {Eisaki}}, \bibinfo {author} {\bibfnamefont
  {S.~I.}\ \bibnamefont {Uchida}}, \bibinfo {author} {\bibfnamefont
  {K.}~\bibnamefont {Takegahara}}, \bibinfo {author} {\bibfnamefont
  {H.}~\bibnamefont {Harima}}, \bibinfo {author} {\bibfnamefont
  {Y.}~\bibnamefont {Saitoh}}, \bibinfo {author} {\bibfnamefont {I.~A.}\
  \bibnamefont {Nekrasov}}, \bibinfo {author} {\bibfnamefont {G.}~\bibnamefont
  {Keller}}, \bibinfo {author} {\bibfnamefont {D.~E.}\ \bibnamefont
  {Kondakov}}, \bibinfo {author} {\bibfnamefont {A.~V.}\ \bibnamefont
  {Kozhevnikov}}, \bibinfo {author} {\bibfnamefont {Th.}\ \bibnamefont
  {Pruschke}}, \bibinfo {author} {\bibfnamefont {K.}~\bibnamefont {Held}},
  \bibinfo {author} {\bibfnamefont {D.}~\bibnamefont {Vollhardt}}, \ and\
  \bibinfo {author} {\bibfnamefont {V.~I.}\ \bibnamefont {Anisimov}},\
  }\bibfield  {title} {\enquote {\bibinfo {title} {Mutual experimental and
  theoretical validation of bulk photoemission spectra of
  {Sr}$_{1-x}${Ca}$_x${VO}$_3$},}\ }\href {\doibase
  10.1103/PhysRevLett.93.156402} {\bibfield  {journal} {\bibinfo  {journal}
  {Phys. Rev. Lett.}\ }\textbf {\bibinfo {volume} {93}},\ \bibinfo {pages}
  {156402} (\bibinfo {year} {2004})}\BibitemShut {NoStop}%
\bibitem [{\citenamefont {Pen}\ \emph {et~al.}(1999)\citenamefont {Pen},
  \citenamefont {Abbate}, \citenamefont {Fujimori}, \citenamefont {Tokura},
  \citenamefont {Eisaki}, \citenamefont {Uchida},\ and\ \citenamefont
  {Sawatzky}}]{pen1999}%
  \BibitemOpen
  \bibfield  {author} {\bibinfo {author} {\bibfnamefont {H.~F.}\ \bibnamefont
  {Pen}}, \bibinfo {author} {\bibfnamefont {M.}~\bibnamefont {Abbate}},
  \bibinfo {author} {\bibfnamefont {A.}~\bibnamefont {Fujimori}}, \bibinfo
  {author} {\bibfnamefont {Y.}~\bibnamefont {Tokura}}, \bibinfo {author}
  {\bibfnamefont {H.}~\bibnamefont {Eisaki}}, \bibinfo {author} {\bibfnamefont
  {S.}~\bibnamefont {Uchida}}, \ and\ \bibinfo {author} {\bibfnamefont {G.~A.}\
  \bibnamefont {Sawatzky}},\ }\bibfield  {title} {\enquote {\bibinfo {title}
  {Electronic structure of {Y}$_{1-x}${Ca}$_x${VO}$_3$ studied by high-energy
  spectroscopies},}\ }\href {\doibase 10.1103/PhysRevB.59.7422} {\bibfield
  {journal} {\bibinfo  {journal} {Phys. Rev. B}\ }\textbf {\bibinfo {volume}
  {59}},\ \bibinfo {pages} {7422--7432} (\bibinfo {year} {1999})}\BibitemShut
  {NoStop}%
\bibitem [{\citenamefont {Laverock}\ \emph {et~al.}(2013)\citenamefont
  {Laverock}, \citenamefont {Chen}, \citenamefont {Smith}, \citenamefont
  {Singh}, \citenamefont {Balakrishnan}, \citenamefont {Gu}, \citenamefont
  {Lu}, \citenamefont {Wolf}, \citenamefont {Qiao}, \citenamefont {Yang},\ and\
  \citenamefont {Adell}}]{laverock2013b}%
  \BibitemOpen
  \bibfield  {author} {\bibinfo {author} {\bibfnamefont {J.}~\bibnamefont
  {Laverock}}, \bibinfo {author} {\bibfnamefont {B.}~\bibnamefont {Chen}},
  \bibinfo {author} {\bibfnamefont {K.~E.}\ \bibnamefont {Smith}}, \bibinfo
  {author} {\bibfnamefont {R.~P.}\ \bibnamefont {Singh}}, \bibinfo {author}
  {\bibfnamefont {G.}~\bibnamefont {Balakrishnan}}, \bibinfo {author}
  {\bibfnamefont {M.}~\bibnamefont {Gu}}, \bibinfo {author} {\bibfnamefont
  {J.~W.}\ \bibnamefont {Lu}}, \bibinfo {author} {\bibfnamefont {S.~A.}\
  \bibnamefont {Wolf}}, \bibinfo {author} {\bibfnamefont {R.~M.}\ \bibnamefont
  {Qiao}}, \bibinfo {author} {\bibfnamefont {W.}~\bibnamefont {Yang}}, \ and\
  \bibinfo {author} {\bibfnamefont {J.}~\bibnamefont {Adell}},\ }\bibfield
  {title} {\enquote {\bibinfo {title} {Resonant soft-x-ray emission as a bulk
  probe of correlated electron behavior in metallic
  {Sr}$_x${Ca}$_{1-x}${VO}$_3$},}\ }\href {\doibase
  10.1103/PhysRevLett.111.047402} {\bibfield  {journal} {\bibinfo  {journal}
  {Phys. Rev. Lett.}\ }\textbf {\bibinfo {volume} {111}},\ \bibinfo {pages}
  {047402} (\bibinfo {year} {2013})}\BibitemShut {NoStop}%
\bibitem [{\citenamefont {Georges}\ \emph {et~al.}(1996)\citenamefont
  {Georges}, \citenamefont {Kotliar}, \citenamefont {Krauth},\ and\
  \citenamefont {Rozenberg}}]{georges1996}%
  \BibitemOpen
  \bibfield  {author} {\bibinfo {author} {\bibfnamefont {A.}~\bibnamefont
  {Georges}}, \bibinfo {author} {\bibfnamefont {G.}~\bibnamefont {Kotliar}},
  \bibinfo {author} {\bibfnamefont {W.}~\bibnamefont {Krauth}}, \ and\ \bibinfo
  {author} {\bibfnamefont {M.~J.}\ \bibnamefont {Rozenberg}},\ }\bibfield
  {title} {\enquote {\bibinfo {title} {Dynamical mean-field theory of strongly
  correlated fermion systems and the limit of infinite dimensions},}\ }\href
  {https://doi.org/10.1103/RevModPhys.68.13} {\bibfield  {journal} {\bibinfo
  {journal} {Rev. Mod. Phys.}\ }\textbf {\bibinfo {volume} {68}},\ \bibinfo
  {pages} {13} (\bibinfo {year} {1996})}\BibitemShut {NoStop}%
\bibitem [{\citenamefont {Nekrasov}\ \emph {et~al.}(2006)\citenamefont
  {Nekrasov}, \citenamefont {Held}, \citenamefont {Keller}, \citenamefont
  {Kondakov}, \citenamefont {Pruschke}, \citenamefont {Kollar}, \citenamefont
  {Andersen}, \citenamefont {Anisimov},\ and\ \citenamefont
  {Vollhardt}}]{nekrasov2006}%
  \BibitemOpen
  \bibfield  {author} {\bibinfo {author} {\bibfnamefont {I.~A.}\ \bibnamefont
  {Nekrasov}}, \bibinfo {author} {\bibfnamefont {K.}~\bibnamefont {Held}},
  \bibinfo {author} {\bibfnamefont {G.}~\bibnamefont {Keller}}, \bibinfo
  {author} {\bibfnamefont {D.~E.}\ \bibnamefont {Kondakov}}, \bibinfo {author}
  {\bibfnamefont {Th.}\ \bibnamefont {Pruschke}}, \bibinfo {author}
  {\bibfnamefont {M.}~\bibnamefont {Kollar}}, \bibinfo {author} {\bibfnamefont
  {O.~K.}\ \bibnamefont {Andersen}}, \bibinfo {author} {\bibfnamefont {V.~I.}\
  \bibnamefont {Anisimov}}, \ and\ \bibinfo {author} {\bibfnamefont
  {D.}~\bibnamefont {Vollhardt}},\ }\bibfield  {title} {\enquote {\bibinfo
  {title} {Momentum-resolved spectral functions of {SrVO}$_3$ calculated by
  {LDA+DMFT}},}\ }\href {\doibase 10.1103/PhysRevB.73.155112} {\bibfield
  {journal} {\bibinfo  {journal} {Phys. Rev. B}\ }\textbf {\bibinfo {volume}
  {73}},\ \bibinfo {pages} {155112} (\bibinfo {year} {2006})}\BibitemShut
  {NoStop}%
\bibitem [{\citenamefont {Byczuk}\ \emph {et~al.}(2007)\citenamefont {Byczuk},
  \citenamefont {Kollar}, \citenamefont {Held}, \citenamefont {Yang},
  \citenamefont {Nekrasov}, \citenamefont {Pruschke},\ and\ \citenamefont
  {Vollhardt}}]{byczuk2007}%
  \BibitemOpen
  \bibfield  {author} {\bibinfo {author} {\bibfnamefont {K.}~\bibnamefont
  {Byczuk}}, \bibinfo {author} {\bibfnamefont {M.}~\bibnamefont {Kollar}},
  \bibinfo {author} {\bibfnamefont {K.}~\bibnamefont {Held}}, \bibinfo {author}
  {\bibfnamefont {Y.-F.}\ \bibnamefont {Yang}}, \bibinfo {author}
  {\bibfnamefont {I.~A.}\ \bibnamefont {Nekrasov}}, \bibinfo {author}
  {\bibfnamefont {Th.}\ \bibnamefont {Pruschke}}, \ and\ \bibinfo {author}
  {\bibfnamefont {D.}~\bibnamefont {Vollhardt}},\ }\bibfield  {title} {\enquote
  {\bibinfo {title} {Kinks in the dispersion of strongly correlated
  electrons},}\ }\href {https://doi.org/10.1038/nphys538} {\bibfield  {journal}
  {\bibinfo  {journal} {Nat. Phys.}\ }\textbf {\bibinfo {volume} {3}},\
  \bibinfo {pages} {168} (\bibinfo {year} {2007})}\BibitemShut {NoStop}%
\bibitem [{\citenamefont {Tomczak}\ \emph {et~al.}(2012)\citenamefont
  {Tomczak}, \citenamefont {Casula}, \citenamefont {Miyake}, \citenamefont
  {Aryasetiawan},\ and\ \citenamefont {Biermann}}]{tomczak2012}%
  \BibitemOpen
  \bibfield  {author} {\bibinfo {author} {\bibfnamefont {J.~M.}\ \bibnamefont
  {Tomczak}}, \bibinfo {author} {\bibfnamefont {M.}~\bibnamefont {Casula}},
  \bibinfo {author} {\bibfnamefont {T.}~\bibnamefont {Miyake}}, \bibinfo
  {author} {\bibfnamefont {F.}~\bibnamefont {Aryasetiawan}}, \ and\ \bibinfo
  {author} {\bibfnamefont {S.}~\bibnamefont {Biermann}},\ }\bibfield  {title}
  {\enquote {\bibinfo {title} {Combined {GW} and dynamical mean-field theory:
  {D}ynamical screening effects in transition metal oxides},}\ }\href
  {https://doi.org/10.1209/0295-5075/100/67001} {\bibfield  {journal} {\bibinfo
   {journal} {Europhys. Lett.}\ }\textbf {\bibinfo {volume} {100}},\ \bibinfo
  {pages} {67001} (\bibinfo {year} {2012})}\BibitemShut {NoStop}%
\bibitem [{\citenamefont {Maiti}\ \emph {et~al.}(2001)\citenamefont {Maiti},
  \citenamefont {Sarma}, \citenamefont {Rozenberg}, \citenamefont {Inoue},
  \citenamefont {Makino}, \citenamefont {Goto}, \citenamefont {Pedio},\ and\
  \citenamefont {Cimino}}]{maiti2001}%
  \BibitemOpen
  \bibfield  {author} {\bibinfo {author} {\bibfnamefont {K.}~\bibnamefont
  {Maiti}}, \bibinfo {author} {\bibfnamefont {D.~D.}\ \bibnamefont {Sarma}},
  \bibinfo {author} {\bibfnamefont {M.~J.}\ \bibnamefont {Rozenberg}}, \bibinfo
  {author} {\bibfnamefont {I.~H.}\ \bibnamefont {Inoue}}, \bibinfo {author}
  {\bibfnamefont {H.}~\bibnamefont {Makino}}, \bibinfo {author} {\bibfnamefont
  {O.}~\bibnamefont {Goto}}, \bibinfo {author} {\bibfnamefont {M.}~\bibnamefont
  {Pedio}}, \ and\ \bibinfo {author} {\bibfnamefont {R.}~\bibnamefont
  {Cimino}},\ }\bibfield  {title} {\enquote {\bibinfo {title} {Electronic
  structure of {Ca}$_{1-x}${Sr}$_x${VO}$_3$: A tale of two energy scales},}\
  }\href {\doibase 10.1209/epl/i2001-00406-6} {\bibfield  {journal} {\bibinfo
  {journal} {EPL (Europhys. Lett.)}\ }\textbf {\bibinfo {volume} {10}},\
  \bibinfo {pages} {246} (\bibinfo {year} {2001})}\BibitemShut {NoStop}%
\bibitem [{\citenamefont {Liebsch}(2003)}]{liebsch2003}%
  \BibitemOpen
  \bibfield  {author} {\bibinfo {author} {\bibfnamefont {A.}~\bibnamefont
  {Liebsch}},\ }\bibfield  {title} {\enquote {\bibinfo {title} {{Surface versus
  Bulk {Coulomb} Correlations in Photoemission Spectra of {SrVO$_3$} and
  {CaVO$_3$}}},}\ }\href {\doibase 10.1103/PhysRevLett.90.096401} {\bibfield
  {journal} {\bibinfo  {journal} {Phys. Rev. Lett.}\ }\textbf {\bibinfo
  {volume} {90}},\ \bibinfo {pages} {096401} (\bibinfo {year}
  {2003})}\BibitemShut {NoStop}%
\bibitem [{\citenamefont {Laverock}\ \emph {et~al.}(2015)\citenamefont
  {Laverock}, \citenamefont {Kuyyalil}, \citenamefont {Chen}, \citenamefont
  {Singh}, \citenamefont {Karlin}, \citenamefont {Woicik}, \citenamefont
  {Balakrishnan},\ and\ \citenamefont {Smith}}]{laverock2015b}%
  \BibitemOpen
  \bibfield  {author} {\bibinfo {author} {\bibfnamefont {J.}~\bibnamefont
  {Laverock}}, \bibinfo {author} {\bibfnamefont {J.}~\bibnamefont {Kuyyalil}},
  \bibinfo {author} {\bibfnamefont {B.}~\bibnamefont {Chen}}, \bibinfo {author}
  {\bibfnamefont {R.~P.}\ \bibnamefont {Singh}}, \bibinfo {author}
  {\bibfnamefont {B.}~\bibnamefont {Karlin}}, \bibinfo {author} {\bibfnamefont
  {J.~C.}\ \bibnamefont {Woicik}}, \bibinfo {author} {\bibfnamefont
  {G.}~\bibnamefont {Balakrishnan}}, \ and\ \bibinfo {author} {\bibfnamefont
  {K.~E.}\ \bibnamefont {Smith}},\ }\bibfield  {title} {\enquote {\bibinfo
  {title} {Enhanced electron correlations at the {Sr}$_x${Ca}$_{1-x}${VO}$_3$
  surface},}\ }\href {\doibase 10.1103/PhysRevB.91.165123} {\bibfield
  {journal} {\bibinfo  {journal} {Phys. Rev. B}\ }\textbf {\bibinfo {volume}
  {91}},\ \bibinfo {pages} {165123} (\bibinfo {year} {2015})}\BibitemShut
  {NoStop}%
\bibitem [{\citenamefont {Ishida}\ \emph {et~al.}(2006)\citenamefont {Ishida},
  \citenamefont {Wortmann},\ and\ \citenamefont {Liebsch}}]{ishida2006}%
  \BibitemOpen
  \bibfield  {author} {\bibinfo {author} {\bibfnamefont {H.}~\bibnamefont
  {Ishida}}, \bibinfo {author} {\bibfnamefont {D.}~\bibnamefont {Wortmann}}, \
  and\ \bibinfo {author} {\bibfnamefont {A.}~\bibnamefont {Liebsch}},\
  }\bibfield  {title} {\enquote {\bibinfo {title} {Electronic structure of
  {SrVO}$_{3}(001)$ surfaces: A local-density approximation plus dynamical
  mean-field theory calculation},}\ }\href {\doibase
  10.1103/PhysRevB.73.245421} {\bibfield  {journal} {\bibinfo  {journal} {Phys.
  Rev. B}\ }\textbf {\bibinfo {volume} {73}},\ \bibinfo {pages} {245421}
  (\bibinfo {year} {2006})}\BibitemShut {NoStop}%
\bibitem [{\citenamefont {Yoshimatsu}\ \emph {et~al.}(2010)\citenamefont
  {Yoshimatsu}, \citenamefont {Okabe}, \citenamefont {Kumigashira},
  \citenamefont {Okamoto}, \citenamefont {Aizaki}, \citenamefont {Fujimori},\
  and\ \citenamefont {Oshima}}]{yoshimatsu2010}%
  \BibitemOpen
  \bibfield  {author} {\bibinfo {author} {\bibfnamefont {K.}~\bibnamefont
  {Yoshimatsu}}, \bibinfo {author} {\bibfnamefont {T.}~\bibnamefont {Okabe}},
  \bibinfo {author} {\bibfnamefont {H.}~\bibnamefont {Kumigashira}}, \bibinfo
  {author} {\bibfnamefont {S.}~\bibnamefont {Okamoto}}, \bibinfo {author}
  {\bibfnamefont {S.}~\bibnamefont {Aizaki}}, \bibinfo {author} {\bibfnamefont
  {A.}~\bibnamefont {Fujimori}}, \ and\ \bibinfo {author} {\bibfnamefont
  {M.}~\bibnamefont {Oshima}},\ }\bibfield  {title} {\enquote {\bibinfo {title}
  {Dimensional-crossover-driven metal-insulator transition in {SrVO}$_3$
  ultrathin films},}\ }\href {\doibase 10.1103/PhysRevLett.104.147601}
  {\bibfield  {journal} {\bibinfo  {journal} {Phys. Rev. Lett.}\ }\textbf
  {\bibinfo {volume} {104}},\ \bibinfo {pages} {147601} (\bibinfo {year}
  {2010})}\BibitemShut {NoStop}%
\bibitem [{\citenamefont {Zhong}\ \emph {et~al.}(2015)\citenamefont {Zhong},
  \citenamefont {Wallerberger}, \citenamefont {Tomczak}, \citenamefont
  {Taranto}, \citenamefont {Parragh}, \citenamefont {Toschi}, \citenamefont
  {Sangiovanni},\ and\ \citenamefont {Held}}]{zhong2015}%
  \BibitemOpen
  \bibfield  {author} {\bibinfo {author} {\bibfnamefont {Z.}~\bibnamefont
  {Zhong}}, \bibinfo {author} {\bibfnamefont {M.}~\bibnamefont {Wallerberger}},
  \bibinfo {author} {\bibfnamefont {J.~M.}\ \bibnamefont {Tomczak}}, \bibinfo
  {author} {\bibfnamefont {C.}~\bibnamefont {Taranto}}, \bibinfo {author}
  {\bibfnamefont {N.}~\bibnamefont {Parragh}}, \bibinfo {author} {\bibfnamefont
  {A.}~\bibnamefont {Toschi}}, \bibinfo {author} {\bibfnamefont
  {G.}~\bibnamefont {Sangiovanni}}, \ and\ \bibinfo {author} {\bibfnamefont
  {K.}~\bibnamefont {Held}},\ }\bibfield  {title} {\enquote {\bibinfo {title}
  {{Electronics with Correlated Oxides: {SrVO$_3$}/{SrTiO$_3$} as a {Mott}
  Transistor}},}\ }\href {\doibase 10.1103/PhysRevLett.114.246401} {\bibfield
  {journal} {\bibinfo  {journal} {Phys. Rev. Lett.}\ }\textbf {\bibinfo
  {volume} {114}},\ \bibinfo {pages} {246401} (\bibinfo {year}
  {2015})}\BibitemShut {NoStop}%
\bibitem [{\citenamefont {Gu}\ \emph {et~al.}(2014)\citenamefont {Gu},
  \citenamefont {Wolf},\ and\ \citenamefont {Lu}}]{gu2014}%
  \BibitemOpen
  \bibfield  {author} {\bibinfo {author} {\bibfnamefont {M.}~\bibnamefont
  {Gu}}, \bibinfo {author} {\bibfnamefont {S.~A.}\ \bibnamefont {Wolf}}, \ and\
  \bibinfo {author} {\bibfnamefont {J.}~\bibnamefont {Lu}},\ }\bibfield
  {title} {\enquote {\bibinfo {title} {Two-dimensional {Mott} insulators in
  {SrVO}$_3$ ultrathin films},}\ }\href {\doibase 10.1002/admi.201300126}
  {\bibfield  {journal} {\bibinfo  {journal} {Adv. Mater. Interfaces}\ }\textbf
  {\bibinfo {volume} {1}},\ \bibinfo {pages} {1300126} (\bibinfo {year}
  {2014})}\BibitemShut {NoStop}%
\bibitem [{\citenamefont {Bhandary}\ \emph {et~al.}(2016)\citenamefont
  {Bhandary}, \citenamefont {Assmann}, \citenamefont {Aichhorn},\ and\
  \citenamefont {Held}}]{bhandary2016}%
  \BibitemOpen
  \bibfield  {author} {\bibinfo {author} {\bibfnamefont {S.}~\bibnamefont
  {Bhandary}}, \bibinfo {author} {\bibfnamefont {E.}~\bibnamefont {Assmann}},
  \bibinfo {author} {\bibfnamefont {M.}~\bibnamefont {Aichhorn}}, \ and\
  \bibinfo {author} {\bibfnamefont {K.}~\bibnamefont {Held}},\ }\bibfield
  {title} {\enquote {\bibinfo {title} {Charge self-consistency in density
  functional theory combined with dynamical mean field theory: $k$-space
  reoccupation and orbital order},}\ }\href {\doibase
  10.1103/PhysRevB.94.155131} {\bibfield  {journal} {\bibinfo  {journal} {Phys.
  Rev. B}\ }\textbf {\bibinfo {volume} {94}},\ \bibinfo {pages} {155131}
  (\bibinfo {year} {2016})}\BibitemShut {NoStop}%
\bibitem [{\citenamefont {Sch{\"u}ler}\ \emph {et~al.}(2018)\citenamefont
  {Sch{\"u}ler}, \citenamefont {Peil}, \citenamefont {Kraberger}, \citenamefont
  {Pordzik}, \citenamefont {Marsman}, \citenamefont {Kresse}, \citenamefont
  {Wehling},\ and\ \citenamefont {Aichhorn}}]{schuler2018}%
  \BibitemOpen
  \bibfield  {author} {\bibinfo {author} {\bibfnamefont {M.}~\bibnamefont
  {Sch{\"u}ler}}, \bibinfo {author} {\bibfnamefont {O.~E.}\ \bibnamefont
  {Peil}}, \bibinfo {author} {\bibfnamefont {G.~J.}\ \bibnamefont {Kraberger}},
  \bibinfo {author} {\bibfnamefont {R.}~\bibnamefont {Pordzik}}, \bibinfo
  {author} {\bibfnamefont {M.}~\bibnamefont {Marsman}}, \bibinfo {author}
  {\bibfnamefont {G.}~\bibnamefont {Kresse}}, \bibinfo {author} {\bibfnamefont
  {T.~O.}\ \bibnamefont {Wehling}}, \ and\ \bibinfo {author} {\bibfnamefont
  {M.}~\bibnamefont {Aichhorn}},\ }\bibfield  {title} {\enquote {\bibinfo
  {title} {Charge self-consistent many-body corrections using optimized
  projected localized orbitals},}\ }\href {\doibase 10.1088/1361-648X/aae80a}
  {\bibfield  {journal} {\bibinfo  {journal} {J. Phys.: Conden. Matter}\
  }\textbf {\bibinfo {volume} {30}},\ \bibinfo {pages} {475901} (\bibinfo
  {year} {2018})}\BibitemShut {NoStop}%
\bibitem [{\citenamefont {Hampel}\ \emph {et~al.}(2019)\citenamefont {Hampel},
  \citenamefont {Beck},\ and\ \citenamefont {Ederer}}]{hampel2019}%
  \BibitemOpen
  \bibfield  {author} {\bibinfo {author} {\bibfnamefont {A.}~\bibnamefont
  {Hampel}}, \bibinfo {author} {\bibfnamefont {S.}~\bibnamefont {Beck}}, \ and\
  \bibinfo {author} {\bibfnamefont {C.}~\bibnamefont {Ederer}},\ }\bibfield
  {title} {\enquote {\bibinfo {title} {Charge self-consistency and
  double-counting in {DFT+DMFT} calculations for complex transition metal
  oxides},}\ }\href {https://arxiv.org/abs/1907.10339} {\bibfield  {journal}
  {\bibinfo  {journal} {arXiv:1907.10339}\ } (\bibinfo {year}
  {2019})}\BibitemShut {NoStop}%
\bibitem [{\citenamefont {Beck}\ \emph {et~al.}(2018)\citenamefont {Beck},
  \citenamefont {Sclauzero}, \citenamefont {Chopra},\ and\ \citenamefont
  {Ederer}}]{beck2018}%
  \BibitemOpen
  \bibfield  {author} {\bibinfo {author} {\bibfnamefont {S.}~\bibnamefont
  {Beck}}, \bibinfo {author} {\bibfnamefont {G.}~\bibnamefont {Sclauzero}},
  \bibinfo {author} {\bibfnamefont {U.}~\bibnamefont {Chopra}}, \ and\ \bibinfo
  {author} {\bibfnamefont {C.}~\bibnamefont {Ederer}},\ }\bibfield  {title}
  {\enquote {\bibinfo {title} {Metal-insulator transition in {CaVO}$_3$ thin
  films: {I}nterplay between epitaxial strain, dimensional confinement, and
  surface effects},}\ }\href {\doibase 10.1103/PhysRevB.97.075107} {\bibfield
  {journal} {\bibinfo  {journal} {Phys. Rev. B}\ }\textbf {\bibinfo {volume}
  {97}},\ \bibinfo {pages} {075107} (\bibinfo {year} {2018})}\BibitemShut
  {NoStop}%
\bibitem [{\citenamefont {Sclauzero}\ \emph {et~al.}(2016)\citenamefont
  {Sclauzero}, \citenamefont {Dymkowski},\ and\ \citenamefont
  {Ederer}}]{sclauzero2016}%
  \BibitemOpen
  \bibfield  {author} {\bibinfo {author} {\bibfnamefont {G.}~\bibnamefont
  {Sclauzero}}, \bibinfo {author} {\bibfnamefont {K.}~\bibnamefont
  {Dymkowski}}, \ and\ \bibinfo {author} {\bibfnamefont {C.}~\bibnamefont
  {Ederer}},\ }\bibfield  {title} {\enquote {\bibinfo {title} {{Tuning the
  metal-insulator transition in $d^1$ and $d^2$ perovskites by epitaxial
  strain: {A} first-principles-based study}},}\ }\href
  {https://doi.org/10.1103/PhysRevB.94.245109} {\bibfield  {journal} {\bibinfo
  {journal} {Phys. Rev. B}\ }\textbf {\bibinfo {volume} {94}},\ \bibinfo
  {pages} {245109} (\bibinfo {year} {2016})}\BibitemShut {NoStop}%
\bibitem [{\citenamefont {Colakerol}\ \emph {et~al.}(2006)\citenamefont
  {Colakerol}, \citenamefont {Veal}, \citenamefont {Jeong}, \citenamefont
  {Plucinski}, \citenamefont {DeMasi}, \citenamefont {Learmonth}, \citenamefont
  {Glans}, \citenamefont {Wang}, \citenamefont {Zhang}, \citenamefont {Piper},
  \citenamefont {Jefferson}, \citenamefont {Fedorov}, \citenamefont {Chen},
  \citenamefont {Moustakas}, \citenamefont {McConville},\ and\ \citenamefont
  {Smith}}]{colakerol2006}%
  \BibitemOpen
  \bibfield  {author} {\bibinfo {author} {\bibfnamefont {L.}~\bibnamefont
  {Colakerol}}, \bibinfo {author} {\bibfnamefont {T.~D.}\ \bibnamefont {Veal}},
  \bibinfo {author} {\bibfnamefont {H.-K.}\ \bibnamefont {Jeong}}, \bibinfo
  {author} {\bibfnamefont {L.}~\bibnamefont {Plucinski}}, \bibinfo {author}
  {\bibfnamefont {A.}~\bibnamefont {DeMasi}}, \bibinfo {author} {\bibfnamefont
  {T.}~\bibnamefont {Learmonth}}, \bibinfo {author} {\bibfnamefont {P.-A.}\
  \bibnamefont {Glans}}, \bibinfo {author} {\bibfnamefont {S.}~\bibnamefont
  {Wang}}, \bibinfo {author} {\bibfnamefont {Y.}~\bibnamefont {Zhang}},
  \bibinfo {author} {\bibfnamefont {L.~F.~J.}\ \bibnamefont {Piper}}, \bibinfo
  {author} {\bibfnamefont {P.~H.}\ \bibnamefont {Jefferson}}, \bibinfo {author}
  {\bibfnamefont {A.}~\bibnamefont {Fedorov}}, \bibinfo {author} {\bibfnamefont
  {T.-C.}\ \bibnamefont {Chen}}, \bibinfo {author} {\bibfnamefont {T.~D.}\
  \bibnamefont {Moustakas}}, \bibinfo {author} {\bibfnamefont {C.~F.}\
  \bibnamefont {McConville}}, \ and\ \bibinfo {author} {\bibfnamefont {K.~E.}\
  \bibnamefont {Smith}},\ }\bibfield  {title} {\enquote {\bibinfo {title}
  {Quantized electron accumulation states in indium nitride studied by
  angle-resolved photoemission spectroscopy},}\ }\href {\doibase
  10.1103/PhysRevLett.97.237601} {\bibfield  {journal} {\bibinfo  {journal}
  {Phys. Rev. Lett.}\ }\textbf {\bibinfo {volume} {97}},\ \bibinfo {pages}
  {237601} (\bibinfo {year} {2006})}\BibitemShut {NoStop}%
\bibitem [{\citenamefont {Chiang}(2000)}]{chiang2000}%
  \BibitemOpen
  \bibfield  {author} {\bibinfo {author} {\bibfnamefont {T.-C.}\ \bibnamefont
  {Chiang}},\ }\bibfield  {title} {\enquote {\bibinfo {title} {Photoemission
  studies of quantum well states in thin films},}\ }\href {\doibase
  10.1016/S0167-5729(00)00006-6} {\bibfield  {journal} {\bibinfo  {journal}
  {Surf. Sci. Rep.}\ }\textbf {\bibinfo {volume} {39}},\ \bibinfo {pages} {181}
  (\bibinfo {year} {2000})}\BibitemShut {NoStop}%
\bibitem [{\citenamefont {Liu}\ \emph {et~al.}(2012)\citenamefont {Liu},
  \citenamefont {Kareev}, \citenamefont {Meyers}, \citenamefont {Gray},
  \citenamefont {Ryan}, \citenamefont {Freeland},\ and\ \citenamefont
  {Chakhalian}}]{liu2012}%
  \BibitemOpen
  \bibfield  {author} {\bibinfo {author} {\bibfnamefont {J.}~\bibnamefont
  {Liu}}, \bibinfo {author} {\bibfnamefont {M.}~\bibnamefont {Kareev}},
  \bibinfo {author} {\bibfnamefont {D.}~\bibnamefont {Meyers}}, \bibinfo
  {author} {\bibfnamefont {B.}~\bibnamefont {Gray}}, \bibinfo {author}
  {\bibfnamefont {P.}~\bibnamefont {Ryan}}, \bibinfo {author} {\bibfnamefont
  {J.~W.}\ \bibnamefont {Freeland}}, \ and\ \bibinfo {author} {\bibfnamefont
  {J.}~\bibnamefont {Chakhalian}},\ }\bibfield  {title} {\enquote {\bibinfo
  {title} {Metal-insulator transition and orbital reconstruction in {M}ott-type
  quantum wells made of {NdNiO$_3$}},}\ }\href {\doibase
  10.1103/PhysRevLett.109.107402} {\bibfield  {journal} {\bibinfo  {journal}
  {Phys. Rev. Lett.}\ }\textbf {\bibinfo {volume} {109}},\ \bibinfo {pages}
  {107402} (\bibinfo {year} {2012})}\BibitemShut {NoStop}%
\bibitem [{\citenamefont {Jeong}\ \emph {et~al.}(2020)\citenamefont {Jeong},
  \citenamefont {Min}, \citenamefont {Woo}, \citenamefont {Kim}, \citenamefont
  {Zhang}, \citenamefont {Cho}, \citenamefont {Son}, \citenamefont {Kim},
  \citenamefont {Han}, \citenamefont {Park}, \citenamefont {Jeong},
  \citenamefont {Ohta}, \citenamefont {Lee}, \citenamefont {Noh}, \citenamefont
  {Lee},\ and\ \citenamefont {Choi}}]{jeong2020}%
  \BibitemOpen
  \bibfield  {author} {\bibinfo {author} {\bibfnamefont {S.~G.}\ \bibnamefont
  {Jeong}}, \bibinfo {author} {\bibfnamefont {T.}~\bibnamefont {Min}}, \bibinfo
  {author} {\bibfnamefont {S.}~\bibnamefont {Woo}}, \bibinfo {author}
  {\bibfnamefont {J.}~\bibnamefont {Kim}}, \bibinfo {author} {\bibfnamefont
  {Y.-Q.}\ \bibnamefont {Zhang}}, \bibinfo {author} {\bibfnamefont {S.~W.}\
  \bibnamefont {Cho}}, \bibinfo {author} {\bibfnamefont {J.}~\bibnamefont
  {Son}}, \bibinfo {author} {\bibfnamefont {Y.-M.}\ \bibnamefont {Kim}},
  \bibinfo {author} {\bibfnamefont {J.~H.}\ \bibnamefont {Han}}, \bibinfo
  {author} {\bibfnamefont {S.}~\bibnamefont {Park}}, \bibinfo {author}
  {\bibfnamefont {H.~Y.}\ \bibnamefont {Jeong}}, \bibinfo {author}
  {\bibfnamefont {H.}~\bibnamefont {Ohta}}, \bibinfo {author} {\bibfnamefont
  {S.}~\bibnamefont {Lee}}, \bibinfo {author} {\bibfnamefont {T.~W.}\
  \bibnamefont {Noh}}, \bibinfo {author} {\bibfnamefont {J.}~\bibnamefont
  {Lee}}, \ and\ \bibinfo {author} {\bibfnamefont {W.~S.}\ \bibnamefont
  {Choi}},\ }\bibfield  {title} {\enquote {\bibinfo {title} {Phase instability
  amid dimensional crossover in artificial oxide crystal},}\ }\href {\doibase
  10.1103/PhysRevLett.124.026401} {\bibfield  {journal} {\bibinfo  {journal}
  {Phys. Rev. Lett.}\ }\textbf {\bibinfo {volume} {124}},\ \bibinfo {pages}
  {026401} (\bibinfo {year} {2020})}\BibitemShut {NoStop}%
\bibitem [{\citenamefont {Kawasaki}\ \emph {et~al.}(2018)\citenamefont
  {Kawasaki}, \citenamefont {Kim}, \citenamefont {Nelson}, \citenamefont
  {Crisp}, \citenamefont {Zollner}, \citenamefont {Biegenwald}, \citenamefont
  {Heron}, \citenamefont {Fennie}, \citenamefont {Schlom},\ and\ \citenamefont
  {Shen}}]{kawasaki2018}%
  \BibitemOpen
  \bibfield  {author} {\bibinfo {author} {\bibfnamefont {J.~K.}\ \bibnamefont
  {Kawasaki}}, \bibinfo {author} {\bibfnamefont {C.~H.}\ \bibnamefont {Kim}},
  \bibinfo {author} {\bibfnamefont {J.~N.}\ \bibnamefont {Nelson}}, \bibinfo
  {author} {\bibfnamefont {S.}~\bibnamefont {Crisp}}, \bibinfo {author}
  {\bibfnamefont {C.~J.}\ \bibnamefont {Zollner}}, \bibinfo {author}
  {\bibfnamefont {E.}~\bibnamefont {Biegenwald}}, \bibinfo {author}
  {\bibfnamefont {J.~T.}\ \bibnamefont {Heron}}, \bibinfo {author}
  {\bibfnamefont {C.~J.}\ \bibnamefont {Fennie}}, \bibinfo {author}
  {\bibfnamefont {D.~G.}\ \bibnamefont {Schlom}}, \ and\ \bibinfo {author}
  {\bibfnamefont {K.~M.}\ \bibnamefont {Shen}},\ }\bibfield  {title} {\enquote
  {\bibinfo {title} {Engineering carrier effective masses in ultrathin quantum
  wells of {IrO$_2$}},}\ }\href {\doibase 10.1103/PhysRevLett.121.176802}
  {\bibfield  {journal} {\bibinfo  {journal} {Phys. Rev. Lett.}\ }\textbf
  {\bibinfo {volume} {121}},\ \bibinfo {pages} {176802} (\bibinfo {year}
  {2018})}\BibitemShut {NoStop}%
\bibitem [{\citenamefont {Yoshimatsu}\ \emph {et~al.}(2011)\citenamefont
  {Yoshimatsu}, \citenamefont {Horiba}, \citenamefont {Kumigashira},
  \citenamefont {Yoshida}, \citenamefont {Fujimori},\ and\ \citenamefont
  {Oshima}}]{yoshimatsu2011}%
  \BibitemOpen
  \bibfield  {author} {\bibinfo {author} {\bibfnamefont {K.}~\bibnamefont
  {Yoshimatsu}}, \bibinfo {author} {\bibfnamefont {K.}~\bibnamefont {Horiba}},
  \bibinfo {author} {\bibfnamefont {H.}~\bibnamefont {Kumigashira}}, \bibinfo
  {author} {\bibfnamefont {T.}~\bibnamefont {Yoshida}}, \bibinfo {author}
  {\bibfnamefont {A.}~\bibnamefont {Fujimori}}, \ and\ \bibinfo {author}
  {\bibfnamefont {M.}~\bibnamefont {Oshima}},\ }\bibfield  {title} {\enquote
  {\bibinfo {title} {Metallic quantum well states in artificial structures of
  strongly correlated oxide},}\ }\href {\doibase 10.1126/science.1205771}
  {\bibfield  {journal} {\bibinfo  {journal} {Science}\ }\textbf {\bibinfo
  {volume} {333}},\ \bibinfo {pages} {319} (\bibinfo {year}
  {2011})}\BibitemShut {NoStop}%
\bibitem [{\citenamefont {Kobayashi}\ \emph {et~al.}(2015)\citenamefont
  {Kobayashi}, \citenamefont {Yoshimatsu}, \citenamefont {Sakai}, \citenamefont
  {Kitamura}, \citenamefont {Horiba}, \citenamefont {Fujimori},\ and\
  \citenamefont {Kumigashira}}]{kobayashi2015}%
  \BibitemOpen
  \bibfield  {author} {\bibinfo {author} {\bibfnamefont {M.}~\bibnamefont
  {Kobayashi}}, \bibinfo {author} {\bibfnamefont {K.}~\bibnamefont
  {Yoshimatsu}}, \bibinfo {author} {\bibfnamefont {E.}~\bibnamefont {Sakai}},
  \bibinfo {author} {\bibfnamefont {M.}~\bibnamefont {Kitamura}}, \bibinfo
  {author} {\bibfnamefont {K.}~\bibnamefont {Horiba}}, \bibinfo {author}
  {\bibfnamefont {A.}~\bibnamefont {Fujimori}}, \ and\ \bibinfo {author}
  {\bibfnamefont {H.}~\bibnamefont {Kumigashira}},\ }\bibfield  {title}
  {\enquote {\bibinfo {title} {Origin of the anomalous mass renormalization in
  metallic quantum well states of strongly correlated oxide {SrVO}$_3$},}\
  }\href {\doibase 10.1103/PhysRevLett.115.076801} {\bibfield  {journal}
  {\bibinfo  {journal} {Phys. Rev. Lett.}\ }\textbf {\bibinfo {volume} {115}},\
  \bibinfo {pages} {076801} (\bibinfo {year} {2015})}\BibitemShut {NoStop}%
\bibitem [{\citenamefont {Adler}\ \emph {et~al.}(2019)\citenamefont {Adler},
  \citenamefont {Kang}, \citenamefont {Yee},\ and\ \citenamefont
  {Kotliar}}]{adler2019}%
  \BibitemOpen
  \bibfield  {author} {\bibinfo {author} {\bibfnamefont {R.}~\bibnamefont
  {Adler}}, \bibinfo {author} {\bibfnamefont {C.-J.}\ \bibnamefont {Kang}},
  \bibinfo {author} {\bibfnamefont {C.-H.}\ \bibnamefont {Yee}}, \ and\
  \bibinfo {author} {\bibfnamefont {G.}~\bibnamefont {Kotliar}},\ }\bibfield
  {title} {\enquote {\bibinfo {title} {Correlated materials design: prospects
  and challenges},}\ }\href {\doibase 10.1088/1361-6633/aadca4} {\bibfield
  {journal} {\bibinfo  {journal} {Rep. Prog. Phys.}\ }\textbf {\bibinfo
  {volume} {82}},\ \bibinfo {pages} {012504} (\bibinfo {year}
  {2019})}\BibitemShut {NoStop}%
\bibitem [{\citenamefont {Dewhurst}\ \emph {et~al.}(2019)\citenamefont
  {Dewhurst}, \citenamefont {Sharma}, \citenamefont {Nordstr\"{o}m},
  \citenamefont {Cricchio}, \citenamefont {Granas},\ and\ \citenamefont
  {Gross}}]{elk}%
  \BibitemOpen
  \bibfield  {author} {\bibinfo {author} {\bibfnamefont {J.~K.}\ \bibnamefont
  {Dewhurst}}, \bibinfo {author} {\bibfnamefont {S.}~\bibnamefont {Sharma}},
  \bibinfo {author} {\bibfnamefont {L.}~\bibnamefont {Nordstr\"{o}m}}, \bibinfo
  {author} {\bibfnamefont {F.}~\bibnamefont {Cricchio}}, \bibinfo {author}
  {\bibfnamefont {O.}~\bibnamefont {Granas}}, \ and\ \bibinfo {author}
  {\bibfnamefont {E.~K.~U.}\ \bibnamefont {Gross}},\ }\href@noop {} {}\bibinfo
  {howpublished} {\url{http://elk.sourceforge.net/}} (\bibinfo {year}
  {2019})\BibitemShut {NoStop}%
\bibitem [{\citenamefont {Gu}\ \emph {et~al.}(2018)\citenamefont {Gu},
  \citenamefont {Wolf},\ and\ \citenamefont {Lu}}]{gu2018}%
  \BibitemOpen
  \bibfield  {author} {\bibinfo {author} {\bibfnamefont {M.}~\bibnamefont
  {Gu}}, \bibinfo {author} {\bibfnamefont {S.~A.}\ \bibnamefont {Wolf}}, \ and\
  \bibinfo {author} {\bibfnamefont {J.}~\bibnamefont {Lu}},\ }\bibfield
  {title} {\enquote {\bibinfo {title} {Transport phenomena in
  {SrVO$_3$}/{SrTiO$_3$} superlattices},}\ }\href {\doibase
  10.1088/1361-6463/aaabac} {\bibfield  {journal} {\bibinfo  {journal} {J.
  Phys. D: Appl. Phys.}\ }\textbf {\bibinfo {volume} {51}},\ \bibinfo {pages}
  {10LT01} (\bibinfo {year} {2018})}\BibitemShut {NoStop}%
\bibitem [{sup()}]{sup}%
  \BibitemOpen
  \bibfield  {title} {\enquote {\bibinfo {title} {Supplementary, [url]},}\
  }\href@noop {} {\ }\BibitemShut {NoStop}%
\bibitem [{\citenamefont {Seth}\ \emph {et~al.}(2016)\citenamefont {Seth},
  \citenamefont {Krivenko}, \citenamefont {Ferrero},\ and\ \citenamefont
  {Parcollet}}]{seth2016}%
  \BibitemOpen
  \bibfield  {author} {\bibinfo {author} {\bibfnamefont {P.}~\bibnamefont
  {Seth}}, \bibinfo {author} {\bibfnamefont {I.}~\bibnamefont {Krivenko}},
  \bibinfo {author} {\bibfnamefont {M.}~\bibnamefont {Ferrero}}, \ and\
  \bibinfo {author} {\bibfnamefont {O.}~\bibnamefont {Parcollet}},\ }\bibfield
  {title} {\enquote {\bibinfo {title} {{TRIQS/CTHYB}: A continuous-time quantum
  monte carlo hybridisation expansion solver for quantum impurity problems},}\
  }\href {\doibase 10.1016/j.cpc.2015.10.023} {\bibfield  {journal} {\bibinfo
  {journal} {Comp. Phys. Commun.}\ }\textbf {\bibinfo {volume} {200}},\
  \bibinfo {pages} {274} (\bibinfo {year} {2016})}\BibitemShut {NoStop}%
\bibitem [{\citenamefont {Parcollet}\ \emph {et~al.}(2015)\citenamefont
  {Parcollet}, \citenamefont {Ferrero}, \citenamefont {Ayral}, \citenamefont
  {Hafermann}, \citenamefont {Krivenko}, \citenamefont {Messio},\ and\
  \citenamefont {Seth}}]{triqs}%
  \BibitemOpen
  \bibfield  {author} {\bibinfo {author} {\bibfnamefont {O.}~\bibnamefont
  {Parcollet}}, \bibinfo {author} {\bibfnamefont {M.}~\bibnamefont {Ferrero}},
  \bibinfo {author} {\bibfnamefont {T.}~\bibnamefont {Ayral}}, \bibinfo
  {author} {\bibfnamefont {H.}~\bibnamefont {Hafermann}}, \bibinfo {author}
  {\bibfnamefont {I.}~\bibnamefont {Krivenko}}, \bibinfo {author}
  {\bibfnamefont {L.}~\bibnamefont {Messio}}, \ and\ \bibinfo {author}
  {\bibfnamefont {P.}~\bibnamefont {Seth}},\ }\bibfield  {title} {\enquote
  {\bibinfo {title} {{TRIQS}: A toolbox for research on interacting quantum
  systems},}\ }\href {\doibase 10.1016/j.cpc.2015.04.023} {\bibfield  {journal}
  {\bibinfo  {journal} {Comp. Phys. Commun.}\ }\textbf {\bibinfo {volume}
  {196}},\ \bibinfo {pages} {398} (\bibinfo {year} {2015})}\BibitemShut
  {NoStop}%
\bibitem [{\citenamefont {Aichhorn}\ \emph {et~al.}(2009)\citenamefont
  {Aichhorn}, \citenamefont {Pourovskii}, \citenamefont {Vildosola},
  \citenamefont {Ferrero}, \citenamefont {Parcollet}, \citenamefont {Miyake},
  \citenamefont {Georges},\ and\ \citenamefont {Biermann}}]{aichhorn2009}%
  \BibitemOpen
  \bibfield  {author} {\bibinfo {author} {\bibfnamefont {M.}~\bibnamefont
  {Aichhorn}}, \bibinfo {author} {\bibfnamefont {L.}~\bibnamefont
  {Pourovskii}}, \bibinfo {author} {\bibfnamefont {V.}~\bibnamefont
  {Vildosola}}, \bibinfo {author} {\bibfnamefont {M.}~\bibnamefont {Ferrero}},
  \bibinfo {author} {\bibfnamefont {O.}~\bibnamefont {Parcollet}}, \bibinfo
  {author} {\bibfnamefont {T.}~\bibnamefont {Miyake}}, \bibinfo {author}
  {\bibfnamefont {A.}~\bibnamefont {Georges}}, \ and\ \bibinfo {author}
  {\bibfnamefont {S.}~\bibnamefont {Biermann}},\ }\bibfield  {title} {\enquote
  {\bibinfo {title} {Dynamical mean-field theory within an augmented plane-wave
  framework: Assessing electronic correlations in the iron pnictide
  {LaFeAsO}},}\ }\href {\doibase 10.1103/PhysRevB.80.085101} {\bibfield
  {journal} {\bibinfo  {journal} {Phys. Rev. B}\ }\textbf {\bibinfo {volume}
  {80}},\ \bibinfo {pages} {085101} (\bibinfo {year} {2009})}\BibitemShut
  {NoStop}%
\bibitem [{\citenamefont {Aichhorn}\ \emph {et~al.}(2011)\citenamefont
  {Aichhorn}, \citenamefont {Pourovskii},\ and\ \citenamefont
  {Georges}}]{aichhorn2011}%
  \BibitemOpen
  \bibfield  {author} {\bibinfo {author} {\bibfnamefont {M.}~\bibnamefont
  {Aichhorn}}, \bibinfo {author} {\bibfnamefont {L.}~\bibnamefont
  {Pourovskii}}, \ and\ \bibinfo {author} {\bibfnamefont {A.}~\bibnamefont
  {Georges}},\ }\bibfield  {title} {\enquote {\bibinfo {title} {Importance of
  electronic correlations for structural and magnetic properties of the iron
  pnictide superconductor {LaFeAsO}},}\ }\href {\doibase
  10.1103/PhysRevB.84.054529} {\bibfield  {journal} {\bibinfo  {journal} {Phys.
  Rev. B}\ }\textbf {\bibinfo {volume} {84}},\ \bibinfo {pages} {054529}
  (\bibinfo {year} {2011})}\BibitemShut {NoStop}%
\bibitem [{\citenamefont {Aichhorn}\ \emph {et~al.}(2016)\citenamefont
  {Aichhorn}, \citenamefont {Pourovskii}, \citenamefont {Seth}, \citenamefont
  {Vildosola}, \citenamefont {Zingl}, \citenamefont {Peil}, \citenamefont
  {Deng}, \citenamefont {Mravlje}, \citenamefont {Kraberger}, \citenamefont
  {Martins}, \citenamefont {Ferrero},\ and\ \citenamefont
  {Parcollet}}]{aichhorn2016}%
  \BibitemOpen
  \bibfield  {author} {\bibinfo {author} {\bibfnamefont {M.}~\bibnamefont
  {Aichhorn}}, \bibinfo {author} {\bibfnamefont {L.}~\bibnamefont
  {Pourovskii}}, \bibinfo {author} {\bibfnamefont {P.}~\bibnamefont {Seth}},
  \bibinfo {author} {\bibfnamefont {V.}~\bibnamefont {Vildosola}}, \bibinfo
  {author} {\bibfnamefont {M.}~\bibnamefont {Zingl}}, \bibinfo {author}
  {\bibfnamefont {O.~E.}\ \bibnamefont {Peil}}, \bibinfo {author}
  {\bibfnamefont {X.}~\bibnamefont {Deng}}, \bibinfo {author} {\bibfnamefont
  {J.}~\bibnamefont {Mravlje}}, \bibinfo {author} {\bibfnamefont {G.~J.}\
  \bibnamefont {Kraberger}}, \bibinfo {author} {\bibfnamefont {C.}~\bibnamefont
  {Martins}}, \bibinfo {author} {\bibfnamefont {M.}~\bibnamefont {Ferrero}}, \
  and\ \bibinfo {author} {\bibfnamefont {O.}~\bibnamefont {Parcollet}},\
  }\bibfield  {title} {\enquote {\bibinfo {title} {{TRIQS/DFTTools}: A {TRIQS}
  application for ab initio calculations of correlated materials},}\ }\href
  {\doibase 10.1016/j.cpc.2016.03.014} {\bibfield  {journal} {\bibinfo
  {journal} {Comp. Phys. Commun.}\ }\textbf {\bibinfo {volume} {204}},\
  \bibinfo {pages} {200} (\bibinfo {year} {2016})}\BibitemShut {NoStop}%
\bibitem [{\citenamefont {Kraberger}\ and\ \citenamefont
  {Zingl}(2019)}]{maxent}%
  \BibitemOpen
  \bibfield  {author} {\bibinfo {author} {\bibfnamefont {G.~J.}\ \bibnamefont
  {Kraberger}}\ and\ \bibinfo {author} {\bibfnamefont {M.}~\bibnamefont
  {Zingl}},\ }\href@noop {} {}\bibinfo {howpublished}
  {\url{https://github.com/TRIQS/maxent}} (\bibinfo {year} {2019})\BibitemShut
  {NoStop}%
\bibitem [{\citenamefont {Bhandary}\ and\ \citenamefont
  {Held}(2019)}]{bhandary2019}%
  \BibitemOpen
  \bibfield  {author} {\bibinfo {author} {\bibfnamefont {S.}~\bibnamefont
  {Bhandary}}\ and\ \bibinfo {author} {\bibfnamefont {K.}~\bibnamefont
  {Held}},\ }\bibfield  {title} {\enquote {\bibinfo {title} {Self-energy
  self-consistent density functional theory plus dynamical mean field
  theory},}\ }\href {https://arxiv.org/abs/1904.02967v1} {\bibfield  {journal}
  {\bibinfo  {journal} {arXiv:1904.02967v1}\ } (\bibinfo {year}
  {2019})}\BibitemShut {NoStop}%
\bibitem [{\citenamefont {Laverock}\ \emph {et~al.}(2020)\citenamefont
  {Laverock}, \citenamefont {Gu}, \citenamefont {Jovic}, \citenamefont {Lu},
  \citenamefont {Wolf}, \citenamefont {Watson}, \citenamefont {Qiao},
  \citenamefont {Yang},\ and\ \citenamefont {Smith}}]{svo_sl_int}%
  \BibitemOpen
  \bibfield  {author} {\bibinfo {author} {\bibfnamefont {J.}~\bibnamefont
  {Laverock}}, \bibinfo {author} {\bibfnamefont {M.}~\bibnamefont {Gu}},
  \bibinfo {author} {\bibfnamefont {V.}~\bibnamefont {Jovic}}, \bibinfo
  {author} {\bibfnamefont {J.~W.}\ \bibnamefont {Lu}}, \bibinfo {author}
  {\bibfnamefont {S.~A.}\ \bibnamefont {Wolf}}, \bibinfo {author}
  {\bibfnamefont {G.}~\bibnamefont {Watson}}, \bibinfo {author} {\bibfnamefont
  {R.~M.}\ \bibnamefont {Qiao}}, \bibinfo {author} {\bibfnamefont
  {W.}~\bibnamefont {Yang}}, \ and\ \bibinfo {author} {\bibfnamefont {K.~E.}\
  \bibnamefont {Smith}},\ }\bibfield  {title} {\enquote {\bibinfo {title}
  {Electronic reconstructions at the interface of {SrVO$_3$}/{SrTiO$_3$}
  superlattices},}\ }\href@noop {} {\bibfield  {journal} {\bibinfo  {journal}
  {unpublished}\ } (\bibinfo {year} {2020})}\BibitemShut {NoStop}%
\bibitem [{\citenamefont {Okamoto}(2011)}]{okamoto2011}%
  \BibitemOpen
  \bibfield  {author} {\bibinfo {author} {\bibfnamefont {S.}~\bibnamefont
  {Okamoto}},\ }\bibfield  {title} {\enquote {\bibinfo {title} {Anomalous mass
  enhancement in strongly correlated quantum wells},}\ }\href {\doibase
  10.1103/PhysRevB.84.201305} {\bibfield  {journal} {\bibinfo  {journal} {Phys.
  Rev. B}\ }\textbf {\bibinfo {volume} {84}},\ \bibinfo {pages} {201305(R)}
  (\bibinfo {year} {2011})}\BibitemShut {NoStop}%
\bibitem [{\citenamefont {Oka}\ \emph {et~al.}(2018)\citenamefont {Oka},
  \citenamefont {Okada}, \citenamefont {Hitosugi},\ and\ \citenamefont
  {Fukumura}}]{oka2018}%
  \BibitemOpen
  \bibfield  {author} {\bibinfo {author} {\bibfnamefont {H.}~\bibnamefont
  {Oka}}, \bibinfo {author} {\bibfnamefont {Y.}~\bibnamefont {Okada}}, \bibinfo
  {author} {\bibfnamefont {T.}~\bibnamefont {Hitosugi}}, \ and\ \bibinfo
  {author} {\bibfnamefont {T.}~\bibnamefont {Fukumura}},\ }\bibfield  {title}
  {\enquote {\bibinfo {title} {Two distinct surface terminations of {SrVO$_3$}
  (001) ultrathin films as an influential factor on metallicity},}\ }\href
  {\doibase 10.1063/1.5051434} {\bibfield  {journal} {\bibinfo  {journal}
  {Appl. Phys. Lett.}\ }\textbf {\bibinfo {volume} {113}},\ \bibinfo {pages}
  {171601} (\bibinfo {year} {2018})}\BibitemShut {NoStop}%
\bibitem [{\citenamefont {Wang}\ \emph {et~al.}(2019)\citenamefont {Wang},
  \citenamefont {Wang}, \citenamefont {Meng}, \citenamefont {Saghayezhian},
  \citenamefont {Chen}, \citenamefont {Chen}, \citenamefont {Guo},
  \citenamefont {Zhu}, \citenamefont {Plummer},\ and\ \citenamefont
  {Zhang}}]{wang2019b}%
  \BibitemOpen
  \bibfield  {author} {\bibinfo {author} {\bibfnamefont {G.}~\bibnamefont
  {Wang}}, \bibinfo {author} {\bibfnamefont {Z.}~\bibnamefont {Wang}}, \bibinfo
  {author} {\bibfnamefont {M.}~\bibnamefont {Meng}}, \bibinfo {author}
  {\bibfnamefont {M.}~\bibnamefont {Saghayezhian}}, \bibinfo {author}
  {\bibfnamefont {L.}~\bibnamefont {Chen}}, \bibinfo {author} {\bibfnamefont
  {C.}~\bibnamefont {Chen}}, \bibinfo {author} {\bibfnamefont {H.}~\bibnamefont
  {Guo}}, \bibinfo {author} {\bibfnamefont {Y.}~\bibnamefont {Zhu}}, \bibinfo
  {author} {\bibfnamefont {E.~W.}\ \bibnamefont {Plummer}}, \ and\ \bibinfo
  {author} {\bibfnamefont {J.}~\bibnamefont {Zhang}},\ }\bibfield  {title}
  {\enquote {\bibinfo {title} {Role of disorder and correlations in the
  metal-insulator transition in ultrathin {SrVO$_3$} films},}\ }\href {\doibase
  10.1103/PhysRevB.100.155114} {\bibfield  {journal} {\bibinfo  {journal}
  {Phys. Rev. B}\ }\textbf {\bibinfo {volume} {100}},\ \bibinfo {pages}
  {155114} (\bibinfo {year} {2019})}\BibitemShut {NoStop}%
\bibitem [{\citenamefont {Matsuda}\ \emph {et~al.}(2002)\citenamefont
  {Matsuda}, \citenamefont {Ohta},\ and\ \citenamefont {Yeom}}]{matsuda2002}%
  \BibitemOpen
  \bibfield  {author} {\bibinfo {author} {\bibfnamefont {I.}~\bibnamefont
  {Matsuda}}, \bibinfo {author} {\bibfnamefont {T.}~\bibnamefont {Ohta}}, \
  and\ \bibinfo {author} {\bibfnamefont {H.~W.}\ \bibnamefont {Yeom}},\
  }\bibfield  {title} {\enquote {\bibinfo {title} {In-plane dispersion of the
  quantum-well states of the epitaxial silver films on silicon},}\ }\href
  {\doibase 10.1103/PhysRevB.65.085327} {\bibfield  {journal} {\bibinfo
  {journal} {Phys. Rev. B}\ }\textbf {\bibinfo {volume} {65}},\ \bibinfo
  {pages} {085327} (\bibinfo {year} {2002})}\BibitemShut {NoStop}%
\bibitem [{\citenamefont {Yoshimatsu}\ \emph {et~al.}(2013)\citenamefont
  {Yoshimatsu}, \citenamefont {Sakai}, \citenamefont {Kobayashi}, \citenamefont
  {Horiba}, \citenamefont {Yoshida}, \citenamefont {Fujimori}, \citenamefont
  {Oshima},\ and\ \citenamefont {Kumigashira}}]{yoshimatsu2013}%
  \BibitemOpen
  \bibfield  {author} {\bibinfo {author} {\bibfnamefont {K.}~\bibnamefont
  {Yoshimatsu}}, \bibinfo {author} {\bibfnamefont {E.}~\bibnamefont {Sakai}},
  \bibinfo {author} {\bibfnamefont {M.}~\bibnamefont {Kobayashi}}, \bibinfo
  {author} {\bibfnamefont {K.}~\bibnamefont {Horiba}}, \bibinfo {author}
  {\bibfnamefont {T.}~\bibnamefont {Yoshida}}, \bibinfo {author} {\bibfnamefont
  {A.}~\bibnamefont {Fujimori}}, \bibinfo {author} {\bibfnamefont
  {M.}~\bibnamefont {Oshima}}, \ and\ \bibinfo {author} {\bibfnamefont
  {H.}~\bibnamefont {Kumigashira}},\ }\bibfield  {title} {\enquote {\bibinfo
  {title} {Determination of the surface and interface phase shifts in metallic
  quantum well structures of perovskite oxides},}\ }\href {\doibase
  10.1103/PhysRevB.88.115308} {\bibfield  {journal} {\bibinfo  {journal} {Phys.
  Rev. B}\ }\textbf {\bibinfo {volume} {88}},\ \bibinfo {pages} {115308}
  (\bibinfo {year} {2013})}\BibitemShut {NoStop}%
\bibitem [{\citenamefont {Ahn}\ \emph {et~al.}(1999)\citenamefont {Ahn},
  \citenamefont {Gariglio}, \citenamefont {Paruch}, \citenamefont {Tybell},
  \citenamefont {Antognazza},\ and\ \citenamefont {Triscone}}]{ahn1999}%
  \BibitemOpen
  \bibfield  {author} {\bibinfo {author} {\bibfnamefont {C.~H.}\ \bibnamefont
  {Ahn}}, \bibinfo {author} {\bibfnamefont {S.}~\bibnamefont {Gariglio}},
  \bibinfo {author} {\bibfnamefont {P.}~\bibnamefont {Paruch}}, \bibinfo
  {author} {\bibfnamefont {T.}~\bibnamefont {Tybell}}, \bibinfo {author}
  {\bibfnamefont {L.}~\bibnamefont {Antognazza}}, \ and\ \bibinfo {author}
  {\bibfnamefont {J.-M.}\ \bibnamefont {Triscone}},\ }\bibfield  {title}
  {\enquote {\bibinfo {title} {Electrostatic modulation of superconductivity in
  ultrathin {GdBa$_2$Cu$_3$O$_{7-x}$} films},}\ }\href {\doibase
  10.1126/science.284.5417.1152} {\bibfield  {journal} {\bibinfo  {journal}
  {Science}\ }\textbf {\bibinfo {volume} {284}},\ \bibinfo {pages} {1152}
  (\bibinfo {year} {1999})}\BibitemShut {NoStop}%
\bibitem [{\citenamefont {Chen}\ \emph {et~al.}(2013)\citenamefont {Chen},
  \citenamefont {Millis},\ and\ \citenamefont {Marianetti}}]{chen2013}%
  \BibitemOpen
  \bibfield  {author} {\bibinfo {author} {\bibfnamefont {H.}~\bibnamefont
  {Chen}}, \bibinfo {author} {\bibfnamefont {A.~J.}\ \bibnamefont {Millis}}, \
  and\ \bibinfo {author} {\bibfnamefont {C.~A.}\ \bibnamefont {Marianetti}},\
  }\bibfield  {title} {\enquote {\bibinfo {title} {Engineering correlation
  effects via artificially designed oxide superlattices},}\ }\href {\doibase
  10.1103/PhysRevLett.111.116403} {\bibfield  {journal} {\bibinfo  {journal}
  {Phys. Rev. Lett.}\ }\textbf {\bibinfo {volume} {111}},\ \bibinfo {pages}
  {116403} (\bibinfo {year} {2013})}\BibitemShut {NoStop}%
\bibitem [{\citenamefont {Samanta}\ \emph {et~al.}(2018)\citenamefont
  {Samanta}, \citenamefont {Mishra},\ and\ \citenamefont
  {Nanda}}]{samanta2018}%
  \BibitemOpen
  \bibfield  {author} {\bibinfo {author} {\bibfnamefont {S.}~\bibnamefont
  {Samanta}}, \bibinfo {author} {\bibfnamefont {S.~B.}\ \bibnamefont {Mishra}},
  \ and\ \bibinfo {author} {\bibfnamefont {B.~R.~K.}\ \bibnamefont {Nanda}},\
  }\bibfield  {title} {\enquote {\bibinfo {title} {Quantum well structure of a
  double perovskite superlattice and formation of a spin-polarized
  two-dimensional electron gas},}\ }\href {\doibase 10.1103/PhysRevB.98.115155}
  {\bibfield  {journal} {\bibinfo  {journal} {Phys. Rev. B}\ }\textbf {\bibinfo
  {volume} {98}},\ \bibinfo {pages} {115155} (\bibinfo {year}
  {2018})}\BibitemShut {NoStop}%
\end{thebibliography}
\end{document}


\title{Supplementary Information for ``Quantum Confinement Induced
Metal-Insulator Transition in Strongly Correlated Quantum Wells of SrVO$_3$
Superlattices''}

\author{A.~D.~N.~James}
\affiliation{H.\ H.\ Wills Physics Laboratory,
University of Bristol, Tyndall Avenue, Bristol, BS8 1TL, United Kingdom}

\author{M.~Aichhorn}
\affiliation{Institute of Theoretical and Computational Physics, TU Graz, NAWI
Graz, Petersgasse 16, 8010 Graz, Austria}

\author{J.~Laverock}
\affiliation{H.\ H.\ Wills Physics Laboratory,
University of Bristol, Tyndall Avenue, Bristol, BS8 1TL, United Kingdom}

\date{\today}

\maketitle

\section{Comparison with experimental quantities}
\begin{figure}[b]
\centerline{\includegraphics[width=0.5\linewidth]{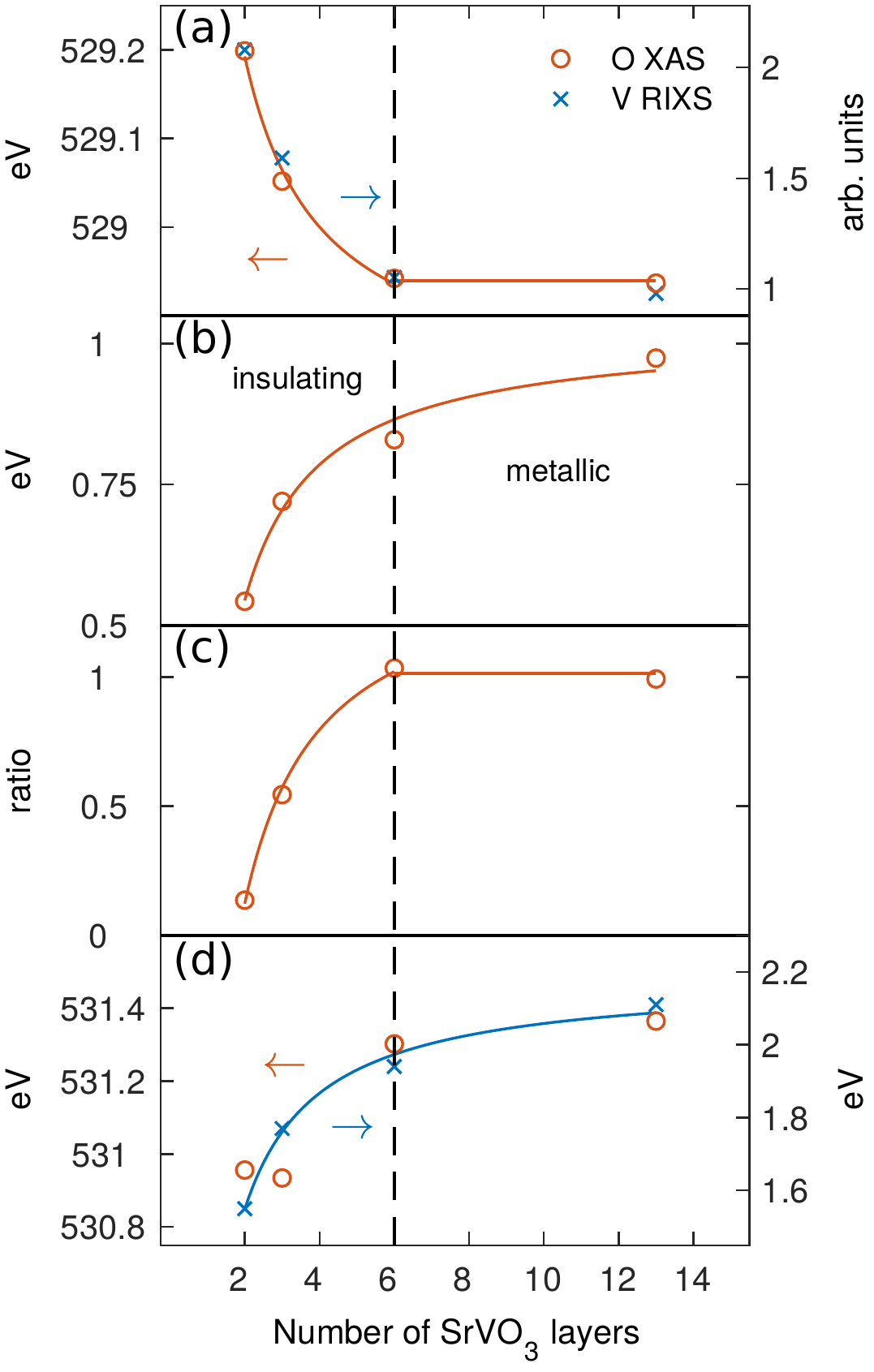}}
\caption{Evolution of correlated electron behavior from experimental x-ray
absorption spectroscopy (XAS) and
resonant inelastic x-ray scattering (RIXS)
measurements of SrVO$_3$/SrTiO$_3$ superlattices, reproduced from
Ref.~\onlinecite{laverock2017}. From top to bottom,
evolution in (a)~metallicity of SLs, (b)~quasiparticle (QP)
bandwidth, (c)~QP spectral weight and (d)~the energy of the upper Hubbard band
are shown.}
\label{f:experiment}
\end{figure}

Figure \ref{f:experiment} illustrates the main experimental spectroscopic
results on the superlattices (SL) \cite{gu2018,laverock2017}, in which
transport measurements established that
the 2:7 and 3:6 SLs were insulating, whereas the 13:4 SL was metallic. The 6:5
SL was found to be
metallic at room temperature, with an metal-insulator transition (MIT)
at low temperature. The experimental
results (performed at room temperature) show the evolution in correlated
electron behavior extracted from x-ray absorption spectroscopy (XAS) and
resonant inelastic x-ray scattering (RIXS).

For completeness, we briefly outline the experimental properties and how they
were extracted here.
The metallicity [Fig.~\ref{f:experiment}(a)] was extracted from both XAS and
RIXS as the leading edge of the O {\em K} edge XAS and from the intensity of the
quasi-elastic peak in V {\em L} edge RIXS.
The quasiparticle (QP)
bandwidth [Fig.~\ref{f:experiment}(b)] was extracted from the SrVO$_3$
layber
contribution to the O {\em K} edge XAS as the full-with at half-maximum of the
QP peak.
The QP spectral weight [Fig.~\ref{f:experiment}(c)] was also extracted from the
SrVO$_3$ layer contribution to the O {\em K} edge XAS as the ratio of the
area under the QP peak to the total area under unoccupied V $3d$ states (i.e.\
the sum of QP, upper Hubbard band (UHB) and $e_g$ spectral weights). Occupied states are not
accessible in O {\em K} edge XAS.
Finally, the UHB energy [Fig.~\ref{f:experiment}(d)] is accessible to both XAS
and RIXS. From O {\em K} edge XAS, the UHB peak is directly observed, and its
center is shown here. From V {\em L} edge RIXS, the UHB energy is available from
transitions from occupied QP states to the unoccupied UHB. Both show equivalent
evolution with SL structure.

\begin{figure}[b!]
\centerline{\includegraphics[width=0.9\linewidth]{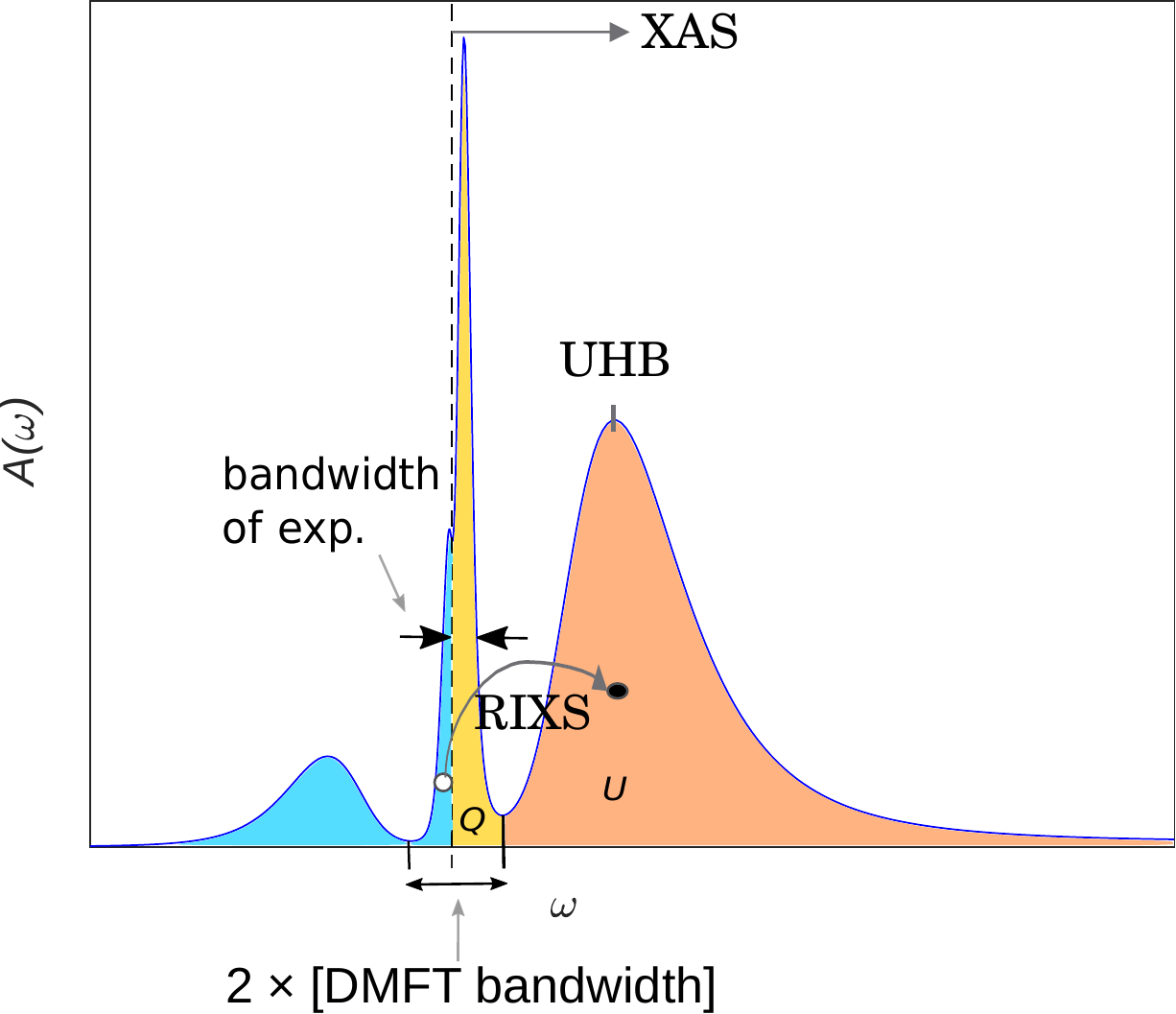}}
\caption{A schematic illustration showing how the variables in the
theory-experimental comparison were determined.}
\label{f:exp_dia}
\end{figure}

Figure \ref{f:exp_dia} shows a schematic illustration comparing the extracted
experimental quantities with their DFT+DMFT (density functional
theory with dynamical mean-field theory) definitions. The QP
bandwidth has been extracted from the DMFT spectral function
by obtaining the width defined by the minima around the central QP peak.
The QP ratio was
determined by taking the ratio of the the quasiparticle
weight (labeled $Q$ in Fig.\ \ref{f:exp_dia}) and the UHB
weight (labeled $U$).
Finally, the energy of the UHB was
obtained by locating the peak in the DMFT spectral function, referenced to
$\omega=0$. In the experimental RIXS process, the UHB peak energy represents the
peak in the joint QP and UHB density of states, and therefore is referenced to
an energy $\omega < 0$. To compare the theoretical and experimental
quantities, we therefore shift all theoretical quantities to match for the 2:7
SL (the shift is $-0.584$~eV).

\begin{figure*}[t]
\centerline{\includegraphics[width=0.8\linewidth]{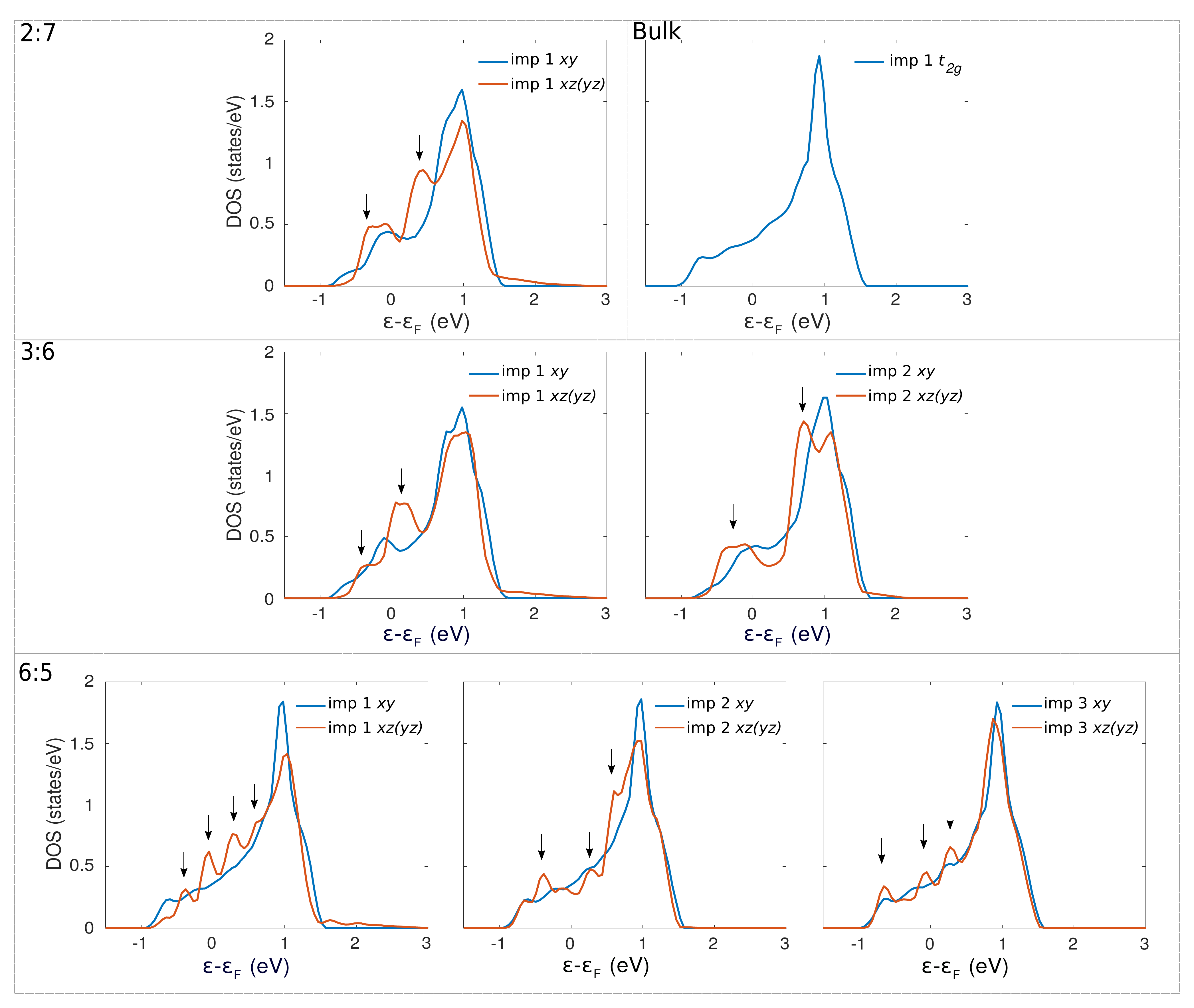}}
\caption{The DFT V $t_{2g}$ partial density of states (PDOS)
of the bulk SrVO$_3$, 2:7, 3:6 and 6:5
SLs. The dashed borders outline which plots belong to the
corresponding structure. Each panel shows the PDOS of the inequivalent
V atoms in each structure, labeled by their impurity number in the DMFT cycles
(imp 1 refers to the interface). The
arrows indicate the contributions of quantized states to each
inequivalent V atom. The greatest contribution are from the relatively flat
bands along $\Gamma$-$X$.}
\label{f:pdos}
\end{figure*}

\section{Density Functional Theory calculations}
DFT calculations were performed using the {\sc Elk}
FP-LAPW (full potential linearized augmented plane wave) code within the local
density approximation (LDA) \cite{elk}. The results are in excellent agreement
with previous pseudopotential calculations within the generalized gradient
approximation (GGA) of the same SLs \cite{laverock2017}.
Previous PES work \cite{yoshimatsu2010}, show how the dimensionality
of SrVO$_3$ influences the MIT. In their results, ten SrVO$_3$ layers
closely resembles bulk behavior. From this,
we approximate the 13:4 SL in Ref.\ \onlinecite{laverock2017} with bulk DFT+DMFT
calculations.

Self-consistency was achieved on a $12 \times 12 \times 4$ mesh in
the full Brillouin zone (BZ) for relatively low computational cost with
sufficient sampling, corresponding to 84 k-points in the irreducible ($1/16^{\rm
th}$) BZ.  To stabilize the DFT self-consistent cycles, small values of mixing
of the new potentials was used, at the cost of computational time. For bulk
SrVO$_3$, a k-mesh of $12 \times 12 \times 12$ was used (84 k-points in the
$1/48^{\rm th}$ irreducible BZ).

The partial densities of states (PDOS) of $t_{2g}$ orbitals are shown in
Fig.~\ref{f:pdos} for the bulk and 2:7, 3:6 and 6:5
SLs. Sharp peaks in the PDOS reflect the quantized electronic
structure along $c$. For the inner layers of the
thicker SLs, the PDOS more closely resembles that of bulk SrVO$_3$, e.g.\
impurity 3 of the 6:5 SL. Near the interface, the $xz(yz)$ PDOS extends to
higher energies as a result of mixing of these states with Ti states in the SrTiO$_3$
layer.

\begin{figure*}[t]
\centerline{\includegraphics[width=0.92\linewidth]{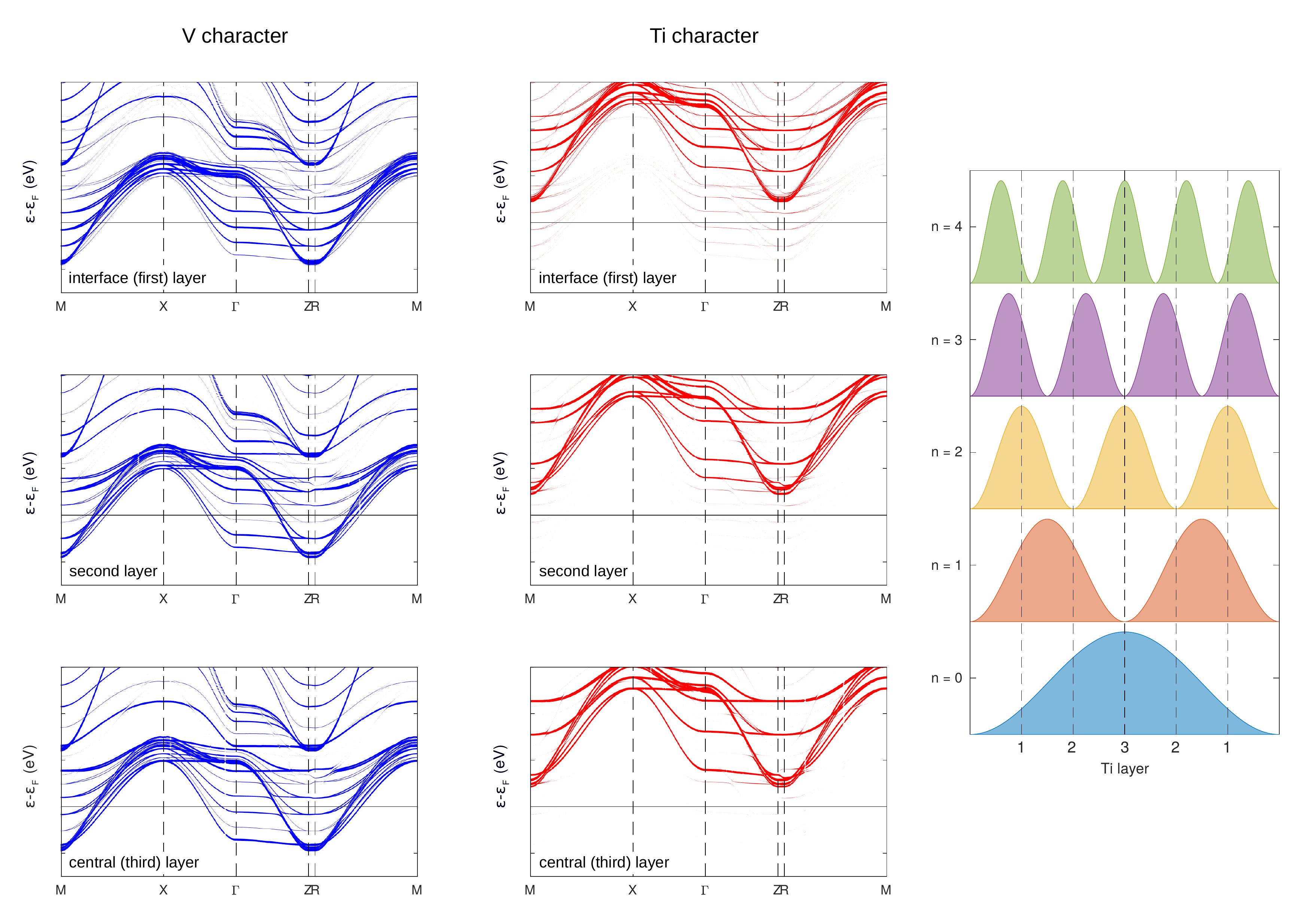}}
\caption{The DFT V and Ti $t_{2g}$ band characters of the 6:5 superlattice (SL).
The thickness of the lines indicates the total character of
each V (left) and Ti (right) site. The top row shows the band characters
at the interface, while the bottom row shows the character in the center of the
each layer. On the right, a schematic illustration of the real-space probability
distribution of the quantized subbands in the out-of-plane direction of the
SrTiO$_3$ layers is shown. The edge of the box does not coincide with the
interface Ti ion or its neighboring SrO layer due to the finite phase
accumulated at the interface.}
\label{f:char}
\end{figure*}

The characters of the subbands are shown in Fig.\ \ref{f:char} for each of the
different V and Ti sites, using the 6:5 SL as an example. As expected, the V
bands dominate the character at the Fermi level, with weak contribution
from interfacial Ti ions. The spatial distribution of the subband wavefunctions
of the SrVO$_3$ quantum wells
can be seen directly in the characters. The V $n=0$ subband, with greatest
amplitude in the centre of the well, has strong character in the central V ion
and weak character at the interface. Correspondingly, the V $n=2$ subband has
strongest character at the interface and is almost absent in the second layer
close to where a node is expected in the quantum well wavefunction. At higher
energies, the quantized V $e_g$ subbands appear above 1~eV.

The Ti band characters shed light on the broadening of the Ti bands in the
DFT+DMFT calculations shown in Fig.\ 3 of the main manuscript. As above, the
central Ti ion contributes strongly to the Ti $n=0$ and $n=2$ subbands. On the
other hand, the interfacial Ti ion contributes significantly to the Ti $n=1$, 2
and 3 subbands, with the largest contribution to the Ti $n=2$ subband. The
interfacial Ti ion mixes most strongly with the V orbitals, which are the
correlated orbitals in the subsequent DMFT cycle. This demonstrates how the
spatial penetration of the Ti $n=1$ and $n=2$ subbands into the correlated
SrVO$_3$ layers leads to substantial broadening of these subbands in the
subsequent DMFT cycle. In contrast, the Ti $n=0$ subband is spatially deep
within the SrTiO$_3$ layer and does not feel the effects of the correlated
SrVO$_3$ orbitals very strongly, remaining reasonably sharp even in the
insulating phase [Fig.\ 3(j) of the main manuscript].

\section{Quantized tight-binding model}
\subsection{Bulk tight-binding bands}
The tight-binding (TB) model was constructed up to 12$^{\rm th}$ nearest
neighbors, consisting of 24 hopping terms, $t_i$, up to $[l,m,n] = [2,2,2]$. For
the $xy$ band the TB dispersion, $\varepsilon_{xy}$, is given by,
\begin{equation}
\varepsilon_{xy} = E_{xy}^{0} + \sum_{lmn}
t_{xy}^{lmn} \cos \left( l k_x + m k_y + n k_z \right),
\label{e:tb}
\end{equation}
where the band energy, $E_{xy}^0$, corresponds to the crystal field energy.
Since the purpose of our model is to accurately describe the bulk 3D DFT band
structure, we do not attempt to analyze individual parameters, as has been done
before \cite{liebsch2003}. Although terms corresponding to the
5$^{\rm th}$ nearest neighbor and higher had a magnitude of less than 10~meV,
these terms were found to be necessary to adequately describe the FP-LAPW band
structure. After fitting this model to the bulk LDA band structure
in the full cubic Brillouin zone, we find the r.m.s.\ difference is
less than 11~meV, with a maximum difference of 70~meV.

\subsection{Quantum confinement}
In order to account for the effects of quantum confinement of the V $3d$
electrons in the SrVO$_3$ layers, we apply the Bohr-Sommerfeld phase accumulation
model \cite{chiang2000},
\begin{equation}
2k_z^n(E)L + \delta(E) = 2\pi n,
\end{equation}
where $n = 0, 1, 2, \dots$ is the quantum number, $2k_z^n(E)L$ is the total
phase accumulated in traveling through the SrVO$_3$ layer and back, $k_z^n(E)$ is the
quantized out-of-plane wavevector, $L = mc$ is the SrVO$_3$ layer thickness ($m$ and
$c$ are the number of SrVO$_3$ layers and $c$-axis lattice parameter of the SrVO$_3$
layers, respectively).  $\delta(E)$ is the total phase acquired due to
reflection at both SrVO$_3$/SrTiO$_3$ interfaces. For asymmetric quantum wells, e.g.\ thin
overlayers with a vacuum interface, $\delta = \phi_1 + \phi_2$ is composed of
different individual phase shifts at each reflection; in our case of symmetric
barriers, $\delta = 2\phi$, where $\phi$ is the phase at a single SrVO$_3$/SrTiO$_3$ 
interface. In general, $\delta = \delta(E)$ is explicitly dependent on the
energy of the confined state. However, in order to simplify the fitting, and
avoid unnecessary degrees of freedom, we instead implicitly include the energy
dependence through different phases for each quantum number, $\delta =
\delta_n$. With this, the quantization condition reduces to,
\begin{equation}
k_z^n = \frac{2 \pi n - \delta_n}{2 m c},
\end{equation}
from which the quantized TB dispersion, $E_n(k_x,k_y,k_z^n)$, may be evaluated.

\begin{figure}[t]
\includegraphics[width=1.0\linewidth]{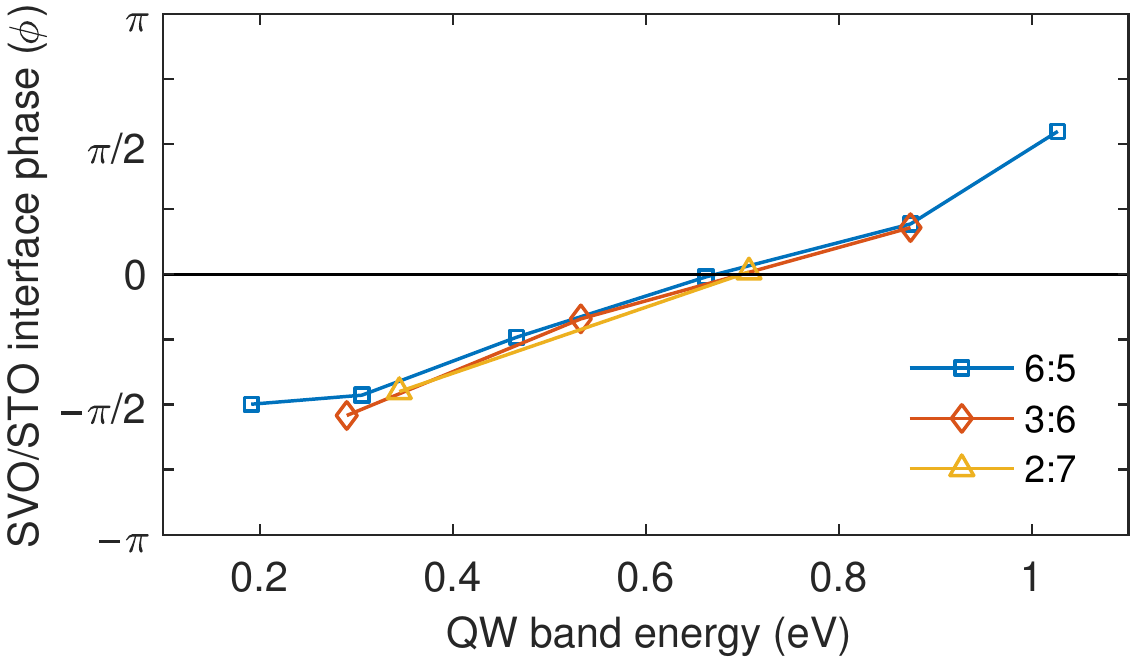}
\caption{The phase shift at the SrVO$_3$/SrTiO$_3$ interface for each quantized state in
the SLs (note $\phi_n=\delta_n/2$ is shown), shown against the mean band
energy.}
\label{f:phases}
\end{figure}

\begin{table}[b]
\begin{tabular}{c|c|ccc|ccc}
SL & CF splitting & \multicolumn{3}{c|}{Intrinsic} & 
   \multicolumn{3}{c}{Quantized bands} \\
\hspace*{0.30in} & (meV) & $W_{xy}$ & $W_{yz}$ & anis. &
                           $W_{xy}$ & $W_{yz}$ & anis. \\
\hline
 6:5 & 33 & 0.966 & 0.959 & 0.993 & 0.950 & 0.900 & 0.948 \\
 3:6 & 40 & 0.971 & 0.960 & 0.989 & 0.935 & 0.798 & 0.853 \\
 2:7 & 51 & 0.963 & 0.952 & 0.988 & 0.911 & 0.713 & 0.782 \\
\end{tabular}
\caption{\label{t:qwbands} Results of fitting the FP-LAPW {\sc Elk} bands to a
quantized tight-binding model. The crystal field (CF)
splitting is the energy difference,
$E^0_{yz} - E^0_{xy}$. The bandwidth (relative to bulk SrVO$_3$), $W_i$,
of the $xy$ and $yz$ bands are shown for both intrinsic bands
(before quantization) and for the quantized bands, alongside their anisotropy
($W_{yz}/W_{xy}$).}
\end{table}

\subsection{Full quantization parameters}
For each SL, four parameters were fitted to describe the ``intrinsic'' band
structure, and $n$ parameters described the confined bands. The quantized TB
dispersion was fitted to the FP-LAPW {\sc Elk} band structure of the SLs.  The
four intrinsic parameters consist of band centers ($E_i^0$ in Eqn.~\ref{e:tb})
and band widths for the $xy$ and $xz(yz)$ bands. The band width
parameter, $W_i$, is a multiplicative factor to the hopping terms, $t_i$ (the
hopping terms themselves were fixed to the cubic bulk parameters determined
above, effectively fixing the shape of the band). In addition to the
intrinsic parameters, the phase shifts for each confined state,
$\delta_n$, were also fitted. The fitted phases are shown in Fig.~\ref{f:phases}
against the mean energy of each state, and closely follow the same roughly
linear relationship with energy for all SLs.

The results of fitting the FP-LAPW bands to the quantized TB model are shown in
Table~\ref{t:qwbands}, separated into contributions from the underlying bulk
``intrinsic'' bands and after quantizing these bands. An example of the fitted
band structure is shown in Fig.\ 1 of the main paper for the 2:7 SL. Since its
wavefunction is perpendicular to the quantization axis, the $xy$ bandwidth
is hardly affected by confinement, but the $xz(yz)$ bands are significantly
narrowed compared with their intrinsic (bulk-like) counterparts. The confinement
leads to the preferential filling of the quantized $xz(yz)$ out-of-plane bands
as their $k_z$ dispersion is suppressed and they become 1D-like, which also
pulls the Fermi level down slightly.

We note that confinement alone is capable of reproducing the SL band structure
to a large extent, correctly describing the narrowing of the quantized bandwidth
and its variation with SrVO$_3$ layer thickness. This has been checked by
restricting the ``intrinsic'' bands in the fit to the bulk bands (i.e.\ setting
$E^0_i$ and $W_i$ to the bulk ones). This provides additional support that the
band narrowing that eventually drives the MIT is primarily due to quantization
effects rather than crystal field (CF) effects.

\begin{figure*}[th!]
\centerline{\includegraphics[width=0.85\linewidth]{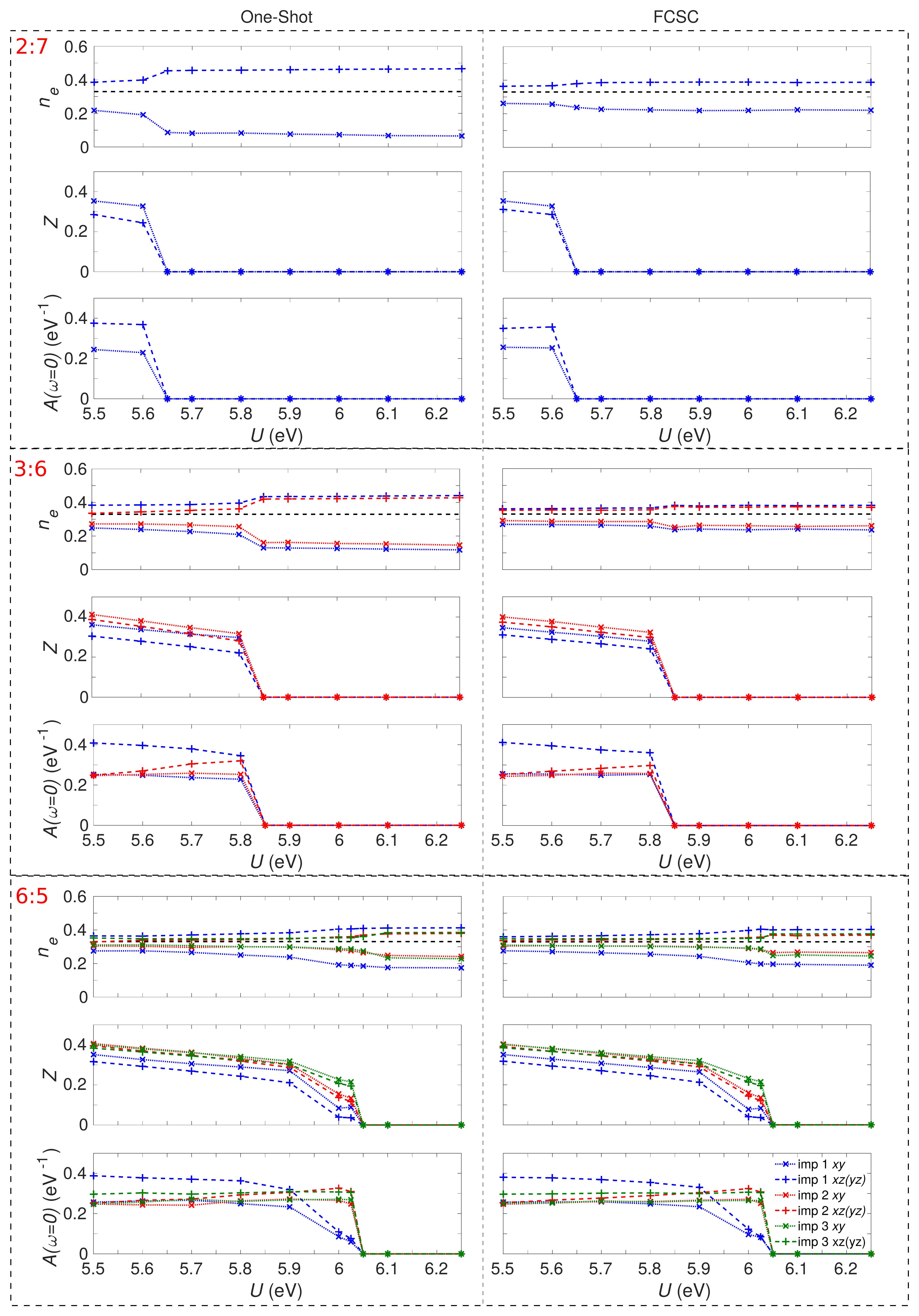}}
\caption{The effect of U on the orbital charge $n_e$ (top), quasiparticle
residue $Z$ (middle) and the spectral function around the Fermi level (bottom)
for each SL one-shot and fully charge self-consistent (FCSC) DFT+DMFT calculation. The dashed line represents the bulk degenerate
orbital charge. }
\label{f:Uimp}
\end{figure*}

\section{Dynamical Mean-Field Theory calculations}

\begin{figure}[t]
\centerline{\includegraphics[width=1.0\linewidth]{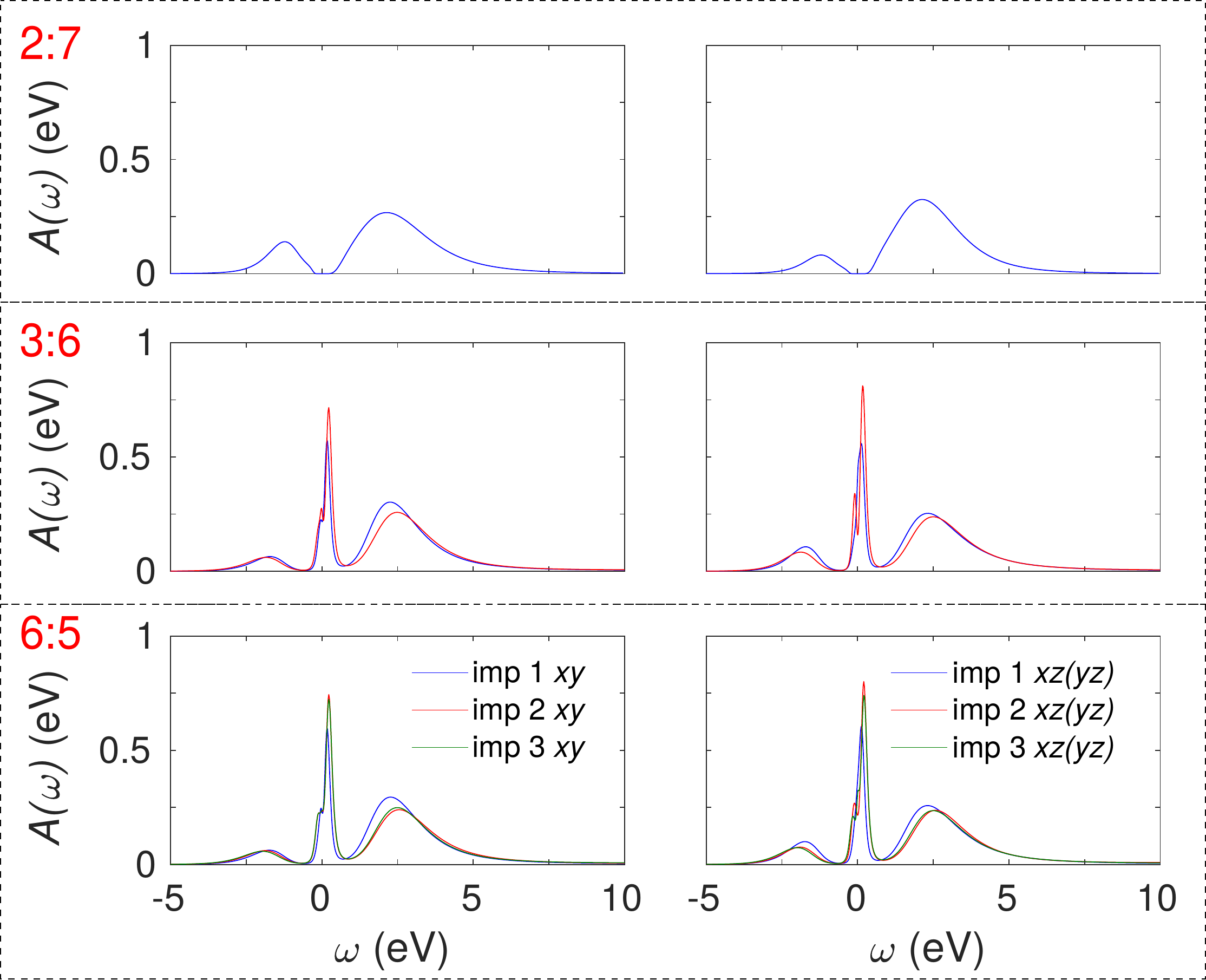}}
\caption{Spectral functions of the 2:7 (top), 3:6 (middle) and 6:5 (bottom) SLs
from fully charge self-consistent calculations, showing $xy$ (left) and $xz(yz)$ (right)
orbitals. The spectra of the insulating 2:7 SL have been shifted such that the
Fermi level lies at the center of the band gap of the $xy$ spectrum.}
\label{f:Aw}
\end{figure}

The output from the {\sc Elk} DFT calculation
was imported to the TRIQS library \cite{triqs} via
an in-house interface with the dmftproj \cite{aichhorn2016} application.
As in the literature \cite{zhong2015,bhandary2016,schuler2018}, only the V
$t_{2g}$ bands were projected (using Wannier projectors) \cite{aichhorn2009}
to construct the LDA Hamiltonian in Wannier space.
These projectors were constructed in the following correlated energy windows:
2:7 SL: $[-1.36, 2.0]$~eV, 3:6 SL: $[-1.29, 2.0]$~eV, 6:5 SL: $[-1.29,
2.0]$~eV and bulk: $[-1.50, 1.90]$~eV. These windows were constructed such that
all of the V t$_{2g}$ bands are included and the valence charge above the lower bound, corresponding to the charge in the
V $t_{2g}$ orbitals, is equal to 1 per V impurity.
Each DMFT cycle calculation
used $84 \times 10^6$ Monte Carlo sweeps.

In order to avoid potential complications from the ill-posed problem of analytic
continuation, quantities were determined from the Green's function and
self-energy on the imaginary time ($\tau$) or frequency axis as much as
possible.  The charge of each orbital ($n_{e}$) was determined by,

\begin{equation}
n_{e} = \sum_n G(i\omega_n)e^{i\omega_n0^{+}} \label{e:chg}
\end{equation}
within the TRIQS library. As there is negligible inter orbital-orbital overlap
on the impurity, $n_{e}$ is diagonal. The spectral function at
the Fermi level, $A(\omega=0)$, is an averaged quantity over a frequency window
approximately equal to $T$ \cite{zhong2015}. Here, $A(\omega=0)$ is
determined directly from the imaginary time Green's function by,
\begin{equation}
A(\omega=0) = \frac{\beta G(\tau=\frac{1}{2}\beta)}{\pi}, \label{e:A0}
\end{equation}
where $\beta$ is the inverse temperature in natural units. 
The value of QP residue $Z$ was determined by 

\begin{equation}
Z = \bigg(1-\frac{\partial \Im[\Sigma(\iu\omega_n)]}{\partial
\iu\omega_n}\bigg|_{\iu\omega_n\rightarrow 0^{+}}\bigg)^{-1}, \label{e:Z}
\end{equation}
where the $Z$ is evaluated from the differential of the imaginary part of the
Matsubara self-energy at $\iu\omega_n\rightarrow 0^{+}$. For $U$ values far from the Fermi liquid regime (namely for the 6:5 $Z$ values close to the MIT), the $Z$ values were approximated by using the differential of the interpolated self-energies at $\iu\omega_n$ = 0. There are two ways to realize the insulating solution. First, by a divergence in $\Im[\Sigma(\iu\omega_n)]$, which comes naturally with $Z = 0$. Second, the combination of the $\Re[\Sigma(\iu\omega_n)]$ and the chemical potential might move the pole position outside of the non-interacting bandwidth, meaning that no QP peak is possible in the Green's function. In the latter case, we have $A(\omega=0)$ vanishing with non-diverging $\Im[\Sigma(\iu\omega_n)]$. In that case, that we also see here, we set $Z$ to zero manually. From this, the MIT $U$ value ($U_{\rm MIT}$) is defined as the lowest $U$ value in which $A(\omega=0)=0$.

The spectral functions, $A(\omega)$, for each impurity were calculated from
$G(\tau)$ using the {\em LineFitAnalyzer} technique of the maximum entropy
analytic continuation method implemented within the {\em MaxEnt} application of
TRIQS \cite{maxent}. The k-resolved spectral functions $A(\textbf{k},\omega)$
were calculated from the analytically continued self-energy.

The effective and correlation subband mass enhancement factors, 1/$Z_{\nu}$ and 1/$Z^{\rm c}_{\nu}$, were calculated from the ratios of the Fermi velocities using 

\begin{equation}
Z^{\rm c}_{\nu} = \frac{v^c_{\rm F}}{v^{\rm QTB}_{\rm F}} 
\end{equation}
and
\begin{equation}
Z_{\nu} = \frac{v^c_{\rm F}}{v^i_{\rm F}}. 
\end{equation}
Here, the Fermi velocities were determined from the gradient of the linearly expanded band dispersions along M-X around ${k}_{\rm F}$ of the DFT+DMFT subbands ($v^c_{\rm F}$), the quantized bands from QTB ($v^{\rm QTB}_{\rm F}$) and the intrinsic (bulk-like) TB bands ($v^i_{\rm F}$). The intrinsic bands were used as they incorporate the effect of renormalization due to strain. Therefore, $Z^{\rm c}_{\nu}$ and $Z_{\nu}$ describe the effect of renormalization from correlations and the combination of correlations and confinement (band) effects respectively.

The DFT+DMFT subband energy centers, $E_{\nu,\mathbf{k}}$, were calculated by using

\begin{equation}
E_{\nu,\mathbf{k}}=\epsilon_{\nu,\mathbf{k}} - \mu + \Re[\Sigma_{\nu}(\mathbf{k},\omega=E_{\nu,\mathbf{k}})], \label{e:Enu}
\end{equation}
where $\epsilon_{\nu,\mathbf{k}}$ is the DFT energy, $\mu$ is the chemical potential and
$\Re[\Sigma_{\nu}(\mathbf{k},\omega)]$ is the real part of the diagonal upfolded
self-energy elements on the real frequency axis.
The QP lifetime in
the inset of Fig.\ 4 of the main manuscript
was determined from the inverse imaginary part of the analytically continued upfolded
self-energy. Finally, the subband energies at the $\Gamma$ high symmetry point in
Fig.\ 4 of the main manuscript were determined from Eqn.~\ref{e:Enu}.

\begin{figure}[t!]
\centerline{\includegraphics[width=1.0\linewidth]{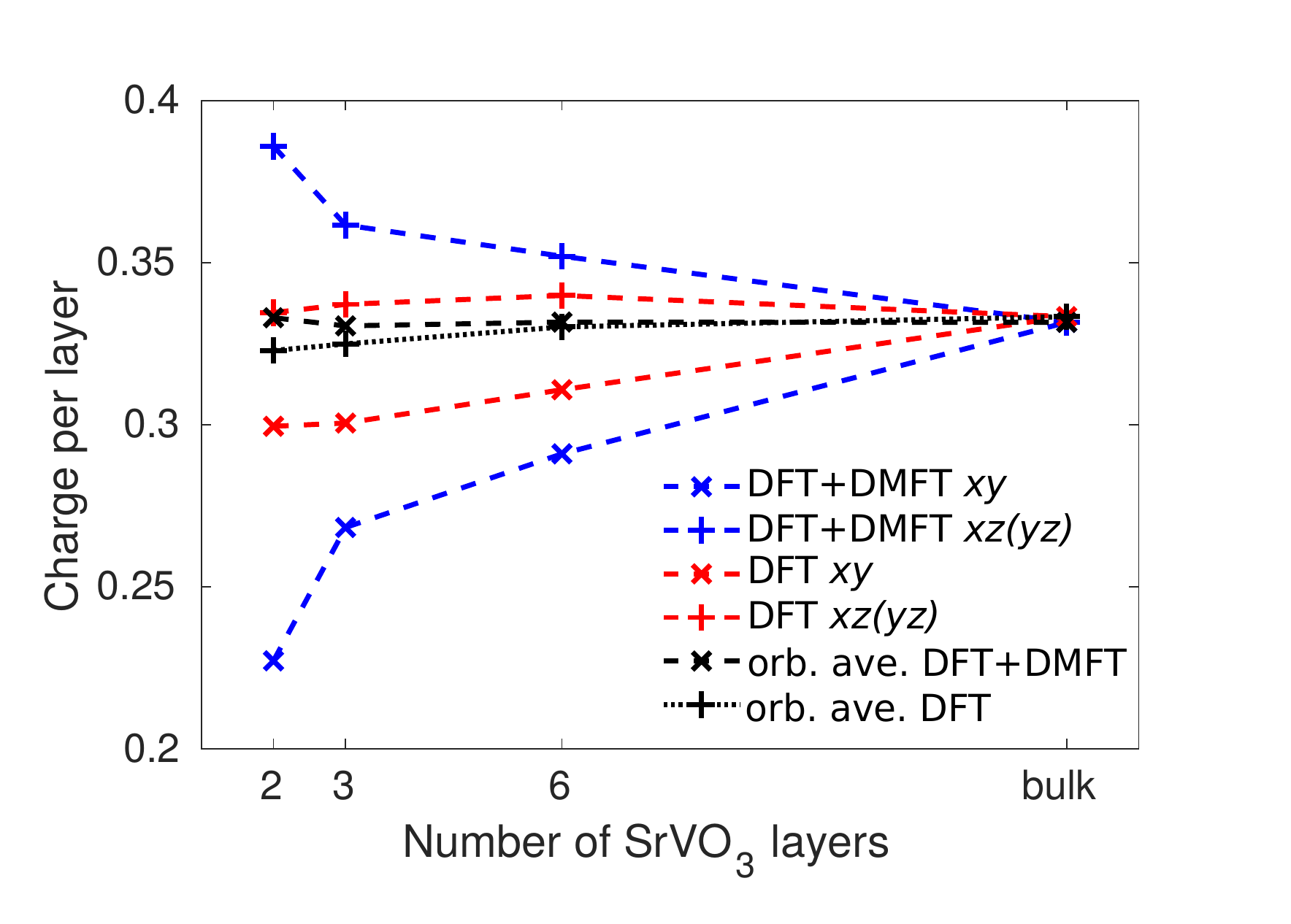}}
\caption{ The averaged orbital charge over all layers from the DFT and DFT+DMFT
Wannier V $xy$ and $xz(yz)$ orbitals for each SL and bulk. This includes charges from DFT, DFT+DMFT and average orbital charge per layer (orb. ave.) for each SL and bulk.}
\label{f:chg}
\end{figure}

\subsection{One-shot and FCSC DFT+DMFT results}

The main manuscript presents fully charge self-consistent (FCSC)
DFT+DMFT calculations.
Here, we present the results of one-shot (OS) DFT+DMFT for comparison. Overall,
charge self-consistency slightly adjusts some details of the results, but the
main conclusions of our study are already present in one-shot calculations.

\begin{figure}[t!]
\centerline{\includegraphics[width=1.0\linewidth]{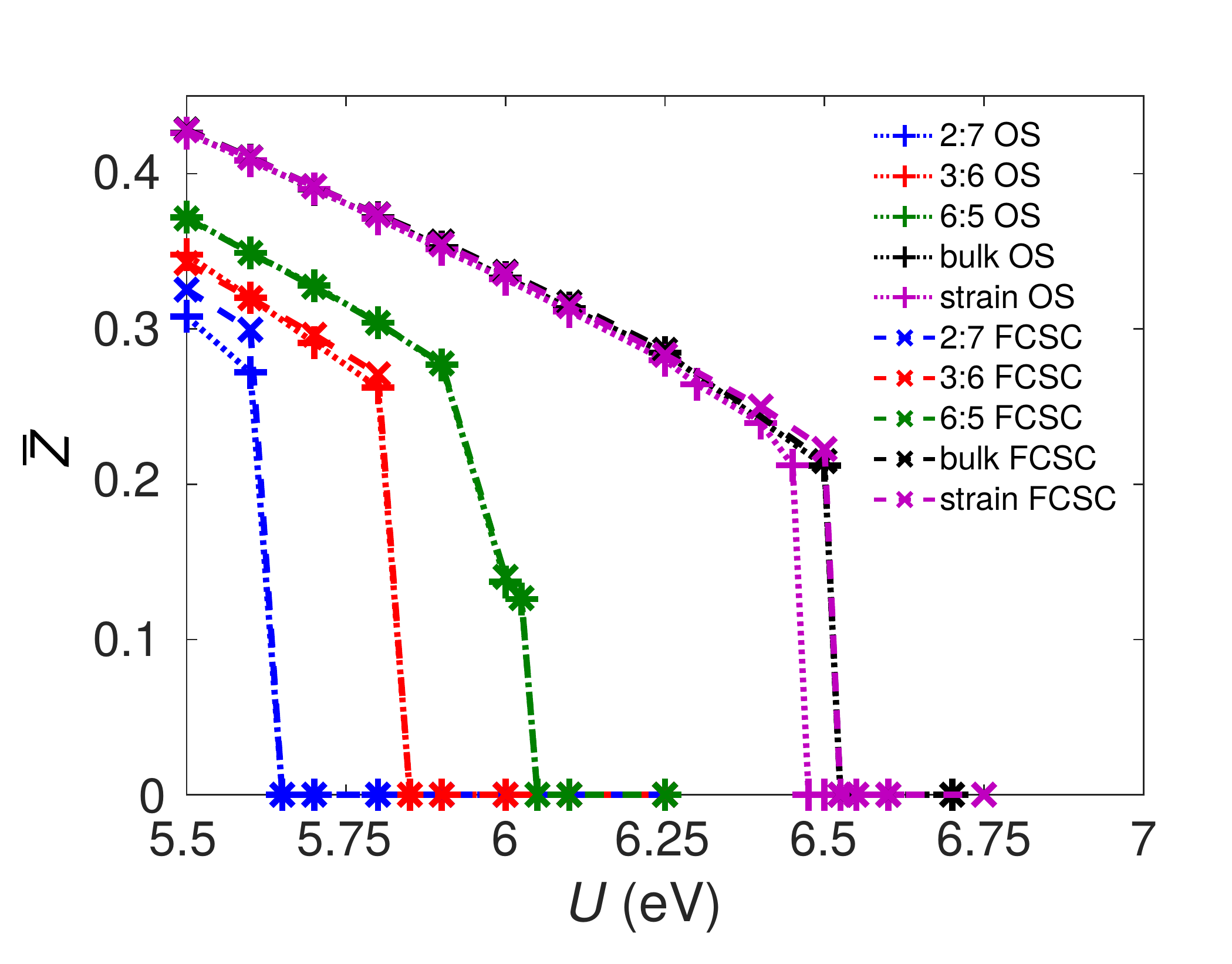}}
\caption{ The comparison of the orbitally-averaged quasiparticle residue,
$\bar{Z}$, between the one-shot (OS and fully charge self-consistent (FCSC) DFT+DMFT
methods. The plot lines are guides to the eye.}
\label{f:Zcomp}
\end{figure}

Figure \ref{f:Uimp} shows the $U$-dependent MIT
for each SL and different DFT+DMFT methods. The behavior of the OS and FCSC
calculations is very similar, exhibiting a similar $U_{\rm MIT}$ with similar
characteristics, e.g.\ $A(\omega=0)$ and $Z$. Some
differences are observed in the orbital polarization between the two methods,
whereby the polarization is somewhat suppressed in the FCSC calculation compared
with OS. This behavior, most notable for the 2:7 SL, is consistent with other
studies \cite{bhandary2016,schuler2018,hampel2019}, and is caused by the charge
redistribution with the rest of the system at the DFT stage. This trend from 2:7
to bulk is also seen in Fig.\ \ref{f:Zcomp} for the orbitally-averaged values of
$Z$, where
there are some differences in $\bar{Z}$ for the 2:7 SL between
OS and FCSC, but the bulk values are very similar. 

An important note to make about Fig.\ \ref{f:Uimp} is that $Z$ at the interface
(imp 1) for the $xz(yz)$ orbitals tends to zero first for each SL. This
suggests that the weight
from the $xz(yz)$ QP peak depletes first. Therefore, when the
interface $xz(yz)$ QP state has been fully depleted, this causes the
SL to transition into the insulating state. From this, the interface between the
oxides has a strong influence on the MIT. The $A(\omega=0)$ for imp 1 of the 6:5
also tends to zero which strengthens the argument for at least that SL.

\begin{figure}[t!]
\centerline{\includegraphics[width=1.0\linewidth]{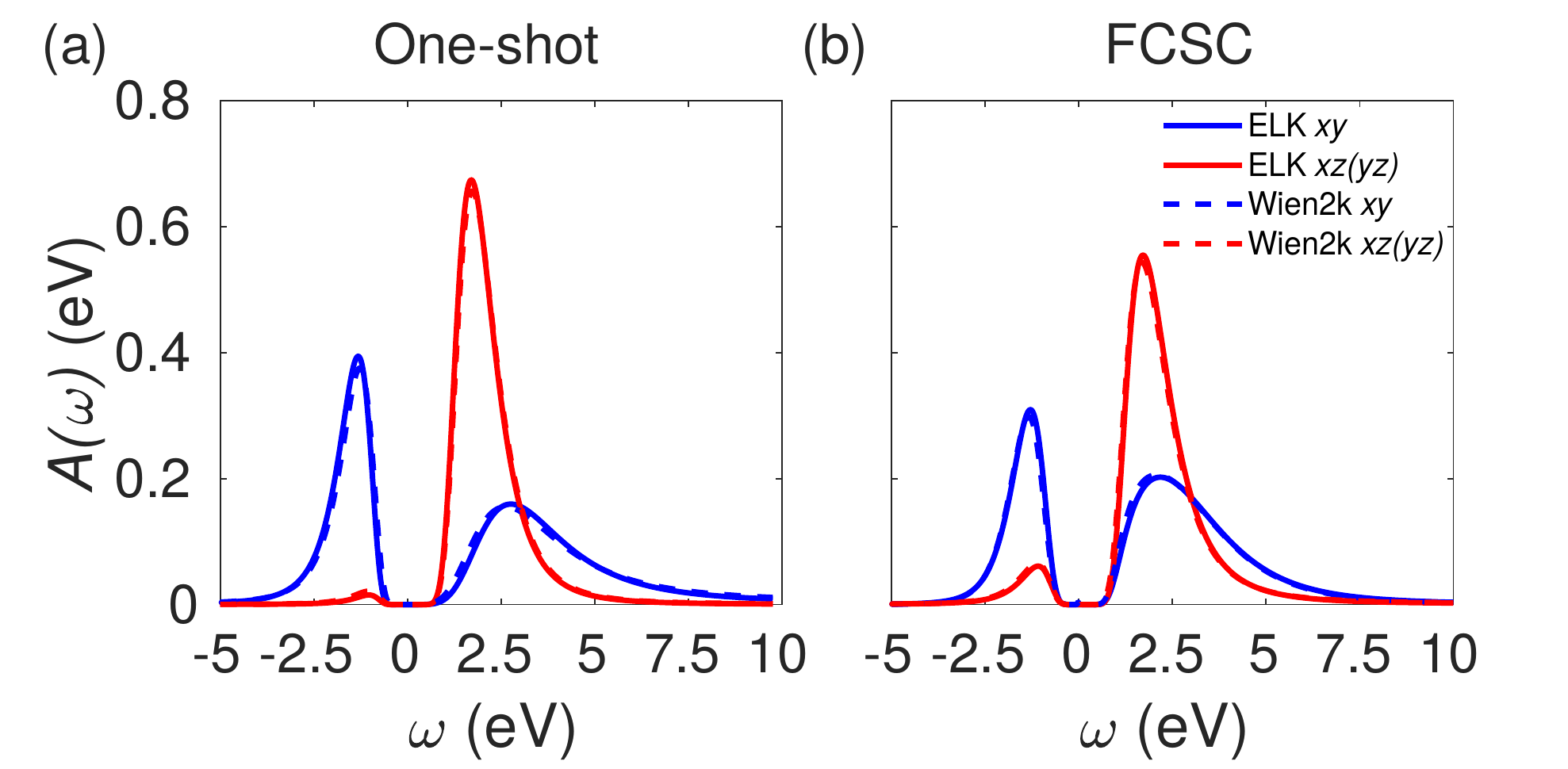}}
\caption{The orbital A($\omega$) comparisons between the one shot and fully
charge self consistent (FCSC) DFT+DMFT methods from different input DFT codes
for mono layer SrVO$_3$.}
\label{f:mono}
\end{figure}

\section{Other Significant SL FCSC Results}

Fig.~\ref{f:Aw} shows the A($\omega$) of each correlated impurity orbital in 
each SL at $U=5.7$ eV (the value used in the theoretical-experimental 
comparisons). It is evident that the 2:7 is insulating and the 3:6 and 6:5 
is metallic from the absence/presence of the QP peak at the Fermi level. 
There are sharp features in the QP peaks around the Fermi level for the 3:6 
and 6:5 SLs. These features are often attributed to spurious noise from the 
analytic continuation procedure, however, it may not be the case here due 
to the quantized bands being present around the Fermi level. The peak 
position of the Hubbard bands (notably the UHB) are closer in energy to the 
Fermi level for the interface layer (impurity 1) compared to the other 
layers for the 3:6 and 6:5. This is another indication that the interface 
layer is more correlated compared to the other layers.

The splitting of the orbital degeneracy strongly
effects the polarization of the orbital
charge as shown in Fig.~\ref{f:chg}. It is interesting to note that
the reduction of the number of layers
significantly increases the charge in the xz(yz) orbitals, which appear to 
tend to half filling (whereas the xy orbitals are tending towards zero 
charge). This is a likely consequence of these orbitals trying to reduce the 
potential energy, analogous to what is seen in the previous mono layer calculations.
The reduction in the orbitally-averaged
DFT charge with lower SrVO$_3$ layers is likely due to hybridization with Ti
at the interface.

\subsection{Elk-TRIQS interface test: monolayer SrVO$_3$}

The results presented
used an in-house interface between {\sc Elk} and TRIQS, so this section presents
comparison results between {\sc Elk} and Wien2k inputs into TRIQS to show that
the interface works for a similar system, namely monolayer SrVO$_3$.  The
monolayer SrVO$_3$ calculation was set up in the same way and using the same
parameters as in Ref.\ \onlinecite{schuler2018}. Figure \ref{f:mono} shows the
comparison between $A(\omega)$ calculated from the different DFT code inputs.
This comparison shows excellent agreement between the different inputs for the
different DFT+DMFT methods.
This test shows that the interface used between Elk
and TRIQS is able to reliably perform DFT+DMFT calculations.

\section{Effects of strain}
We performed volume conserving strain calculations on bulk SrVO$_3$ to
investigate the effect of CF has on the MIT while the bandwidths of
the $t_{2g}$ orbitals are approximately unchanged. Compressive strain of 1\% was
applied along the $c$-axis; the other axes were tensively strained to
conserve volume compared with the bulk. This strain was chosen to yield a CF
splitting of 53~meV, slightly larger than but comparable to the CF splitting
of the 2:7 SL. The strained FCSC $U_{\text{MIT}}$ is
approximately 6.525~eV, the same as for the bulk. The OS
strained calculation had a slightly lower
$U_{\text{MIT}}$ of 6.475 eV. Due to the small
change on $U_{\text{MIT}}$, the CF splitting is insufficient to
cause the MIT in these SLs.

\bibliography{ref}